\documentclass[5p,twocolumn,preprint,times]{elsarticle}
\usepackage{natbib}
\usepackage{graphicx}
\usepackage{epstopdf, epsfig}
\usepackage{amsmath}
\usepackage{dutchcal}
\usepackage{tikz}
\usepackage{babel}[english]
\usetikzlibrary{shapes,arrows}
\usepackage{array}
\usetikzlibrary{shapes.geometric}
\newcolumntype{L}[1]{>{\raggedright\let\newline\\\arraybackslash\hspace{0pt}}m{#1}}
\newcolumntype{C}[1]{>{\centering\let\newline\\\arraybackslash\hspace{0pt}}m{#1}}
\newcolumntype{R}[1]{>{\raggedleft\let\newline\\\arraybackslash\hspace{0pt}}m{#1}}
\usepackage{enumerate}
\usepackage{dcolumn}
\usepackage{bm}
\usepackage{xcolor}
\usepackage{makecell}

\usepackage{multicol}

\usepackage[colorlinks=true,urlcolor=blue]{hyperref} 

\usepackage{xspace}
\usepackage{booktabs}
\usepackage{multirow}
\usepackage{adjustbox}
\usepackage{titlesec}
\usepackage{relsize}
\usepackage{colortbl}
\usepackage{steinmetz}
\usepackage{scalerel}
\usepackage{amssymb}
\usepackage{float}
\usepackage{subcaption}
\usepackage{caption}

\usepackage{lineno}

\usepackage{siunitx}

\usepackage{nicefrac}

\journal{Energy}

\newcommand*{\Reynolds}{\mathrm{Re}}

\newcommand*{\Rayleigh}{\mathrm{Ra}}

\newcommand*{\Jakob}{\mathrm{Ja}}

\usepackage{nomencl} 
\makenomenclature
\renewcommand*\nompreamble{\begin{multicols}{2}}
	\renewcommand*\nompostamble{\end{multicols}}
\usepackage{framed} 
\usepackage{etoolbox} 
\renewcommand\nomgroup[1]{%
	\item[\bfseries
	\ifstrequal{#1}{A}{Letter symbols}{%
		\ifstrequal{#1}{B}{Greek symbols}{%
			\ifstrequal{#1}{C}{Subscripts and superscripts}{}}}%
	]}
\usepackage{etoolbox}
\renewcommand\nomgroup[1]{%
  \item[\bfseries
  \ifstrequal{#1}{L}{Letter symbols}{%
  \ifstrequal{#1}{G}{Greek symbols}{%
  \ifstrequal{#1}{N}{Non-dimensional numbers}{%
  \ifstrequal{#1}{A}{Abbreviations}{%
  \ifstrequal{#1}{S}{Subscripts and superscripts}{}}}}}%
]}

\begin{document}

\begin{frontmatter}

    \title{Physics-integrated neural network modeling of heat and mass transfer in cryogenic liquid storage under static and dynamic conditions}
    \author[1,2]{P. A. Marques\corref{cor1}%
    }
    \ead{pedro.marques@vki.ac.be}
    
    \author[1,3]{S. A. Ahizi}
    \author[1,3,4]{M. A. M\'endez}
    \cortext[cor1]{Corresponding author}
    \address[1]{von Karman Institute for Fluid Dynamics, Waterloosesteenweg 72, Sint-Genesius-Rode, Belgium}
    \address[2]{Transferts, Interfaces Et Procédés (TIPs), Université Libre de Bruxelles, Av. Franklin Roosevelt 50, Brussels, 1050, Belgium}
    \address[3]{Aerospace Engineering Research Group, Universidad Carlos III de Madrid, Av. de la Universidad 30, 28911 Leganés, Spain}
    \address[4]{Aero-Thermo-Mechanics Laboratory, École Polytechnique de Bruxelles, Université Libre de Bruxelles, Av. Franklin Roosevelt 50, Brussels, 1050, Belgium}
    \date{\today}
    
    \begin{abstract}
Accurate system-level prediction of cryogenic liquid storage remains challenging because reduced-order models rely on regime-dependent closures for unresolved heat and mass transfer, particularly under sloshing. We present a physics-integrated neural-network framework that combines a conservation-based zero-dimensional nodal model with data-driven closures for wall and interfacial heat transfer, phase change, pressurant quality, and liquid thermal-boundary-layer evolution. Thermal stratification and mixing within the liquid are represented through a first-order dynamical model for the boundary-layer thickness, while four operating-regime-specific neural networks infer the closure parameters for self-pressurization and relaxation, active pressurization, venting, and lateral sloshing.

The framework was identified and evaluated using a dedicated database of 48 multi-stage cryogenic-tank experiments conducted in an optically accessible facility operated with liquid nitrogen. The experiments combine controlled wall heating, vapor injection and evacuation, and forced lateral sloshing, thereby covering both slowly evolving thermal states and strongly transient operating conditions. Generalization was assessed through experiment-level \(K\)-fold cross-validation, while an entropy-production penalty was included in the loss function to discourage violations of the second law of thermodynamics.

Across the investigated cross-validation configurations, the global normalized root-mean-square error remained between \(2.5\%\) and \(2.8\%\), with sloshing representing the most demanding regime. A final model trained on the complete database reconstructed the experiments with a global error of \(1.6\%\). These results demonstrate that the proposed framework can extract physically interpretable and computationally efficient closure laws from a limited but information-rich experimental database, providing a promising model for system-level prediction and control of cryogenic storage tanks.
    \end{abstract}

\end{frontmatter}


\nomenclature[L002]{$A$}{area, ${\rm m^2}$}
\nomenclature[L003]{$A_e$}{forcing (excitation) amplitude, ${\rm m}$}
\nomenclature[L004]{$A_s$}{measured interface (wave) amplitude, ${\rm m}$}
\nomenclature[L005]{$\mathcal{A}$}{augmented loss functional}
\nomenclature[L006]{$\mathcal{b}_i$}{frequency offset parameter, ${-}$}
\nomenclature[L007]{$b^{(k)}$}{Bernoulli sampling variable at iteration $k$, ${-}$}
\nomenclature[L008]{$c_p$}{isobaric specific heat, ${\rm J/(kgK)}$}
\nomenclature[L009]{$c_w$}{specific heat of the solid nodes, ${\rm J/(kgK)}$}
\nomenclature[L010]{$c_\delta$}{thermal boundary-layer growth closure parameter, ${-}$}
\nomenclature[L011]{$D_p$}{pressurant pipe diameter, ${\rm m}$}
\nomenclature[L012]{$f$}{forward (physics-based) model function}
\nomenclature[L013]{$f_e$}{excitation frequency, ${\rm Hz}$}
\nomenclature[L014]{$f_{1,1}$}{first asymmetric natural frequency, ${\rm Hz}$}
\nomenclature[L015]{$g$}{closure parameter function}
\nomenclature[L016]{$\bm{g}$}{gravitational acceleration, ${\rm m/s^2}$}
\nomenclature[L017]{$\bm{g}_k$}{gradient of the augmented loss at iteration $k$}
\nomenclature[L018]{$H$}{internal tank height, ${\rm m}$}
\nomenclature[L019]{$H_l$}{liquid fill level, ${\rm m}$}
\nomenclature[L020]{$h$}{heat transfer coefficient, ${\rm W/(m^2K)}$}
\nomenclature[L021]{$\mathcal{h}$}{mass-specific enthalpy, ${\rm J/kg}$}
\nomenclature[L022]{$\mathcal{J}$}{loss functional}
\nomenclature[L023]{$K$}{number of cross-validation folds, ${-}$}
\nomenclature[L024]{$K_p$}{entropy-production penalty weighting factor}
\nomenclature[L025]{$K_{q,u},K_{q,s}$}{heater contact accommodation coefficients, ${-}$}
\nomenclature[L026]{$L_{j-k}$}{characteristic length between nodes $j$, $k$, ${\rm m}$}
\nomenclature[L027]{$L_p$}{pressurant pipe length, ${\rm m}$}
\nomenclature[L030]{$\ell,\ \ell_a$}{integrands of the loss and augmented loss functionals}
\nomenclature[L031]{$\mathcal{L}_v$}{latent heat of vaporization, ${\rm J/kg}$}
\nomenclature[L032]{$m$}{mass, ${\rm kg}$}
\nomenclature[L033]{$\dot{m}$}{mass flow rate, ${\rm kg/s}$}
\nomenclature[L034]{$\bm{m}_k,\bm{v}_k$}{AdamW first- and second-moment estimates}
\nomenclature[L035]{$\mathcal{M}_j^{(K)}$}{model trained on fold $j$ of the $K$-fold partition}
\nomenclature[L036]{$N_b$}{number of shorter segments sampled per time series, ${-}$}
\nomenclature[L037]{$N_d$}{number of time series per mini-batch, ${-}$}
\nomenclature[L038]{$N_h$}{fill-level correction iterations, ${-}$}
\nomenclature[L039]{$N_{\mathcal{B}}$}{number of trajectories integrated per mini-batch, ${-}$}
\nomenclature[L040]{$n_s,n_w,n_x,n_\vartheta,n_{\mathcal{z}}$}{dimensions of the state, weight, exogenous-input, closure-parameter and feature vectors, ${-}$}
\nomenclature[L041]{$n_t$}{number of time samples, ${-}$}
\nomenclature[L042]{$p$}{pressure, ${\rm Pa}$}
\nomenclature[L043]{$\mathcal{p}_b$}{probability of sampling shorter segments, ${-}$}
\nomenclature[L044]{$\mathcal{p}_d$}{dropout probability, ${-}$}
\nomenclature[L045]{$\mathcal{P}_\sigma$}{entropy-production penalty term}
\nomenclature[L046]{$\dot{Q}$}{heat transfer rate, ${\rm W}$}
\nomenclature[L047]{$R$}{internal tank radius, ${\rm m}$}
\nomenclature[L049]{$r,\theta,z$}{cylindrical coordinates, ${\rm m}$, ${\rm rad}$, ${\rm m}$}
\nomenclature[L050]{$\bm{s},\check{\bm{s}},\hat{\bm{s}}$}{virtual, measured and normalized state vectors}
\nomenclature[L051]{$\mathcal{S}$}{entropy, ${\rm J/K}$}
\nomenclature[L052]{$\mathcal{s}$}{mass-specific entropy, ${\rm J/(kgK)}$}
\nomenclature[L053]{$s_w$}{wall thickness, ${\rm m}$}
\nomenclature[L054]{$T$}{temperature, ${\rm K}$}
\nomenclature[L055]{$t$}{time, ${\rm s}$}
\nomenclature[L056]{$t_o,t_e$}{episode start and end times, ${\rm s}$}
\nomenclature[L057]{$\mathcal{T}_o$}{episode duration, ${\rm s}$}
\nomenclature[L058]{$\Delta t_b,\Delta t_{\mathrm{tot}}$}{segment and total experiment durations, ${\rm s}$}
\nomenclature[L059]{$\mathcal{U}$}{internal energy, ${\rm J}$}
\nomenclature[L060]{$\mathcal{u}$}{mass-specific internal energy, ${\rm J/kg}$}
\nomenclature[L061]{$V$}{volume, ${\rm m^3}$}
\nomenclature[L062]{$\dot{W}$}{work rate, ${\rm W}$}
\nomenclature[L063]{$\bm{W}$}{diagonal state weighting matrix}
\nomenclature[L064]{$\bm{w}$}{trainable closure-parameter weights}
\nomenclature[L065]{$\bm{\mathcal{X}}$}{vector of exogenous inputs}
\nomenclature[L066]{$\bm{\mathcal{z}}$}{scenario-specific network input feature vector}
\nomenclature[L067]{$\Delta z_{h,s}$}{vertical extent of the lateral heating element, ${\rm m}$}
 
\nomenclature[G001]{$\alpha$}{thermal diffusivity, ${\rm m^2/s}$}
\nomenclature[G002]{$\beta$}{volumetric thermal expansion coefficient, ${\rm K^{-1}}$}
\nomenclature[G003]{$\beta_1,\beta_2$}{AdamW moment decay rates, ${-}$}
\nomenclature[G004]{$\gamma_P,\gamma_B,\gamma_S,\gamma_V$}{operating-regime indicator functions, ${-}$}
\nomenclature[G005]{$\delta_T$}{thermal boundary-layer thickness, ${\rm m}$}
\nomenclature[G007]{$\delta_{j-k}$}{conductive wall thickness between nodes $j$ and $k$, ${\rm m}$}
\nomenclature[G008]{$\bm{\varepsilon}$}{normalized prediction error}
\nomenclature[G009]{$\varepsilon_{\mathrm{rel}},\varepsilon_{\mathrm{abs}}$}{relative and absolute solver tolerances, ${-}$}
\nomenclature[G011]{$\eta$}{learning rate}
\nomenclature[G012]{$\bm{\vartheta}$}{vector of closure parameters}
\nomenclature[G013]{$\kappa$}{thermal conductivity, ${\rm W/(mK)}$}
\nomenclature[G014]{$\bm{\lambda}$}{adjoint variable}
\nomenclature[G015]{$\lambda_w$}{weight-decay coefficient}
\nomenclature[G016]{$\mu$}{dynamic viscosity, ${\rm Pa\,s}$}
\nomenclature[G017]{$\nu$}{kinematic viscosity, ${\rm m^2/s}$}
\nomenclature[G019]{$\rho$}{density, ${\rm kg/m^3}$}
\nomenclature[G020]{$\sigma$}{surface tension, ${\rm N/m}$}
\nomenclature[G021]{$\dot{\sigma}_s$}{entropy production rate, ${\rm W/K}$}
\nomenclature[G022]{$\varphi_w,\varphi_f$}{internodal solid and fluid flags, ${-}$}
\nomenclature[G023]{$\varphi_j$}{phase sign function ($\varphi_v=1$, $\varphi_l=-1$), ${-}$}
\nomenclature[G024]{$\chi_p$}{pressurant vapor quality (fraction), ${-}$}
\nomenclature[G025]{$\psi_\delta$}{boundary-layer disruption parameter, ${-}$}
\nomenclature[G026]{$\Omega$}{total thermal resistance, ${\rm K/W}$}
\nomenclature[G027]{$\omega_e$}{excitation angular frequency, ${\rm rad/s}$}
\nomenclature[G028]{$\omega_{1,1}$}{first asymmetric natural angular frequency, ${\rm rad/s}$}
 
 
\nomenclature[S001]{$a$}{ambient}
\nomenclature[S002]{$b$}{boundary crossed by a mass flux}
\nomenclature[S006]{eq.}{equilibrium}
\nomenclature[S007]{$f$}{fluid}
\nomenclature[S008]{$h$}{heating element}
\nomenclature[S009]{$i$}{interface}
\nomenclature[S010]{$j,k$}{node indices}
\nomenclature[S011]{$l$}{liquid}
\nomenclature[S013]{$p$}{pressurant}
\nomenclature[S014]{ph}{phase change}
\nomenclature[S017]{sat}{saturation}
\nomenclature[S018]{$sr$}{cryostat sample space (residual conditions)}
\nomenclature[S020]{$v$}{vapor}
\nomenclature[S021]{$w,u$}{tank upper cover}
\nomenclature[S022]{$w,s$}{tank side-wall}
\nomenclature[S004]{$w,d$}{tank downward cover}

\renewcommand*\nompreamble{}
\renewcommand*\nompostamble{}
\printnomenclature




\section{Introduction}
\label{sec:introduction}
The long-term storage and handling of cryogenic propellants are central to future space missions and to decarbonization pathways in aviation and shipping \cite{hansen_cryogenic_2020, atilhan_green_2021, baroutaji_comprehensive_2019}. However, their very low operating temperatures, on the order of 20~K for LH$_2$ and 120~K for LCH$_4$, together with the strongly coupled heat and mass-transfer processes inside the tank, make thermal management and predictive modeling particularly challenging.

Modeling approaches for cryogenic propellant tanks range from high-fidelity computational fluid dynamics (CFD) to reduced-order lumped-parameter models. Classical finite-volume (FV) CFD approaches can resolve the coupled transport of momentum, heat, and mass, including phase change and free-surface dynamics \cite{van_foreest_modeling_2014, konopka_analysis_2016, kassemi_validation_2018, scheufler_twophaseflow_2023}. However, their computational cost limits long-duration and system-level applications, while uncertainties in geometry, boundary conditions, turbulence modeling, and mesh resolution often require empirical accommodation factors to recover experimental observations \cite{kartuzova_modeling_2011, stewart_self-pressurization_2016, kassemi_effect_2016, agui_modeling_2015}. Meshless approaches, including Smooth Particle Hydrodynamics (SPH), Moving Particle Semi-implicit (MPS), Lagrangian Differencing Dynamics (LDD), and Finite Pointset Methods (FPM), offer attractive and more readily parallelizable alternatives \cite{lind_2020, zhang_2022, MPS_2018, basic_2022, suchde_2018}, but their non-isothermal extension to cryogenic applications remains at an early stage \cite{lee_modeling_2023, wang_2020, diaz_2023, liu_2021}.

At the opposite end of the computational-cost spectrum, nodal models represent the tank through coupled 0D, 1D, or 2D control volumes and are therefore well suited to long-duration prediction, parametric analysis, and control-oriented applications \cite{bolshinskiy_tank_2017, daigle_model-based_2011, majumdar_two-dimensional_2025}. They have been used to predict boil-off, thermal stratification, and support component design \cite{petitpas_boil_off_2018, migliore_non-equilibrium_2017, kalikatzarakis_model_2022, zheng_parametric_2021, osipov_dynamic_2008, van_foreest_modeling_2010, seo_analysis_2010, daigle_temperature_2013, joseph_effect_2017}. Their main limitation lies in the closure of unresolved heat and mass-transfer processes: reliable correlations remain scarce under dynamic conditions involving sloshing, and empirical correction factors are often required even in static regimes \cite{grotle_dynamic_2018, marques_experimental_analysis_2023, al_ghafri_modelling_2022, wang_dynamic_2021, hartwig_numerical_2016}. This motivates data-driven closure strategies that retain the efficiency and conservation structure of nodal models while learning the unresolved transfer processes from observations.

Advances in machine learning, optimization, and data assimilation have enabled data-driven methods to reconcile imperfect models with observations. Their impact is already visible in applications ranging from weather forecasting, where neural forecasters can rival state-of-the-art numerical models at substantially lower computational cost \cite{pathak2023fourcastnet, price_probabilistic_2025, bodnar_foundation_2025}, to turbulence modeling, where they have been used to correct Reynolds stresses, construct subgrid closures, and combine specialized models across flow regimes \cite{wang2017piml, wu2018piml, duraisamy2019age, oulghelou2025_data_driven_turbulence,Fiore2022,Dominique2022}.

In cryogenic propellant management, data-driven methods have similarly been used to augment, correct, and replace conventional thermo-hydraulic models \cite{cheng_adaptive_2025, alipour_bonab_artificial_2025, zhu_theory_informed_2023}. Hybrid approaches retain a low-order physical model while learning state-dependent closure relations from data \cite{marques_real_time_2024}. Other strategies instead learn correction mappings between reduced-order and higher-fidelity predictions. For example, Cheng et al. \cite{cheng_data_driven_2025} trained an artificial neural network on concurrent CFD and lumped-model simulations, then used it to correct the inexpensive model during long-duration predictions. Purely data-driven models have also been developed, including neural-network predictors of pressure evolution during LH$_2$ self-pressurization based on tank geometry and operating conditions \cite{rahman_prediction_2024}. To mitigate the scarcity of experimental data, Qu et al. \cite{qu_data_driven_2024} pre-trained a neural network on synthetic numerical data before fine-tuning it on a limited experimental dataset. These studies demonstrate the potential of machine learning for cryogenic storage, while also highlighting the need for approaches that preserve physical consistency and remain effective when data is limited.

The present work builds on \cite{marques_real_time_2024}, where an artificial neural network (ANN) was embedded in a thermal nodal model of a cryogenic tank to represent unresolved heat and mass-transfer processes as functions of the tank state, with its parameters identified through adjoint-based optimization \cite{errico_what_1997, cao_adjoint_2003}. The present contribution extends this framework in three main directions. First, the closure parameters are inferred from experimental data collected in a cryogenic facility across operating conditions spanning self-pressurization, active-pressurization, venting, and sloshing. Second, the physical model is augmented to represent liquid thermal stratification and destratification through the evolution of the thermal boundary-layer thickness beneath the gas-liquid interface. Third, an entropy-production penalty is introduced in the loss function to discourage negative entropy production and promote consistency with the second law of thermodynamics.


Relying on real-world observations, this work assesses the model's learning performance and generalizability under varying data availability. To quantify the out-of-fold performance, we employ a $K$-fold cross-validation approach with varying subsets ($K=2, 3, 4, 5$). Finally, a comprehensive model is trained on the full database to reconstruct all available experimental data. The performance of this final model is assessed through global error metrics and its ability to reproduce individual thermodynamic state variables and capture the distinct physical responses associated with each operating scenario.

The remainder of the article is organized as follows. Section \ref{sec:problem_set} introduces the problem, modeling assumptions, and variables of interest. Section \ref{sec:modeling} presents the thermal model, the parametrization of its closure relations, and their identification from experimental observations. Section \ref{sec:dataset_description} describes the experimental setup, database, validation strategy, and data-processing procedures. The results are presented and discussed in Section \ref{sec:results_and_discussion}, followed by the conclusions and future perspectives in Section \ref{sec:conclusions}.

\section{Problem statement}
\label{sec:problem_set}
\begin{figure}[!htb]
	\centering
	\includegraphics[width=1\linewidth]{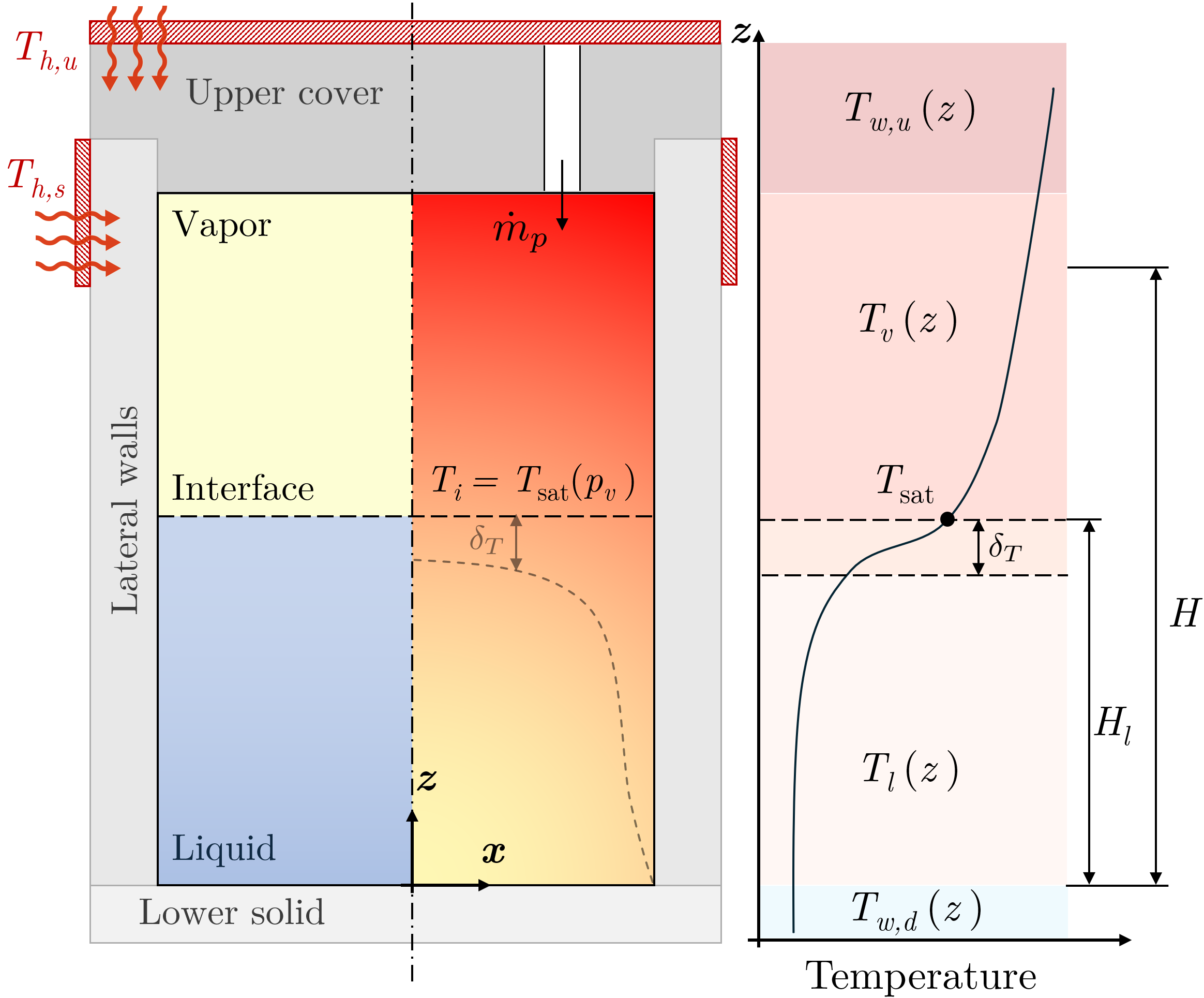}
	\caption{Schematic of the cryogenic fuel tank and the corresponding lumped control volumes used in the thermal model. The environmental heat fluxes are denoted by $\dot q_a$, while the pressurant flow rate is $\dot m_p$. This quantity is positive during pressurization and negative during venting. The right-hand side shows a representative thermal profile along the tank centerline ($r=0$).}
	\label{fig:problem_set}
\end{figure}

We consider a single-species system comprising a cryogenic liquid and its vapor in an upright cylindrical tank subject to an acceleration field $\bm{g}=(g_r,g_\theta,g_z)$. The configuration and main variables are illustrated in Figure \ref{fig:problem_set}. The tank has internal radius $R$ and height $H$, and is partially filled with liquid to a depth $H_l$. Its side-wall, of thickness $s_{w,s}$, is sealed by flat upper and downward covers of thicknesses $s_{w,u}$ and $s_{w,d}$, respectively. The upper cover is connected to a pipeline through which the vessel can be pressurized with superheated vapor at mass flow rate $\dot{m}_p>0$, upstream pressure $p_u$, and temperature $T_u$, or vented at an evacuation flow rate $\dot{m}_p<0$. The downward cover remains closed.

The tank is decomposed into fluid and solid control volumes. Two control volumes represent the vapor and liquid phases, while the solid structure is divided into the upper and bottom covers and the upper and lower portions of the side-wall. The subscripts ($v$), ($l$), ($w,u$), ($w,d$), ($w,s,u$), and ($w,s,d$) denote quantities associated with the vapor, liquid, upper cover, bottom cover, and upper and lower lateral-wall portions, respectively. Ambient quantities are identified by the subscript ($a$).

\begin{figure*}[!htb]
    \centering
    \includegraphics[width=0.8\linewidth]{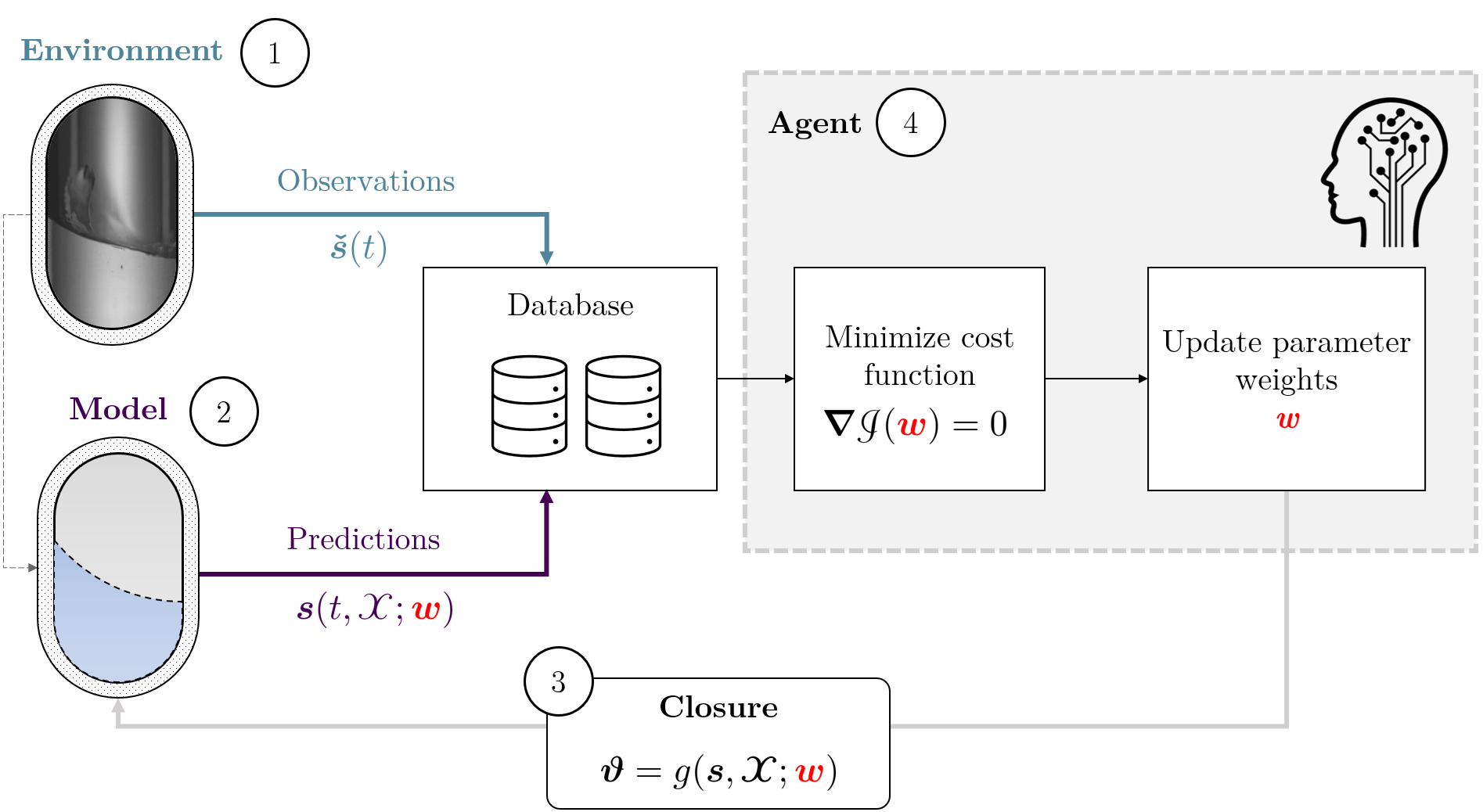}
    \caption{Overview of the proposed digital-twinning framework: (1) the environment, representing the physical system from which observations are collected; (2) the digital twin, which predicts the system evolution through a set of governing equations; (3) the closure parameters, expressed as parametric functions of the state, tank inputs, and trainable weights; and (4) the optimization agent, which links the environment and the digital twin by adjusting the closure parameters to minimize the discrepancy between model predictions and observations.}
    \label{fig:framework_overview}
\end{figure*}

The tank pressure is denoted by $p_v$, while the temperature distributions in the vapor and liquid phases are $T_v(r,z,\theta)$ and $T_l(r,z,\theta)$, respectively. For a single-species system, the interface temperature corresponds to the saturation temperature, $T_i = T_{\mathrm{sat}}(p_v)$, as dictated by thermodynamic equilibrium \cite{moran_fundamentals_1995}. The temperatures of the upper cover, side-wall, and downward cover are denoted by $T_{w,u}(r,z,\theta)$, $T_{w,s}(r,z,\theta)$, and $T_{w,d}(r,z,\theta)$, respectively.

The tank is installed inside a vacuum-insulated cryostat at residual pressure $p_{sr}$ and average temperature $T_{sr}$. The external heat flux $\dot{q}_a$, arising from residual-gas conduction through the vacuum insulation and radiative exchange between the tank and cryostat walls, enters through the tank walls and progressively induces thermal stratification, with the upper fluid regions becoming warmer than the lower ones \cite{chin_analytical_1965}. This stratification results primarily from buoyancy-driven convection and the much lower thermal capacity of the vapor relative to the liquid, $m_v c_{p,v} \ll m_l c_{p,l}$, where $m_v$ and $m_l$ are the vapor and liquid masses, and $c_{p,v}$ and $c_{p,l}$ are their specific heat capacities at constant pressure. A vertical liquid-side thermal layer of thickness $\delta_{T,z}$ consequently develops beneath the saturated free surface, separating the warmer interfacial region from the subcooled bulk \cite{hasan_self-pressurization_1991}. In addition, a radial thermal boundary layer of thickness $\delta_{T,r}$ develops from the inner wall through conductive and buoyancy-driven heat transfer. Since this work focuses predominantly on the impact of the vertical thermal boundary layer, we write $\delta_T \equiv \delta_{T,z}$ for the remainder of the manuscript.

The outer surface of the upper cover is maintained at a uniform temperature $T_{h,u}(t)$ by a temperature-controlled heater. Similarly, the upper portion of the lateral wall, spanning a vertical extent $\Delta z_{h,s}$, is maintained at a prescribed temperature $T_{h,s}(t)$. With the main variables defined, the following section introduces the operating scenarios considered in this work.

\subsection{Scenarios of interest}
\label{sec:scenarios_of_interest}
The thermo-hydraulic evolution of cryogenic storage tanks depends strongly on the operating conditions. Four representative scenarios are considered: (I) self-pressurization, (II) active-pressurization, (III) free venting, and (IV) lateral sloshing. These are briefly summarized as follows:

\begin{enumerate}[I.]
    \item Self-pressurization: The environmental heat flux $\dot{q}_a$ penetrates the solid and fluid phases. The tank pressure rises due to free-surface evaporation and the increase in vapor temperature \cite{timmerhaus_experimental_1960}. As heat is transported in the system, both thermal layers $\delta_{T,z}$ and $\delta_{T,r}$ grow, yielding a thermally-stratified system warmer in the upper and outer regions of the tank \cite{haug_analytical_1968}.
    \item Active-pressurization: Superheated vapor enters the system with a flow rate $\dot{m}_p$, upstream pressure $p_u$, and upstream temperature $T_u>T_i$. The vapor pressure increases both due to the additional ullage mass injected into the tank and due to the larger enthalpy carried by the incoming pressurant gas \cite{stochl_gaseous_helium_5ft_1970}. These combined mechanisms ensure that the active-pressurization occurs at much smaller time scales than the self-pressurization \cite{petitpas_boil_off_2018}. Thus, the liquid is characterized by a thin $\delta_{T,z}$ with sharp thermal gradients \cite{ludwig_investigations_2014}.
    \item Direct venting: When the tank pressure rises above ambient conditions, either by self- or active-pressurization, free venting allows the ullage to be evacuated until $p_v \approx p_\mathrm{amb}$. During the venting transient, the drop in pressure leaves the liquid superheated with respect to the new saturation temperature, and it evaporates until saturation is recovered. The system thus tends toward a steady state with
    $T_\mathrm{eq.}=T_\mathrm{sat}(p_\mathrm{amb})$, in which the persistent ambient heat input $\dot{q}_a$ drives a continuous evaporation that gradually decreases the fill level.
    
    \item Lateral sloshing: External accelerations, e.g., vibrations, translation, and rotation maneuvers, dynamically excite the free surface \cite{dodge_new_2000} and disrupt the thermo-hydraulic state of the tank \cite{moran_experimental_1994}. The interface motion, with instantaneous wave amplitude $A_s(t)$, induces convective fluxes in the fluid, which trigger thermal mixing and localized phase change. Sloshing waves rising over superheated wall regions promote evaporation and boiling effects \cite{arndt_sloshing_2011}, while condensation occurs when subcooled liquid bulk extracts energy from the warmer fluid near the interface, decreasing $T_\mathrm{sat}$ and, consequently, $p_v(T_\mathrm{sat})$ \cite{moran_experimental_1994, ludwig_pressure_2013, lacapere_experimental_2009}.
\end{enumerate}

While in the first three scenarios the system is studied in static conditions at $1g$, in (IV) the tank experiences a lateral harmonic excitation, with imposed acceleration
\begin{equation}
    a_r = A_e \omega_e^2\sin(\omega_e t)
\end{equation}
\noindent where $A_e$ is the forcing amplitude, and $\omega_e$ is the corresponding frequency. The forcing frequency $\omega_e$ is chosen in the vicinity of the reservoir's first asymmetrical eigenfrequency \cite{dodge_new_2000}
\begin{equation}
    \omega_{1,1} 
    \approx 
    \sqrt{
        \left(\frac{1.841 g}{R} + \frac{\sigma}{\rho_l}\frac{6.242}{R^3}\right)
        \tanh{\left(\frac{1.841 H_l}{R}\right)}
    }\,,
\end{equation}
\noindent
where $\sigma$ is the surface tension and $\rho_l$ is the liquid's density. By exciting the system around $\omega_{1,1}$, three sloshing regimes are observed: (1) planar waves, (2) chaotic motion, and (3) swirl sloshing \cite{miles_resonantly_1984}. The boundaries between these regimes, in upright cylindrical tanks, are expressed through
\begin{equation}
\label{eq:Miles_Beta}
    \frac{A_e}{R}
    \approx
    \frac{1}{1.684}\left(
        \frac{\left(\omega_e/\omega_{1,1}\right)^2 - 1}{\mathcal{b}_i}
    \right)^{3/2} \,,
\end{equation}
\noindent where $\mathcal{b}_i$ with $i\in\{1,2,3\}$ is the frequency offset parameter indicating the boundaries separating the sloshing regimes. Here, $\mathcal{b}_1 = -0.36$ separates the chaotic and swirl regimes, $\mathcal{b}_2=-1.55$ divides the planar and chaotic ones, and $\mathcal{b}_3 = 0.735$ denotes the end of the swirl behavior \cite{royon_lebeaud_liquid_2007}.

\section{Physics-based digital twinning}
\label{sec:modeling}

The proposed framework, illustrated in Figure \ref{fig:framework_overview}, combines a reduced-order thermo-hydraulic model of the cryogenic tank with adaptive closure laws inferred from experimental observations. It consists of four main components: (1) the \textit{physical system}, from which observations $\check{\bm{s}}(t)$ are collected; (2) the \textit{digital twin}, which predicts the system evolution through a reduced-order physical model; (3) the adaptive closure laws $\bm{\vartheta}$, which represent the unresolved heat- and mass-transfer processes; and (4) the optimization procedure, which adjusts the trainable weights $\bm{w}$ to minimize the discrepancy between model predictions and observations. In the present context, the term \textit{digital twin} denotes this observation-linked model: measurements from the physical tank are assimilated to adapt its closure laws and align the virtual state with the measured one, following the framework introduced in \cite{marques_real_time_2024}.

The digital twin consists of a system of ordinary differential equations governing the evolution of the tank's thermo-hydraulic state. Following \cite{marques_real_time_2024}, the closure parameters $\bm{\vartheta}=g(\bm{s},\bm{\mathcal{X}};\bm{w})$ are expressed as parametric functions of the state $\bm{s}$, the exogenous inputs $\bm{\mathcal{X}}$, and a set of trainable weights $\bm{w}\in\mathbb{R}^{n_w}$. The complete model is cast as the initial-value problem (IVP):
\begin{equation}
\label{eq:dsdt}
\left\{
\begin{aligned}
\frac{d\bm{s}}{dt} &= f(\bm{s},\bm{\mathcal{X}},t;\bm{\vartheta}), \\
\bm{\vartheta}     &= g(\bm{s},\bm{\mathcal{X}};\bm{w}), \\
\bm{s}(0)          &= \bm{s}_0 ,
\end{aligned}
\right.
\end{equation}
where
$f:\mathbb{R}^{n_s}\times\mathbb{R}^{n_x}\times\mathbb{R}
\times\mathbb{R}^{n_\vartheta}\rightarrow\mathbb{R}^{n_s}$
encodes the physics-based nodal model of the cryogenic tank, and
$g:\mathbb{R}^{n_s}\times\mathbb{R}^{n_x}\times
\mathbb{R}^{n_w}\rightarrow\mathbb{R}^{n_\vartheta}$
is the data-driven closure function. The objective is to identify the weights $\bm{w}$ from experimental data so that the resulting closure parameters yield a predicted trajectory $\bm{s}(t,\bm{\mathcal{X}};\bm{w})$ that reproduces the observations $\check{\bm{s}}(t)\in\mathbb{R}^{n_s}$.

The closure laws are identified through an iterative optimization procedure. At iteration $k$, the weights $\bm{w}^{(k)}$ define the closure terms and therefore the state trajectory over a time interval $t\in[t_o,t_e]$, hereafter referred to as an \emph{episode}. The virtual trajectory $\bm{s}(t;\bm{w}^{(k)})$ is obtained by integrating the initial-value problem in \eqref{eq:dsdt} and is compared with the experimental observations through the cost functional
\begin{equation}
\label{eq:loss}
\mathcal{J}(\bm{w}^{(k)})
=
\frac{1}{\mathcal{T}_o}
\int_{t_o}^{t_e}
\bm{\varepsilon}(t;\bm{w}^{(k)})^T
\bm{W}
\bm{\varepsilon}(t;\bm{w}^{(k)})
\,dt ,
\end{equation}
where
\begin{equation}
\bm{\varepsilon}(t;\bm{w}^{(k)})
=
\hat{\bm{s}}(t;\bm{w}^{(k)})
-
\hat{\check{\bm{s}}}(t)
\end{equation}
is the normalized prediction error. Here, $\hat{\bm{s}}$ and $\hat{\check{\bm{s}}}$ denote the predicted and measured state vectors, respectively, normalized through min-max scaling using the global extrema of each state variable over the full experimental database. Furthermore, $\bm{W}\in\mathbb{R}^{n_s\times n_s}$ is a diagonal weighting matrix controlling the relative contribution of each state variable, and $\mathcal{T}_o=t_e-t_o$ is the episode duration. The integrand of the cost functional is denoted by $\ell(t;\bm{w})$.

The remainder of this section presents the thermo-hydraulic model, corresponding to the function $f$, in Subsection~\ref{sec:physical_model}; the parametrization of the closure laws, represented by $g$, in Subsection~\ref{sec:closure}; and the adjoint-based training procedure in Subsection~\ref{sec:model_training}.

\subsection{Thermo-hydraulic model}
\label{sec:physical_model}

\begin{figure}[h]
	\centering
	\includegraphics[width=0.8\linewidth]{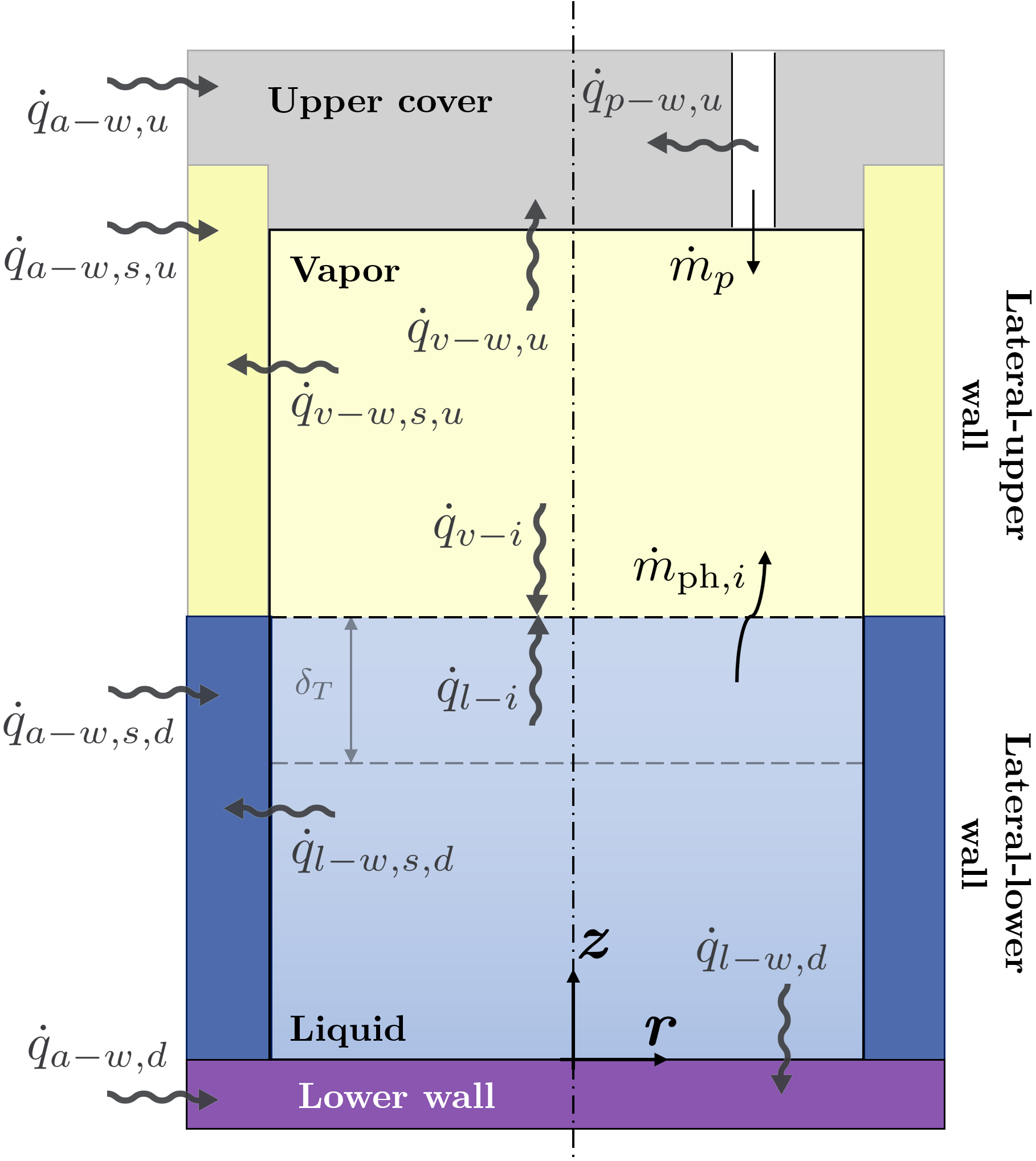}
	\caption{Schematic of the cryogenic fuel tank and the corresponding lumped control volumes: ullage, liquid, top cover, upper side-wall, lower side-wall and bottom cover. The heat fluxes exchanged between each node are represented with solid arrows and denoted as $\dot{Q}$, while the mass fluxes are shown with dotted arrows and identified as $\dot{m}$}
	\label{fig:thermal_model_overview}
\end{figure}

The physical model of the cryogenic propellant tank follows a multi-node framework composed of three main control volume groups: the vapor, the liquid, and the tank's solid shell.
The reduced-order representation employed in this work is illustrated in Figure \ref{fig:thermal_model_overview}. 

The model comprises a system of ordinary differential equations in time derived from conservation balances applied to each node $j$. The mass conservation balance is valid only for the fluid nodes and yields
\begin{align}
\label{eq:dmvdt}
    \frac{d m_{v}}{dt} &=
    \dot{m}_{p,v} +
    \dot{m}_{\mathrm{ph},i} \\[0.5em]
\label{eq:dmldt}
    \frac{d m_{l}}{dt} &=
    \dot{m}_{p,l} -
    \dot{m}_{\mathrm{ph},i}
    \,,
\end{align}
where $\dot{m}_{p,v}$ and $\dot{m}_{p,l}$ are exogenous inputs referring to incoming and outgoing flow rates from outside the tank to the gas or liquid phases, respectively (e.g., due to venting or pressurization), and $\dot{m}_{\mathrm{{ph},i}}$ is the net phase change rate between the fluid nodes.

During venting, mass leaves only through the vapor phase, such that $\dot{m}_{p,v} < 0$ and $\dot{m}_{p,l} = 0$. During pressurization, the injected fluid is typically in the vapor phase, yielding $\dot{m}_{p,v} > 0$. However, if condensation occurs within the pipeline, some liquid may also enter the tank, resulting in $\dot{m}_{p,l} \geq 0$. The split of the total pressurant flow $\dot{m}_p$ between phases is governed by the pressurant quality $\chi_p$:
\begin{align}
    \dot{m}_{p,v} &= \chi_p \, \dot{m}_p \\
	\dot{m}_{p,l} &= (1 - \chi_p) \, \dot{m}_p,
\end{align}
where $\chi_p \in [0, 1]$ is the vapor fraction of the incoming pressurant, and $\dot{m}_p(t)$ is the total pressurant mass flow rate. Since the pipeline is not modeled explicitly, $\chi_p$ is treated as a closure parameter.

Heat transfer between two connecting nodes, $j$ and $k$, is modeled using Newton's law of cooling:
\begin{equation}
\label{eq:Q}
    \dot{Q}_{j-k} 
    = 
    \frac{T_j - T_k}{\Omega_{j-k}} \,,
\end{equation}
where $T_j$ and $T_k$ are the respective node temperatures, and $\Omega_{j-k}$ denotes the total thermal resistance. This resistance accounts for both solid and fluid heat transfer and is given by
\begin{equation}
\label{eq:omega}
    \Omega_{j-k} =
    \varphi_w\frac{\delta_{j-k}}{\kappa_w A_{j-k}} 
    +
    \varphi_f\frac{L_{j-k}}{\left(\mathrm{Nu}_{j-k} + 1\right)\kappa_f A_{j-k}}
    \,,
\end{equation}
where $\varphi_w \in \{0,1\}$ and $\varphi_f \in \{0,1\}$ are internodal flags: $\varphi_w = 1$ if at least one node is a solid and $\varphi_f = 1$ if at least one node is a fluid. Here, $A_{j-k}$ is the heat exchange area, $\delta_{j,k}$ is the wall thickness, $L_{j-k}$ is the convective characteristic length, $\kappa_f$ and $\kappa_w$ are the thermal conductivities of the fluid and solid nodes, respectively, and $\mathrm{Nu}_{j-k}$ is the Nusselt number, defined as
\begin{equation}
\label{eq:Nusselt}
    \mathrm{Nu}_{j-k} = \frac{h_{j-k} L_{j-k}}{\kappa_j} \,,
\end{equation}
with $h_{j-k}$ representing the convective heat transfer coefficient. The convective resistance associated with the fluid, i.e., the second term on the right-hand side of \eqref{eq:omega}, uses $\mathrm{Nu}_{j-k}$ to denote the convective enhancement above the conductive baseline, such that $\mathrm{Nu}_{j-k} = 0$ recovers the purely conductive limit and $1+\mathrm{Nu}_{j-k}$ is the total dimensionless conductance. Since the Nusselt numbers encapsulate complex and strongly scenario-dependent transport mechanisms, they are incorporated into the closure parameter vector $\bm{\vartheta}$.

As shown in Figure \ref{fig:thermal_model_overview}, $\dot{Q}_{v-i}$ and $\dot{Q}_{l-i}$ denote the heat transfer rates from the vapor and liquid nodes to the interface, respectively, driving phase change across the free surface. The corresponding interfacial phase-change rate follows directly from the interfacial energy balance:
\begin{equation}
    \dot{m}_{\mathrm{ph},i} = \frac{\dot{Q}_{v-i} + \dot{Q}_{l-i}}{\mathcal{L}_v}\,.
\end{equation}

The energy conservation balance for a fluid node $j\in\{v,l\}$ is
\begin{equation}
\label{eq:dUdt}
\begin{split}
    \frac{d\mathcal{U}_{j}}{dt}
    &=
    \dot{m}_{p, j}\mathcal{h}_{p,j}
    +
    \varphi_j \dot{m}_{\mathrm{ph},i} \mathcal{h}_{\mathrm{sat},j} \\
    &
    - \sum_{k \neq j} \dot{Q}_{j-k}
    + \dot{W}_{j}
\end{split}
\end{equation}
where $k \neq j$ denotes all nodes connected to $j$, $\mathcal{U}_j$ is the internal energy, $\dot{W}_j$ is the work term, and $\varphi_j$ is a sign function such that $\varphi_v=1$ and $\varphi_l=-1$. We rewrite \eqref{eq:dUdt} with respect to the node's temperature by expressing the internal energy as $\mathcal{U}_{j} = m_{j} \mathcal{u}_{j}(\rho_j, T_j)$, with $\mathcal{u}_{j}$ the specific internal energy, and taking the time derivative with the chain rule
\begin{equation}
\label{eq:dU}
    \frac{d \mathcal{U}_{j}}{dt}  
    = \frac{d m_{j}}{dt} \mathcal{u}_{j}
    + m_{j}
    \left[ 
        \left. \frac{\partial \mathcal{u}_{j} }{\partial \rho} \right|_T \frac{d\rho_j}{dt}
        +
        \left. \frac{\partial \mathcal{u}_{j} }{\partial T} \right|_\rho \frac{d T_j}{dt}
    \right].
\end{equation}
Combining \eqref{eq:dUdt} and \eqref{eq:dU}, and introducing the specific enthalpy $\mathcal{h} = \mathcal{u} + p/\rho$, yield the energy conservation balance as a function of the node's temperature:
\begin{equation}
\label{eq:dTdt}
\begin{split}
    \frac{dT_{j}}{dt}
    &=
    \frac{\dot{m}_{p, j} \left( \mathcal{h}_{p, j} - \mathcal{h}_{j} \right)}
    {m_{j} \left(\partial_T \mathcal{u}\right)_\rho} 
    \\[0.5em]
    &
    +
    \frac{\varphi_j \dot{m}_{\mathrm{ph},i} \left( \mathcal{h}_{\mathrm{sat},j} - \mathcal{h}_{j} \right)}
    {m_{j} \left(\partial_T \mathcal{u}\right)_\rho}
    - 
    \sum_k \frac{\dot{Q}_{j-k}}
    {m_{j} \left(\partial_T \mathcal{u}\right)_\rho}
    \\[0.5em]
    &
    -\frac{\left( p_v - \rho_j^2 (\partial_\rho \mathcal{u}_j)_T\right)}{m_{j} \left(\partial_T \mathcal{u}\right)_\rho}
    \left(\frac{dV_j}{dt} - \frac{1}{{\rho}_j}\frac{dm_j}{dt}\right) \,,
\end{split}
\end{equation} having used the short hand notation $\partial_T$ for the partial derivative with respect to temperature and computing the volume rate of change by considering the tank as rigid, and that the liquid phase is incompressible, i.e., $\rho_l(p_v,T_l) \approx \rho_l(T_l)$ \cite{van_foreest_modeling_2014}. 

The control volume variation becomes
\begin{equation}
\label{eq:dVdt}
    \frac{dV_l}{dt}=-\frac{dV_v}{dt}
    \approx
    \frac{1}{\rho_l}\frac{dm_l}{dt}
    -
    \frac{m_l}{\rho_l^2}
    \left.\frac{\partial \rho}{\partial T}\right|_p
    \frac{dT_l}{dt}.
\end{equation} 

The energy balances of the solid nodes are a simplification of \eqref{eq:dTdt}, keeping only the sensible heat exchanges:
\begin{align}
\label{eq:dTwdt}
    \frac{d T_{w,u}}{dt} &= \frac{1}{m_{w,u} c_{w,u}} \Big(
        -\dot{Q}_{w,u-v} - \dot{Q}_{w,u-w,s,u} \nonumber\\
        &\qquad + \dot{Q}_{p-w,u} + \dot{Q}_{h,u} + \dot{Q}_{a-w,u}
    \Big)
    \\[0.5em]
    \frac{d T_{w,s,u}}{dt} &= \frac{1}{m_{w,s,u} c_{w,s,u}} \Big(
        -\dot{Q}_{w,s,u-v} - \dot{Q}_{w,s,u-l} + \dot{Q}_{w,u-w,s,u} \nonumber\\
        &\qquad  - \dot{Q}_{w,s,u-w,s,d} + \dot{Q}_{h,s} + \dot{Q}_{a-w,s,u}
    \Big)
    \\[0.5em]
    \frac{d T_{w,s,d}}{dt} &= \frac{1}{m_{w,s,d} c_{w,s,d}} \Big(
        -\dot{Q}_{w,s,d-v} - \dot{Q}_{w,s,d-l} \nonumber\\
        &\qquad + \dot{Q}_{w,s,u-w,s,d} - \dot{Q}_{w,s,d-w,d} + \dot{Q}_{a-w,s,d}
    \Big)
    \\[0.5em]
    \frac{d T_{w,d}}{dt} &= \frac{1}{m_{w,d} c_{w,d}} \Big(
        -\dot{Q}_{w,d-l} + \dot{Q}_{w,s,d-w,d} + \dot{Q}_{a-w,d}
    \Big)
    \,,
\end{align}
where $\dot{Q}_{w,u-v}$, $\dot{Q}_{w,{s,u}-v}$ and $\dot{Q}_{w,{s,d}-v}$ correspond to the solid-vapor exchanges, $\dot{Q}_{w,d-l}$, $\dot{Q}_{w,{s,u}-l}$ and $\dot{Q}_{w,{s,d}-l}$ are the solid-liquid exchanges, and $\dot{Q}_{w,u-w,{s,u}}$, $\dot{Q}_{w,{s,u}-w,{s,d}}$ and $\dot{Q}_{w,{s,u}-w,{d}}$ are the conductive heat transfer rates between connected wall nodes. The environmental heat fluxes $\dot{Q}_{a-w,u}$, $\dot{Q}_{a-w,{s,u}}$, $\dot{Q}_{a-w,{s,d}}$ and $\dot{Q}_{a-w,{d}}$ are computed using a thermal resistance circuit which accounts for conductive and convective exchanges between the wall nodes and the vacuum insulation at pressure and temperature $p_{sr}$ and $T_{sr}$, as well as radiative exchanges with the outer environment at temperature $T_a$.

The exchanges with the heating elements, i.e., $\dot{Q}_{h,u}$ and $\dot{Q}_{h,s}$, are computed as:
\begin{equation}
\begin{split}
    \dot{Q}_{h,u} &= K_{q,u} \frac{A_{h,u} \kappa_{w,u} \left( T_{h,u}(t) - T_{w,u} \right) }{ s_{w,u} } \\[0.5em]
    \dot{Q}_{h,s} &= K_{q,s} \frac{A_{h,s} \kappa_{w,s} \left( T_{h,s}(t) - T_{w,{s,u}} \right) }{  s_{w,s}}\,,
\end{split}
\end{equation}
with $A_{h,u}$ and $A_{h,s}$ the contact area between the heating elements and the corresponding solid regions, and $K_{q,u}\approx0.003$ and $K_{q,s}\approx0.060$ are accommodation coefficients representing imperfect contact between the heating elements and the tank walls, determined from experimental measurements of the wall temperature evolution during self-pressurization. 
Lastly, $\dot{Q}_{p,w,u}$ corresponds to heat transfer rate between the pressurant flow during pressurization scenarios, i.e, when $\dot{m}_p > 0$. We compute this contribution as:
\begin{equation}
    \dot{Q}_{p-w,u} = \frac{A_{p-w,u}\kappa_{w,u}\mathrm{Nu}_p}{D_p}\left( T_u(t) - T_{w,u} \right)
    \,,
\end{equation}
where $A_{p-w,u}$ is the exchange area between the pressurant stream and the upper cover, $D_p$ is the pressurant pipe diameter, and $\mathrm{Nu}_p$ is the Nusselt number for forced internal convection. The latter is evaluated using the correlation
\begin{equation}
    \mathrm{Nu}_p = 3.66 + \frac{0.0668 \dfrac{D_p}{L_p} \mathrm{Re}_p \mathrm{Pr}_v}{1 + 0.04\left( \dfrac{D_p}{L_p} \mathrm{Re}_p \mathrm{Pr}_v\right)^{2/3}}
    \,,
\end{equation}
with $\mathrm{Re}_p = 4\dot{m}_p / (\pi D_p \mu_p)$ the Reynolds number and $L_p$ the pipe length \cite{incropera_fundamentals_1996}.


On the vapor side, the pressure rate of change results from differentiating $p_v = p(\rho_v,T_v)$ with respect to time:
\begin{equation}
\label{eq:dpdt}
    \frac{dp_v}{dt}
    =
    \frac{1}{V_v}
    \left.\frac{\partial p_v}{\partial \rho}\right|_{T}
    \frac{dm_v}{dt}
    +
    \left.\frac{\partial p_v}{\partial T}\right|_{\rho}
    \frac{dT_v}{dt} 
    -
    \frac{m_v}{V_v^2}
    \left.\frac{\partial p_v}{\partial \rho}\right|_{T}
    \frac{dV_v}{dt}.
\end{equation}

To account for the effects of thermal stratification, known to play a critical role in the thermodynamic evolution of the tank under both static and dynamic conditions \cite{osipov_dynamic_2008, joseph_effect_2017, arndt_sloshing_2011}, we introduce a simplified treatment within the nodal framework. Specifically, the thickness of the thermal boundary layer on the liquid side is modeled as a first-order system:
\begin{equation}
\label{eq:ddeltadt}
    \frac{d\delta_{T}}{dt} = 
    \frac{\alpha_l}{R}\left( 
    c_\delta\frac{(T_{\mathrm{sat}}-T_l)}{T_{\mathrm{sat}}}\left( 1 - \frac{\delta_T}{R} \right) - \psi_\delta\frac{\delta_T}{R}
    \right)
    \,,
\end{equation}
where $c_\delta$ drives the growth of the thermal boundary layer, and $\psi_\delta$ is a forcing term accounting for sudden changes in $\delta_T$  due to external disturbances such as sloshing, pressurization, or venting. Both $c_\delta$ and $\psi_\delta$ are incorporated into the closure parameter vector $\bm{\vartheta}$, to be identified by the modeling agent.

All thermodynamic properties and their corresponding partial derivatives are obtained from a surrogate of REFPROP 10, following the methodology presented in \cite{marques_real_time_2024}.

The expressions above are sufficient to determine the evolution of the thermodynamic state of each node. However, they do not account for the irreversibility of the transfer processes. Writing an entropy balance for each node $j$, the corresponding entropy production rate $\dot{\sigma}_{s,j}$ is the difference between the entropy change predicted by the state equations and the contributions from heat and mass transfer:
\begin{equation}
\label{eq:sigma_s}
    \dot{\sigma}_{s,j} 
    = 
    \frac{d\mathcal{S}_j}{dt} + 
    \sum_{k \neq j} \frac{\dot{Q}_{j-k}}{T_k} - 
    \sum_{b} \dot{m}_{b} s_{b}
    \,,
\end{equation}
where $k$ denotes all neighboring nodes to $j$, and the subscript $b$ represents all boundaries across which there is a mass flux. 
For wall nodes, since there is no mass flux across solid boundaries, only the heat-flux term applies.
The entropy rate of change is given by
\begin{equation}
    \label{eq:dentropydt}
    \frac{d\mathcal{S}_j}{dt} = 
    \mathcal{s}_j \frac{dm_j}{dt} + 
    m_j 
    \left( 
        \left.\frac{\partial \mathcal{s}}{\partial T}\right|_p \frac{dT_j}{dt} 
        +
        \left.\frac{\partial \mathcal{s}}{\partial p}\right|_T \frac{dp_v}{dt}
    \right)
    \,,
\end{equation}
which follows from the chain rule, with the entropy of node $j$ defined as $\mathcal{S}_j=m_j \,,\mathcal{s}(p,T_j)$.
For wall nodes, the mass term vanishes in \eqref{eq:dentropydt}, and the entropy is treated as a function of temperature alone, i.e., $\mathcal{s}_j \approx \mathcal{s}(T_j)$.

Since the Nusselt numbers driving the heat transfer rates $\dot{Q}_{j,k}$ are data-driven parameters, the current formulation does not guarantee non-negative entropy production, i.e., $\dot{\sigma}_s \geq 0$, as required by the second law of thermodynamics \cite{moran_fundamentals_1995}.
To address this limitation, the loss function is augmented with a penalty term that discourages negative entropy production. While this approach promotes compliance with the second law of thermodynamics, it does not strictly enforce non-negative entropy production, and local violations may still occur depending on the weighting of the penalty term.


For a given set of closure parameters $\bm{\vartheta}(s,\bm{\mathcal{X}};\bm{w})$, the governing equations are integrated in time, and the corresponding entropy production rate is reconstructed \emph{a posteriori}. Since only violations of the second law are of interest, the penalty term is designed to act exclusively on negative entropy production. To account for contributions from all nodes, the penalty is defined as
\begin{equation}
\mathcal{P}_\sigma(\bm{w}^{(k)})
=
\frac{K_p}{\mathcal{T}_o}
\sum_{j}
\int_{t_o}^{t_e}
\left(\min\left(0, \dot{\sigma}_{s,j}(t;\bm{w}^{(k)})\right)\right)^2 \,
dt \,,
\end{equation}
where $j$ represents the vapor and liquid nodes as well as the upper-cover, lateral-wall, and bottom-cover nodes, and $K_p$ is the user-defined penalty factor.

Thus, the loss functional defined in \eqref{eq:loss} is augmented to include the penalty term, yielding
\begin{equation}
\label{eq:augmented_loss}
    \mathcal{A}(\bm{w}^{(k)}) = 
    \mathcal{J}(\bm{w}^{(k)}) + 
    \mathcal{P}_\sigma(\bm{w}^{(k)})
    \,.
\end{equation}

In summary, Equations \eqref{eq:dmvdt}-\eqref{eq:ddeltadt} constitute the physical modeling framework employed in this work, with state vector 
\begin{equation}
\label{eq:state_vector}
\begin{split}
    \bm{s} = &\left[p_v,\, T_v,\, T_l,\, T_{w,u},\, T_{w,{s,u}},\, \right.
    \\
    & \qquad \left. T_{w,{s,d}},\, T_{w,d}, m_v,\, m_l,\, V_l,\, \delta_{T}\right]\in\mathbb{R}^{11} \,,
\end{split}
\end{equation} time-dependent exogenous inputs
\begin{equation}
\label{eq:exogenous}
\begin{split}
    \bm{\mathcal{X}} = 
    & \left[ \dot{m}_p,\, T_u,\, p_u,\, T_{h,u},\, T_{h,{s}}, T_{sr},\, p_{sr},\, T_a, \,\right. \\
    & \quad  \left. A_e, f_e,\, A_s,\, \gamma_P,\, \gamma_B,\, \gamma_S,\, \gamma_V \right] \in \mathbb{R}^{15}\,,
\end{split}
\end{equation}
where $\gamma_P$, $\gamma_B$, $\gamma_S$, $\gamma_V$ are indicator functions specifying the operating scenario: $\gamma_P = 1$ during active pressurization, $\gamma_V = 1$ during venting, $\gamma_S = 1$ during sloshing, and $\gamma_B = 1$ when all other scenarios are inactive.

Lastly, the vector of closure parameters is
\begin{equation}
\label{eq:theta}
\begin{split}
    \bm{\vartheta} = 
    & 
    \left[ 
        \mathrm{Nu}_{v-i},\, \mathrm{Nu}_{l-i},\, 
        \mathrm{Nu}_{w,u-v},\, \mathrm{Nu}_{w,{s,u}-v},\, \mathrm{Nu}_{w,{s,d}-v},\, \right. 
        \\ 
        & \quad \quad \left. 
        \mathrm{Nu}_{w,d-l},\, \mathrm{Nu}_{w,{s,u}-l},\, \mathrm{Nu}_{w,{s,d}-l},\, 
        c_\delta,\, \psi_\delta,\, \chi_p 
    \right] \in \mathbb{R}^{11}
\end{split}
\end{equation}
corresponding to the Nusselt numbers that drive thermal and interfacial exchanges in the system, the parameters governing the growth of the liquid-side thermal boundary layer, and the fraction of vapor pressurant entering the tank. 

\subsection{Neural network closure}
\label{sec:closure}

As defined in \eqref{eq:dsdt}, the closure parameters $\bm{\vartheta}$ are modeled as parametric functions of the tank state, exogenous inputs, and trainable weights. In the present work, these functions are represented by fully connected feed-forward artificial neural networks.

Rather than using a single network across all operating conditions, four specialized networks are associated with the scenarios introduced in Section~\ref{sec:scenarios_of_interest}. Accordingly, the closure vector is decomposed as

\begin{equation}
\label{eq:theta_decomp}
    \bm{\vartheta} = 
    \bm{\vartheta}_{S} \gamma_S +
    \bm{\vartheta}_{P} \gamma_P +
    \bm{\vartheta}_{V} \gamma_V +
    \bm{\vartheta}_{B} \gamma_B\,,
\end{equation}
where $\bm{\vartheta}_{S}$, $\bm{\vartheta}_{P}$, and $\bm{\vartheta}_{V}$ correspond to sloshing, active-pressurization, and venting scenarios respectively, while $\bm{\vartheta}_{B}$ describes the baseline regime comprising self-pressurization and post-pressurization relaxation (see section \ref{sec:scenarios_of_interest}). The indicator functions $\gamma_S$, $\gamma_P$, $\gamma_V$, and $\gamma_B$, included in the exogenous-input vector $\bm{\mathcal{X}}$, define mutually exclusive operating regimes. This hard switching could be replaced in future implementations by fuzzy memberships to represent gradual or ambiguous transitions between scenarios \cite{singh_real_life_2013}.

Following \eqref{eq:theta_decomp}, the scenario-specific closure parameters are defined as
\begin{equation}
\begin{aligned}
    \bm{\vartheta}_{S} &= g_S(\bm{\mathcal{z}}_S; \bm{w}_S),\quad g_S : \mathbb{R}^{n_{\mathcal{z}_S}} \times \mathbb{R}^{n_{w_S}} \rightarrow \mathbb{R}^{11} \\
    \bm{\vartheta}_{P} &= g_P(\bm{\mathcal{z}}_P; \bm{w}_P),\quad g_P : \mathbb{R}^{n_{\mathcal{z}_P}} \times \mathbb{R}^{n_{w_P}} \rightarrow \mathbb{R}^{11} \\
    \bm{\vartheta}_{V} &= g_V(\bm{\mathcal{z}}_V; \bm{w}_V),\quad g_V : \mathbb{R}^{n_{\mathcal{z}_V}} \times \mathbb{R}^{n_{w_V}} \rightarrow \mathbb{R}^{11} \\
    \bm{\vartheta}_{B} &= g_B(\bm{\mathcal{z}}_B; \bm{w}_B),\quad g_B : \mathbb{R}^{n_{\mathcal{z}_B}} \times \mathbb{R}^{n_{w_B}} \rightarrow \mathbb{R}^{11}\,,
\end{aligned}
\end{equation}
where $\bm{\mathcal{z}}_k(\bm{s},\bm{\mathcal{X}})$, with $k\in\{ S, P, V, B\}$, denotes the scenario-specific input feature vector, derived from the thermo-hydraulic tank state and the exogenous inputs, and $\bm{w}_S$, $\bm{w}_P$, $\bm{w}_V$, and $\bm{w}_B$ represent the tunable weights associated with each network. The total number of input features is given by $n_{\mathcal{z}_{\mathrm{tot}}} = n_{\mathcal{z}_S}+n_{\mathcal{z}_P}+n_{\mathcal{z}_V}+n_{\mathcal{z}_B}$, while the number of training weights is $n_w = n_{w_S} + n_{w_P} + n_{w_V} + n_{w_B}$.

\begin{table}[!htb]
\centering
\renewcommand{\arraystretch}{1.2}
\caption{
Overview of the inputs for each parametric function that models the closure parameters in the four major scenarios: baseline ($B$), pressurization ($P$), venting ($V$) and sloshing ($S$). Each row presents the total number of features $n_{\mathcal{z}}$, together with the corresponding feature representation.
}
\resizebox{\columnwidth}{!}{

\begin{tabular}{c|cl}
\hline
Scenario & $n_{\mathcal{z}}$ & Input features \\ \hline \\[-1em]
\makecell[c]{Base\\($B$)} & 15 & 
\makecell[l]{
$\Jakob_{v-w,u}^B$, $\Jakob_{v-w,{s,u}}^B$, $\Jakob_{l-w,{s,d}}^B$, $\Jakob_{l-w,{d}}^B$, $\Rayleigh_{v-i}^B$, $\Rayleigh_{l-i}^B$, \\
$\Rayleigh_{v-w,{u}}^B$, $\Rayleigh_{v-w,{s,u}}^B$, $\Rayleigh_{v-w,{s,d}}^B$, $\Rayleigh_{l-w,{s,u}}^B$, $\Rayleigh_{l-w,{s,d}}^B$, $\Rayleigh_{l-w,{d}}^B$, \\
$\Reynolds_{s,l}$, $V_l/V$, $\delta_T/R$
} \\[2em]
\makecell[c]{Press.\\($P$)} & 16 & 
\makecell[l]{
$\Jakob_{v-w,u}^P$, $\Jakob_{v-w,{s,u}}^P$, $\Jakob_{l-w,{s,d}}^P$, $\Jakob_{l-w,{d}}^P$, $\Rayleigh_{v-i}^P$, $\Rayleigh_{l-i}^P$, \\
$\Rayleigh_{v-w,{u}}^P$, $\Rayleigh_{v-w,{s,u}}^P$, $\Rayleigh_{v-w,{s,d}}^P$, $\Rayleigh_{l-w,{s,u}}^P$, $\Rayleigh_{l-w,{s,d}}^P$, $\Rayleigh_{l-w,{d}}^P$, \\
$\Reynolds_{s,l}$, $\Pi_{p}$, $V_l/V$, $\delta_T/R$
}
\\[2em]
\makecell[c]{Vent\\($V$)} & 16 &
\makecell[l]{
$\Jakob_{v-w,u}^V$, $\Jakob_{v-w,{s,u}}^V$, $\Jakob_{l-w,{s,d}}^V$, $\Jakob_{l-w,{d}}^V$, $\Rayleigh_{v-i}^V$, $\Rayleigh_{l-i}^V$, \\
$\Rayleigh_{v-w,{u}}^V$, $\Rayleigh_{v-w,{s,u}}^V$, $\Rayleigh_{v-w,{s,d}}^V$, $\Rayleigh_{l-w,{s,u}}^V$, $\Rayleigh_{l-w,{s,d}}^V$, $\Rayleigh_{l-w,{d}}^V$, \\
$\Reynolds_{s,l}$, $\Pi_{p}$, $V_l/V$, $\delta_T/R$
}
\\[2em]
\makecell[c]{Slosh\\($S$)} & 16 & 
\makecell[l]{
$\Jakob_{v-w,u}^S$, $\Jakob_{v-w,{s,u}}^S$, $\Jakob_{l-w,{s,d}}^S$, $\Jakob_{l-w,{d}}^S$, $\Rayleigh_{v-i}^S$, $\Rayleigh_{l-i}^S$, \\
$\Rayleigh_{v-w,{u}}^S$, $\Rayleigh_{v-w,{s,u}}^S$, $\Rayleigh_{v-w,{s,d}}^S$, $\Rayleigh_{l-w,{s,u}}^S$, $\Rayleigh_{l-w,{s,d}}^S$, $\Rayleigh_{l-w,{d}}^S$, \\
$\Reynolds_{s,l}$, $\Reynolds_{e,l}$, $V_l/V$, $\delta_T/R$
}
\\[1.5em]
\hline
\end{tabular}

}
\label{tab:ann_details}
\end{table}

The networks which predict $\bm{\vartheta}_S$, $\bm{\vartheta}_P$, $\bm{\vartheta}_V$, and $\bm{\vartheta}_B$ are expressed through feed-forward ANNs, each with two hidden layers with 16 neurons each, using SiLU as the activation function. The Nusselt numbers, $c_\delta$, and $psi_\delta$ are produced through a softplus output to ensure only positive values, whereas $\chi_p$ is mapped through a sigmoid to $[0,1]$.

Table \ref{tab:ann_details} summarizes the inputs for each parametric function. All networks share a common set of input features, which include the liquid fill-ratio $V_l/V$, the thermal boundary layer thickness scaled by the tank inner radius, $\delta_T/R$. In addition, the models incorporate the Jakob and Rayleigh numbers, defined as
\begin{equation}
\label{eq:Ra_and_Ja}
    \Jakob_{j-k} = \frac{ \kappa_j (T_j - T_k) A_{j,k} [t]}{ \mathcal{L}_v  (\rho_l - \rho_v) V_j [L]}
    \ \ \mathrm{and}\ \ 
    \Rayleigh_{j-k} = \frac{g\beta_j (T_j - T_k) [L]^3}{\nu_j\alpha_j}
    \,,
\end{equation}
where $\beta_j$ is the volumetric thermal expansion coefficient, $\alpha_j$ the thermal diffusivity, and $\nu_j$ the kinematic viscosity of node $j$.
The formulation of these dimensionless groups follows the scaling framework previously introduced by the authors in \cite{marques_scaling_2025}. The characteristic length scale $[L]$ and characteristic time $[t]$, depend on the operating scenario considered, i.e., baseline, pressurization, venting, or sloshing. The corresponding definitions used for each scenario are summarized in Table \ref{tab:ann_scales}.

\begin{table}[!htb]
\centering
\renewcommand{\arraystretch}{1.2}
\caption{
Summary of the reference scales defined for each operating scenario. These scales are employed to define the dimensionless groups in \eqref{eq:Ra_and_Ja}.
}

\begin{tabular}{c|cc}
\hline
Scenario & $[L]$ & $[t]$ \\
\hline \\[-1em]
Base ($B$)   & $H_l$ & $\dfrac{[L]^2}{\alpha}$                                   \\
Press. ($P$) & $\dfrac{V_{v,0}}{A_i}$ & $\dfrac{[L]^2}{\alpha}$                  \\
Vent ($V$)   & $\delta_{T,0}$ & $\dfrac{[L]^2}{\alpha}$                          \\
Slosh ($S$)  & $\dfrac{A_s}{\delta_{T,0}} R$ & $\dfrac{f_{11}[L]^2}{f_e \alpha}$ \\[1em]
\hline
\end{tabular}

\label{tab:ann_scales}
\end{table}

Furthermore, all scenarios include, as an additional input, a liquid-based Reynolds number evaluated using the measured experimental interface displacement amplitude $A_s$, defined as
\begin{equation}
    \Reynolds_{s,l} = \frac{\sqrt{g R} A_s}{\nu_l} \,.
\end{equation}
This quantity characterizes the intensity of the free-surface motion and its potential contribution to enhanced interfacial heat and mass transfer, independently of the origin of the disturbance. For sloshing cases, the present model thus uses the measured interface amplitude $A_s(t)$ as an exogenous input; predicting this quantity from the imposed tank motion ($A_e$, $f_e$) remains outside the present scope.

For the sloshing scenario, a second Reynolds number accounts explicitly for the externally imposed excitation amplitude and frequency:

\begin{equation}
    \Reynolds_{e,l} = \left( \frac{f_e}{f_{1,1}} \right) \left( \frac{A_e}{R} \right) \frac{\sqrt{g R} [L]_S}{\nu_l}
\end{equation}
where $[L]_S$ is the characteristic length scale associated with sloshing, reported in Table~\ref{tab:ann_scales}. This expression follows the scaling analysis in \cite{marques_scaling_2025} and characterizes the externally imposed tank motion.

For active-pressurization and venting, the corresponding networks receive the additional dimensionless input

\begin{equation}
    \Pi_{p} = \frac{\dot{m}_pH_l^2}{\alpha_l(\rho_{l} - \rho_{v})V_v} \,.
\end{equation}

This dimensionless group can be interpreted as the ratio between the active-pressurization or venting rate imposed and the intrinsic self-pressurization rate of the system, i.e., the evaporation rate that would arise if thermal diffusion were the dominant source of heat transfer. As shown in the scaling analysis of \cite{marques_scaling_2025}, this parameter is strongly correlated with the resulting tank pressurization dynamics.

All input features listed in Table~\ref{tab:ann_details} are min--max normalized using bounds computed exclusively from the corresponding training data. During evaluation, normalized feature values are not clipped and may therefore fall outside the training interval $[0,1]$ when operating conditions extend beyond those represented in the training set.

\subsection{Model training}
\label{sec:model_training}

The data-assimilation framework uses experimental observations to identify the model parameters by minimizing the augmented loss functional defined in \eqref{eq:augmented_loss}. Each training iteration consists of three steps: (1) sampling experimental trajectories, (2) integrating the nodal model forward in time, and (3) computing the loss gradients and updating the network weights $\bm{w}$ through adjoint-based optimization.

\subsubsection*{Cross-validation}

To assess the generalization capability of the proposed framework, the database is partitioned at the experiment level into $K$ mutually exclusive folds, each containing representative cases from the operating scenarios introduced in Section~\ref{sec:scenarios_of_interest}. At each cross-validation iteration, one fold is held out for validation, while the remaining $K-1$ folds are used for training. Repeating this procedure for all folds produces $K$ independently trained models, with each experiment appearing once in a validation fold.

The held-out experiments are not used for gradient-based optimization of the closure laws. Their predictions therefore provide out-of-fold estimates of model performance. Validation is performed over complete experimental trajectories, initialized from the reconstructed state at the beginning of each experiment and integrated without intermediate state resets. The evaluation therefore measures open-loop trajectory prediction conditional on the experimentally reconstructed initial state, rather than autonomous state estimation. The reported cross-validated metrics are obtained by aggregating the prediction errors over all held-out experiments. Different values of $K$ are considered in a sensitivity analysis to evaluate the influence of the available training-data volume and the variability among folds.


\subsubsection*{Data sampling}

At each optimization iteration, a mini-batch of $N_d$ experimental time series is sampled exclusively from the training folds of the current cross-validation iteration. Because each time series may span several successive operating scenarios, two complementary sampling strategies are employed.

In the first strategy, the complete time series is integrated, preserving the full sequence of operating conditions and the cumulative evolution of the system. In the second, $N_b$ shorter segments of duration $\Delta t_b<\Delta t_{\mathrm{tot}}$ are sampled from each selected experiment. Each segment beginning at time $t_{j,0}$ is initialized from the corresponding reconstructed experimental state, i.e. $
    \bm{s}(t_{j,0})=\check{\bm{s}}(t_{j,0}).$

The sampling strategy at iteration $k$ is determined by the Bernoulli variable
\begin{equation}
    b^{(k)} \sim \operatorname{Bernoulli}(\mathcal{p}_b),
\end{equation}
with $\mathcal{p}_b=0.5$. Complete trajectories are used when $b^{(k)}=0$, whereas shorter segments are sampled when $b^{(k)}=1$.

Complete trajectories promote long-horizon consistency and preserve transitions between successive operating scenarios. By contrast, shorter segments initialize the model from intermediate experimental states rather than systematically from the beginning of each experiment. This limits the repeated propagation of errors from early transients and exposes the optimization to a broader range of states and local transitions.

\subsubsection*{Forward model evaluation}

The forward model in \eqref{eq:dsdt} is integrated using the \texttt{diffrax} package \cite{kidger2021_diffrax} within the JAX ecosystem \cite{jax2018github}. Time integration is performed with the explicit fifth-order Runge--Kutta method of Tsitouras \cite{tsitouras2011runge}. The time-step size is adapted using a proportional--integral--derivative controller with coefficients $(p,i,d)=(0.2,0.4,0)$ and relative and absolute tolerances $\varepsilon_{\mathrm{rel}}=\varepsilon_{\mathrm{abs}}=10^{-7}$.

Because the state variables differ substantially in magnitude, each component $k$ of the state vector is normalized as

\begin{equation}
\label{eq:s_MinMax}
    \hat{s}_k(t) = \frac{s_k(t) - s_k^{\min}}{s_k^{\max} - s_k^{\min}}
\end{equation}
where $s_k^{\min}$ and $s_k^{\max}$ are computed exclusively from the training folds of the current cross-validation iteration. The same bounds are applied unchanged to the corresponding held-out fold. The solver operates on the normalized state $\hat{\bm{s}}(t)\in\mathbb{R}^{n_s}$, while the physical state is recovered through

\begin{equation}
    s_k(t)
    =
    \hat{s}_k(t)
    \left(s_k^{\max}-s_k^{\min}\right)
    +
    s_k^{\min}.
\end{equation}
Accordingly, the normalized rate of change is
\begin{equation}
    \frac{d\hat{s}_k}{dt}
    =
    \frac{1}{s_k^{\max}-s_k^{\min}}
    \frac{ds_k}{dt}.
\end{equation}
This scaling improves the numerical conditioning of the integration and ensures that the prescribed tolerances act on state components of comparable magnitude. Normalized values are not clipped and may therefore fall outside the interval $[0,1]$ when a held-out experiment extends beyond the range represented in its training folds.

\subsubsection*{Adjoint-based gradient computation}

The gradient of the augmented loss functional is computed using an adjoint-based method. For compactness, let $q\in\{1,\ldots,N_{\mathcal{B}}\}$ index the complete trajectories or shorter segments integrated within the current mini-batch. The gradient is then written as
\begin{equation}
\label{eq:dJdw}
    \nabla_{\bm{w}}\mathcal{A}(\bm{w})
    =
    \sum_{q=1}^{N_{\mathcal{B}}}
    \int_{t_{q,0}}^{t_{q,e}}
    \left[
        \nabla_{\bm{w}}\ell_{a,q}
        +
        \left(
            \frac{\partial f}{\partial\bm{w}}
        \right)_{q}^{T}
        \bm{\lambda}_{q}(t)
    \right]
    dt,
\end{equation}
where $\ell_a$ denotes the integrand of the augmented loss functional and $\bm{\lambda}_q(t)\in\mathbb{R}^{n_s}$ is the adjoint variable associated with trajectory segment $q$. It is obtained by solving the terminal-value problem
\begin{equation}
\label{eq:adjoint}
\left\{
\begin{aligned}
    \frac{d\bm{\lambda}_q}{dt}
    &=
    -\nabla_{\bm{s}}\ell_{a,q}
    -
    \left(
        \frac{\partial f}{\partial\bm{s}}
    \right)_{q}^{T}
    \bm{\lambda}_q,
    \\
    \bm{\lambda}_q(t_{q,e})
    &=\bm{0}.
\end{aligned}
\right.
\end{equation}
The adjoint equations are integrated backward in time using the same fifth-order Runge--Kutta method employed for the forward model. The derivatives of $\ell_a$ and $f$ with respect to $\bm{s}$ and $\bm{w}$ are evaluated through JAX automatic differentiation using the computational graphs of the augmented loss and the forward model
$f(\bm{s},\bm{\mathcal{X}},t;\bm{\vartheta}(\bm{w}))$.

\subsubsection*{Optimization}

The network weights are updated using the AdamW optimizer \cite{loshchilov_decoupled_2019}. Denoting the gradient at iteration $k$ by
\begin{equation}
    \bm{g}_k
    =
    \nabla_{\bm{w}}\mathcal{A}(\bm{w}_k),
\end{equation}
the first- and second-moment estimates are
\begin{equation}
\begin{aligned}
    \bm{m}_k
    &=
    \beta_1\bm{m}_{k-1}
    +
    (1-\beta_1)\bm{g}_k,
    \\
    \bm{v}_k
    &=
    \beta_2\bm{v}_{k-1}
    +
    (1-\beta_2)
    \bm{g}_k\odot\bm{g}_k,
\end{aligned}
\end{equation}
with bias-corrected values
\begin{equation}
    \hat{\bm{m}}_k
    =
    \frac{\bm{m}_k}{1-\beta_1^k},
    \qquad
    \hat{\bm{v}}_k
    =
    \frac{\bm{v}_k}{1-\beta_2^k},
\end{equation}
where $\beta_1,\beta_2\in[0,1)$ and $\odot$ denotes element-wise multiplication. The weights are updated according to
\begin{equation}
\label{eq:w_update_wd}
    \bm{w}_{k+1}
    =
    \bm{w}_k
    -
    \eta_k
    \left[
        \frac{\hat{\bm{m}}_k}
        {\sqrt{\hat{\bm{v}}_k}+\epsilon}
        +
        \lambda_w\bm{w}_k
    \right],
\end{equation}
where $\eta_k$ is the learning rate and $\lambda_w$ is the weight-decay coefficient.

The optimizer uses an initial learning rate $\eta_{0}=10^{-3}$ and a weight decay $\lambda_w=10^{-4}$. During training, relative perturbations with an initial amplitude of $5\%$ are added to the experimental data and decay exponentially at a rate of $10^{-4}$ over the optimization iterations. As discussed in \cite{marques_real_time_2024}, this stochastic perturbation promotes exploration of the parameter space and improves robustness to measurement uncertainty.

Dropout is also applied during training with probability $\mathcal{p}_d=0.1$ \cite{srivastava2014_dropout}. At each iteration, $10\%$ of the hidden activations are randomly set to zero, while the retained activations are rescaled by $1/(1-\mathcal{p}_d)$ to preserve their expected magnitude. Artificial noise and dropout are disabled during validation, where the held-out experiments are evaluated deterministically using the complete networks.

\section{Dataset description}
\label{sec:dataset_description}

\subsection{Experimental setup}
\label{sec:experimental_setup}

\begin{figure}[!htb]
	\centering
	\includegraphics[width=\linewidth]{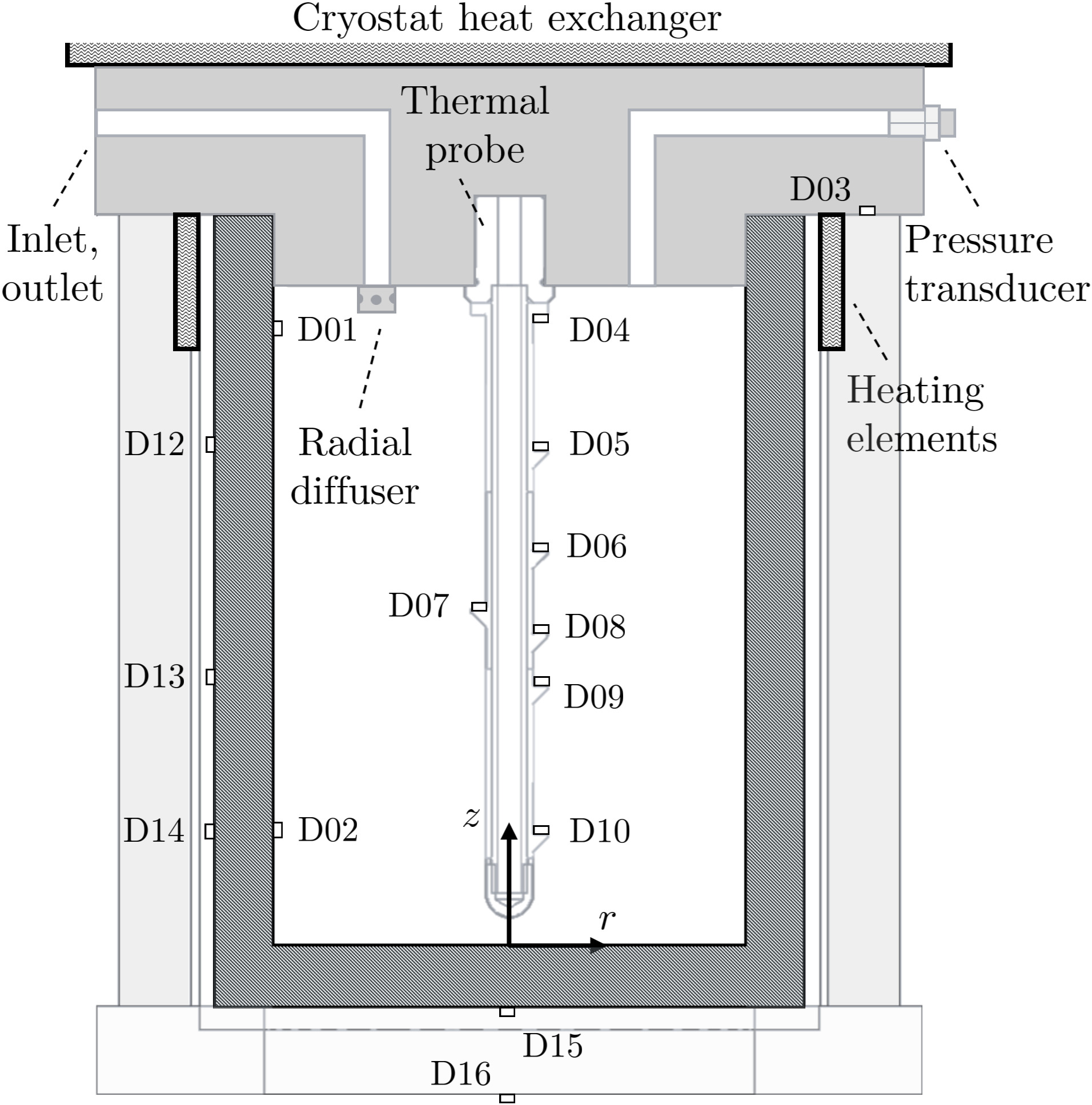}
	\caption{Schematic representation of the real environment cryogenic tank. The temperature sensors attached to the central probe and tank walls are identified according to their identification numbers. The test section's upper cover is attached to the cryostat heat exchanger.}
	\label{fig:exp_schematic}
\end{figure}

The experimental data is sourced from experimental campaigns carried out at the von Karman Institute's CryME cryostat facility operated with LN$_2$ \cite{marques_experimental_icmfht_2024}. The test section has a square external cross-section (100 \unit{\milli\metre} $\times$ 100 \unit{\milli\metre}) and an outer height of 134 \unit{\milli\metre}, with a cylindrical internal cavity of diameter 83 \unit{\milli\metre} and internal height 124 \unit{\milli\metre}. The tank, schematically illustrated in Figure \ref{fig:exp_schematic}, is made of quartz with upper and bottom covers made of copper, compressed with titanium rods. The setup is mounted in the cryostat's sample space, which is kept at low vacuum pressures, i.e., $\sim 0.8$ \unit{\kilo\pascal}. Additionally, the cryostat is equipped with optical viewports that provide visual access to the propellant inside the tank.

\begin{table}[!htb]
\centering
\renewcommand{\arraystretch}{1.2}
\caption{Spatial coordinates of the thermo-diodes sorted by their sensor identification numbers. The positions are defined in cylindrical coordinates $(r,\theta,z)$ according to the reference frame sketched in Figure \ref{fig:exp_schematic}.}
\begin{tabular}{ccccc}
\hline
\multicolumn{1}{l|}{Sensor} & $z$ [mm]   & $r$ [mm] & $\theta$ [deg] & Sensing surface \\ \hline
\multicolumn{1}{l|}{D01}    & 102        & 40       & 180            & Inner wall    \\
\multicolumn{1}{l|}{D02}    & 10         & 40       & 180            & Inner wall    \\
\multicolumn{1}{l|}{D03}    & 134        & 58       & -90            & Top cover     \\
\multicolumn{1}{l|}{D04}    & 106        & 10       & -90            & Fluid         \\
\multicolumn{1}{l|}{D05}    & 85         & 10       & -90            & Fluid         \\
\multicolumn{1}{l|}{D06}    & 68         & 10       & -90            & Fluid         \\
\multicolumn{1}{l|}{D07}    & 58         & 10       & 90             & Fluid         \\
\multicolumn{1}{l|}{D08}    & 54         & 10       & -90            & Fluid         \\
\multicolumn{1}{l|}{D09}    & 45         & 10       & -90            & Fluid         \\
\multicolumn{1}{l|}{D10}    & 20         & 10       & -90            & Fluid         \\
\multicolumn{1}{l|}{D11}    & 102        & 50       & 0              & Heater        \\
\multicolumn{1}{l|}{D12}    & 85         & 50       & 180            & Outer wall    \\
\multicolumn{1}{l|}{D13}    & 45         & 50       & 180            & Outer wall    \\
\multicolumn{1}{l|}{D14}    & 10         & 50       & 180            & Outer wall    \\
\multicolumn{1}{l|}{D15}    & -10        & 60       & 0              & Outer wall    \\
\multicolumn{1}{l|}{D16}    & -25        & 20       & -90            & Bottom cover  \\
\hline
\end{tabular}
\label{tab:sensor_coordinates}
\end{table}

The tank's upper cover is machined with channels for the inlet and outlet ports, and for a Kulite CTL-190 pressure transducer. The upper cover is bolted to the cryostat's heat exchanger, allowing its temperature to be controlled and acquired during the experiments. In addition, the side-wall temperature is controlled through a set of polyimide thermofoil heating elements covering the upper portion of the tank, with $\Delta z_{h_s} = 30$ \unit{\milli\metre}, and connected to a control loop that sets $T_{h,s}(t)$. The temperature distribution within the setup is monitored using sixteen Lakeshore DT-670-SD thermo-diodes, with their positions and sensing surfaces detailed in Table \ref{tab:sensor_coordinates}.
During pressurization and venting scenarios, the vapor flow rates are measured using a Bronkhorst EL-FLOW F-201CV mass flow controller. Moreover, sloshing excitations are generated by prescribing the desired motion signals to the SHAKESPEARE (SHaking Apparatus for Kinetic Experiments of Sloshing Projects with EArthquake Reproduction) facility, on which the entire setup is mounted. The resulting acceleration field is continuously monitored via a triaxial Endevco Model 7298 accelerometer.

The optical access provided by the facility enables flow visualization through backlight illumination, supporting both interface tracking and bulk flow observation. Flow snapshots are acquired at 75 Hz using a JAI SP-1200M-CXP4 camera.

\subsection{Data processing}
\label{sec:data_processing}

This section describes the processing steps used to extract the state variables required by the thermal model from the raw experimental data.

The acquired flow snapshots are processed to extract the interface position and shape over time \cite{marques_experimental_icmfht_2024}. We compute the instantaneous fill level, $H_l(t)$, for each snapshot as the mean interface height. This method suits static and low-amplitude planar sloshing, where the interface is flat and motion is confined to a two-dimensional plane. In more dynamic cases, however, the free surface can become highly curved and move three-dimensionally outside the camera’s acquisition plane.

\begin{figure}[!htb]
	\centering
    \includegraphics[width=\linewidth]{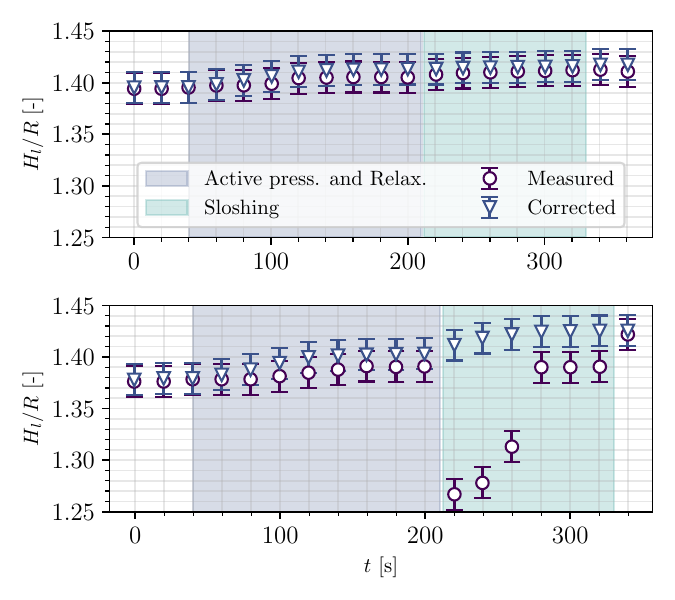}
	\caption{Comparison of the image-based fill level measurements (purple circular markers) with the corrected values (blue triangular markers) for two test cases. The figure background is colored according to the active scenario, with light blue denoting pressurization and light green denoting sloshing. Upper plot: pressurization followed by low-amplitude planar sloshing, i.e., test case T01. Lower plot: pressurization followed by chaotic sloshing, i.e., T04 (see Table \ref{app:database_training}).}
	\label{fig:fill_correction}
\end{figure}

An iterative algorithm is used to correct optical fill level measurements during significant interface motion. Our approach is as follows. First, we combine image-based fill level measurements, pointwise temperature data from thermo-diodes D04–D10, and vapor pressure readings to obtain spatial averages of vapor and liquid densities as
\begin{equation}
\label{eq:rho_mean}
    \begin{split}
        \overline{\rho_v}(t) &= \frac{1}{H-H_l(t)} \int_{H_l(t)}^{H}\rho_v\left( p_v(t),\, \bm{T}_f(t,z) \right) dz
        \\[0.5em]
        \overline{\rho_l}(t) &= \frac{1}{H_l(t)} \int_{0}^{H_l(t)}\rho_l\left( p_v(t),\, \bm{T}_f(t,z) \right) dz,
    \end{split}
\end{equation}
where $H$ is the internal tank height, $\rho_v$ and $\rho_l$ are the vapor and liquid densities evaluated at pressure $p_v(t)$ and temperature $\bm{T}_f(t,z)$. The vector $\bm{T}_f(t,z)$ is constructed by concatenating the values measured by thermo-diodes D04–D10 at their respective coordinates (Table \ref{tab:sensor_coordinates}), along with the saturation temperature $T_\mathrm{sat}$ at $z=H_l$. The integrals in \eqref{eq:rho_mean} are evaluated numerically using the trapezoidal rule. Using the average densities, we estimate the initial fluid mass as
\begin{equation}
\label{eq:m_f0}
    m_{f,0} = \overline{\rho_{v,0}}(V - V_{l,0}) + \overline{\rho_{l,0}} V_{l,0}.
\end{equation}
Since this mass only changes due to pressurization and venting operations, we retrieve
\begin{equation}
\label{eq:m_f}
    m_f(t) = m_{f,0} + \int_0^t \dot{m}_{p}(t) dt \,,
\end{equation}
where the mass flow rate $\dot{m}_p$ is positive during pressurization and negative during venting. The corrected liquid volume is then obtained as
\begin{equation}
    V_l(t)=\frac{m_f(t) - \overline{\rho_v}(t) V}{\overline{\rho_l}(t) - \overline{\rho_v}(t)},
\end{equation}
and the fill level as $H_l(t)=V_l(t)/\pi R^2$. We then return to \eqref{eq:rho_mean} with the updated $H_l(t)$ and repeat the algorithm for $N_h=5$ iterations. Figure \ref{fig:fill_correction} compares the image-based fill level measurements with the corrected values for two representative cases: T01, with low-amplitude sloshing, and T04, with chaotic sloshing (see Table \ref{app:database_training}). For T01, the measured and corrected levels remain within each other's uncertainty bounds. For T04, the correction effectively suppresses the unphysical fluctuations between $t=220$ and $t=280$~\unit{\second}.

Once the fill-level corrections are applied, the vapor and liquid masses are estimated as $m_v(t)=\overline{\rho_v}(t)(V-V_l(t))$ and $m_l(t)=\overline{\rho_l}(t)V_l(t)$, respectively. Moreover, we estimate the mass-averaged temperature in each fluid node as
\begin{equation}
\begin{split}
    \overline{T}_v(t) &= \frac{\pi R^2}{m_v(t)} \int_{H_l(t)}^H \rho_v(t,z) \cdot \bm{T}_f(t,z) dz
    \\[0.5em]
    \overline{T}_l(t) &= \frac{\pi R^2}{m_l(t)} \int_{0}^{H_l(t)} \rho_l(t,z) \cdot \bm{T}_f(t,z) dz.
\end{split}
\end{equation}
Similarly, the temperatures of the nodes along the lateral walls are obtained from spatial averages as follows: $T_{w,{s,u}}$ is computed by averaging $T_\mathrm{D01}$ and $T_\mathrm{D12}$, while $T_{w,{s,d}}$ uses $T_\mathrm{D13}$, $T_\mathrm{D14}$, and $T_\mathrm{D02}$. For the nodes corresponding to the upper cover and lower tank region, we assign $T_{w,u} = T_\mathrm{D03}$ and $T_{w,d} = T_\mathrm{D15}$, respectively.
The limited number of sensors assigned to each solid region introduces an uncertainty in the reconstructed bulk node temperatures. Nevertheless, these estimates provide a consistent indication of the evolution of wall temperature across its main regions, which is deemed sufficient for the lumped model adopted in this work. 

Lastly, $\delta_T$ is estimated by applying a piecewise linear interpolation to the temperature profile $\bm{T}_f(t,z)$ at each time step and identifying the thickness of the region where the temperature drops by 90\% from the saturated interface toward the subcooled bulk measured by $T_\mathrm{D10}$.


\subsection{Training database and validation strategy}

The extracted dataset comprises forty-eight time series, each describing the evolution of the cryogenic tank through multiple operating scenarios. During each experiment, the tank described in Section \ref{sec:experimental_setup} undergoes a sequence of operating conditions, including self-pressurization, active-pressurization, venting, and sloshing, with varying durations and operating parameters.

Table \ref{app:database_training} summarizes the operating conditions of the experiments comprising the experimental database. Each case is assigned a number, and labeled with a `T', e.g., T00-T47. 
For each experiment, the table summarizes the initial pressure, node temperatures, liquid fill ratio $H_l/R$, and thermal boundary layer thickness $\delta_T$, along with the main exogenous inputs driving each scenario and their durations. For self-pressurization, the time-averaged heated surface temperatures, $T_{h,s}$ and $T_{h,u}$, are provided. For active-pressurization and venting, the pressurant mass flow rate $\dot{m}_p$, upstream temperature $T_u$ and pressure $p_u$ are listed. For sloshing, the excitation amplitude scaled by the tank radius $A_e/R$, the measured wave amplitude ratio $A_s/R$, and the excitation frequency scaled by the natural frequency $f_e/f_{1,1}$ are reported.
In addition to the table,\ref{app:database} provides a brief overview of the experimental database and the different operating scenarios considered.

To assess the generalization capability of the digital twin, the complete database is evaluated using a $K$-fold cross-validation approach, as described in Section \ref{sec:model_training}.
Four values of $K$ are considered, namely $K=2$, $K=3$, $K=4$, and $K=5$, corresponding to different training-validation ratios. For each configuration, the time series are assigned pseudo-randomly to the folds while ensuring that each fold contains representative cases from all operating scenarios. This allows for assessing the sensitivity of the model performance with respect to the available training data.

Following the cross-validation analysis, the model is retrained using the complete database. This final model is evaluated across all available operating conditions to characterize the overall predictive capability of the digital twin.


\section{Results and discussion}
\label{sec:results_and_discussion}

This section is organized in two parts. Section \ref{sec:res_sensitivity} examines the sensitivity of the framework to the amount of available training data, focusing on how the
out-of-fold error and its variability across folds evolve with the size of the training set.
Section \ref{sec:full_model_performance} then reports the reconstruction accuracy of the model trained on the full experimental database.

In all cases, the optimization of the closure parameters is performed using mini-batches that randomly sample $N_d=12$ out of the available training time series at each iteration. Moreover, $N_b=2$ segments are sampled from each time series, each of duration $\Delta t_b/\Delta t_\mathrm{tot} = 0.3$ of the total experiment time. The entropy-based penalty factor is set to $K_p = 10^{5}$, based on a preliminary sensitivity analysis that varied $K_p$ between $10^{2}$ and $10^{6}$. The value $K_p = 10^{5}$ was the smallest value that reduced the penalty term without degrading the state reconstruction. Lastly, the diagonal of the weighting matrix in \eqref{eq:loss} is imposed as
\begin{equation}
    \mathrm{diag}(\bm{W}) = 
    \begin{aligned}[t]
        [ 2.0, \, 1.0, \, 1.0, \, & 0.2, \, 0.2, \, 0.2,  \\
          & \, 0.2, 1.0, \, 0.6, \, 0.2, \, 0.1 ]
    \end{aligned}
\end{equation}
where each entry corresponds, in order, to the state components defined in \eqref{eq:state_vector}: the pressure receives weight $2.0$; the vapor and liquid temperatures and the vapor mass receive weight $1.0$; the liquid mass receives weight $0.6$; the wall temperatures and the liquid volume receive weight $0.2$; and the thermal boundary layer thickness receives weight $0.1$. 
This choice increases the penalty on mismatches in the predicted pressure, vapor and liquid temperatures, and vapor mass. Larger deviations are tolerated in the wall temperatures, liquid mass, liquid volume, and thermal boundary layer thickness, for which the higher experimental uncertainty justifies a lower weighting.

\subsection{Sensitivity to training data availability}
\label{sec:res_sensitivity}

For each value of $K$, the model was retrained independently on the corresponding $K-1$ folds, and the resulting training and out-of-fold losses were aggregated across folds to obtain their mean and standard deviation across folds (population definition).

\begin{figure}[!htb]
    \centering
    \includegraphics[width=\linewidth]{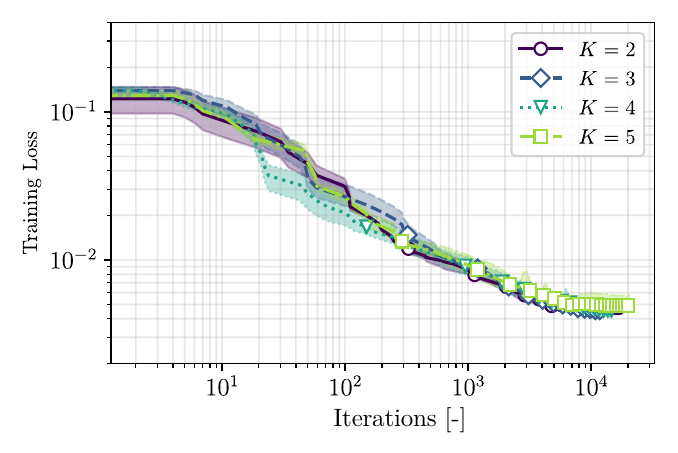}
    \caption{
    Training loss for $K \in \{2,3,4,5\}$-fold cross-validation. Lines show the mean loss across the folds, while the shaded regions show the fold-wise standard deviation. 
    }
    \label{fig:loss_train_fold}
\end{figure}

Figure \ref{fig:loss_train_fold} reports the training loss for $K=2$ (solid line, circular markers), $K=3$ (dashed line, diamond markers), $K=4$ (dotted line, downward-triangular markers), and $K=5$ (dash-dotted line, square markers), with the shaded area around each curve indicating the $\pm$ one standard deviation band. Training performance is essentially identical across all values of $K$ and across folds, with all models converging to values of approximately $(4.6 \pm 0.5) \times 10^{-3}$. This indicates that the closure parametrization is sufficiently general to accommodate additional training data without any loss in fitting capability: increasing the size of the training dataset does not force a trade-off, but simply provides more information for the model to leverage. The convergence rate is likewise similar across all values of $K$, with no significant slowdown observed as the training set size increases.

\begin{figure}[!htb]
    \centering
    \includegraphics[width=\linewidth]{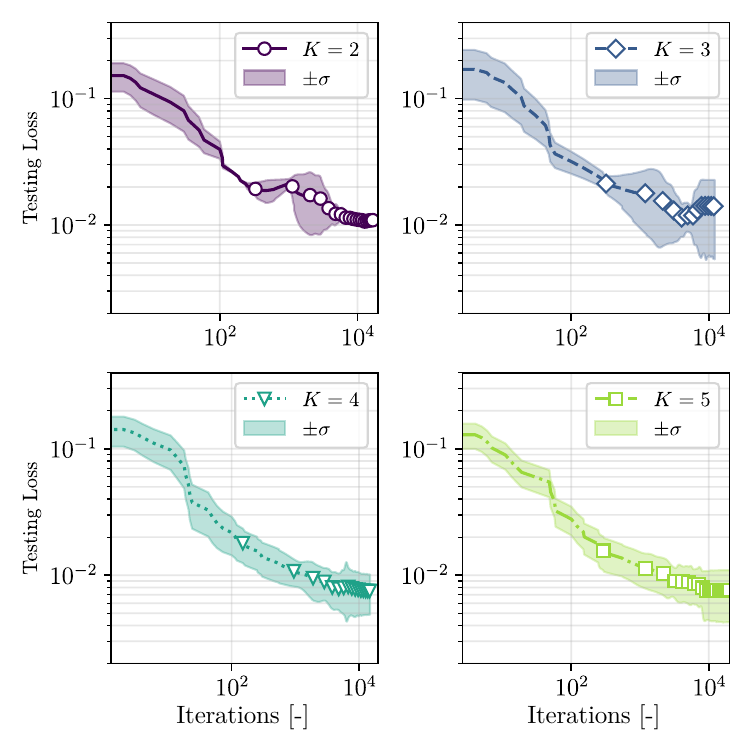}
    \caption{
    Out-of-fold validation loss for $K \in \{2,3,4,5\}$-fold cross-validation. Lines show the mean loss across the folds, while the shaded regions show the fold-wise standard deviation. 
    }
    \label{fig:loss_test_fold}
\end{figure}

The loss evolution on the out-of-fold validation dataset is presented in Figure \ref{fig:loss_test_fold}. The figure is divided into four subplots, corresponding to the different values of $K$. In each subplot, the line represents the loss curve averaged over all folds, while the shaded region indicates the variability between folds, defined as $\pm$ one standard deviation. 
To avoid spurious fluctuations in the loss curves, the training and validation losses are smoothed using a Gaussian filter with a standard deviation of 5 iterations. Each model is evaluated at the iteration corresponding to the onset of overfitting, identified as the point where the (smoothed) validation loss begins to increase \cite{goodfellow2016deep}.

\begin{table}[!htb]
\centering
\renewcommand{\arraystretch}{1.2}
\caption{
Mean and standard deviation of the training and out-of-fold validation loss functionals, defined in \eqref{eq:augmented_loss}, across $K$ folds.  For each fold, the reported validation loss corresponds to the minimum of the loss curve, smoothed with a Gaussian kernel of standard deviation $\sigma = 5$ iterations to reduce sensitivity to noise. The average loss is evaluated at the iteration that minimizes the mean curve across all folds.
}

\begin{tabular}{cc|cc|cc}
\hline
\multirow{2}{*}{$K$} & \multirow{2}{*}{Fold} & \multicolumn{2}{c|}{Training loss $(\times 10^{-3})$} & \multicolumn{2}{c}{Out-of-fold loss $(\times 10^{-3})$} \\
 &  & Per fold & Avg. & Per fold & Avg. \\ \hline
\multirow{2}{*}{2} & 1 & 4.4 & \multirow{2}{*}{$4.5 \pm 0.1$} & 10.4 & \multirow{2}{*}{$10.3 \pm 0.1$} \\
 & 2 & 4.6 &  & 10.3 &  \\ \hline
\multirow{3}{*}{3} & 1 & 4.6 & \multirow{3}{*}{$4.5 \pm 0.2$} & 8.1 & \multirow{3}{*}{$8.7 \pm 0.5$} \\
 & 2 & 4.7 &  & 9.0 &  \\
 & 3 & 4.4 &  & 8.9 &  \\ \hline
\multirow{4}{*}{4} & 1 & 4.1 & \multirow{4}{*}{$4.6 \pm 0.4$} & 8.0 & \multirow{4}{*}{$7.5 \pm 2.7$} \\
 & 2 & 4.9 &  & 6.3 &  \\
 & 3 & 5.0 &  & 4.6 &  \\
 & 4 & 4.3 &  & 11.0 &  \\ \hline
\multirow{5}{*}{5} & 1 & 4.6 & \multirow{5}{*}{$4.7 \pm 0.2$} & 10.9 & \multirow{5}{*}{$7.3 \pm 3.2$} \\
 & 2 & 4.9 &  & 4.1 &  \\
 & 3 & 4.6 &  & 5.6 &  \\
 & 4 & 5.0 &  & 5.3 &  \\
 & 5 & 4.5 &  & 10.4 &  \\ \hline
\end{tabular}

\label{tab:fold_loss_info}
\end{table}

For all cases, the out-of-fold loss remains consistently higher than the training loss reported in Figure \ref{fig:loss_train_fold}. Nevertheless, these curves follow a similar decreasing trend to the training loss, indicating that the model can generalize the learned relationships beyond the training dataset. However, this improvement is accompanied by an increase in the variability between folds, indicating that the model performance becomes more sensitive to the specific data partitioning. 
Moreover, when accounting for the variability between folds, the differences in minimum validation loss across values of $K$ are small relative to the observed fold-to-fold variability, suggesting that the framework's predictive capability is already well-established even with a limited amount of training data.

\begin{table*}[!htb]
\centering
\renewcommand{\arraystretch}{1.4}
\caption{
Global normalized root-mean-square error (NRMSE) between out-of-fold predicted and reference state variables, evaluated per fold and averaged across folds for each value of $K \in {2,3,4,5}$. The global NRMSE is computed as the average over all components of the state vector, and is reported both for the overall out-of-fold validation dataset (row `Overall') and for the specific operating scenarios: self-pressurization/relaxation, active-pressurization, venting, and sloshing. 
}
\resizebox{\textwidth}{!}{

\begin{tabular}{cc|cc|ccc|cccc|ccccc}
\hline
\multirow{2}{*}{} & \multirow{2}{*}{\makecell{Global\\{NRMSE $[\%]$}}} & \multicolumn{2}{c|}{$K=2$} & \multicolumn{3}{c|}{$K=3$} & \multicolumn{4}{c|}{$K=4$} & \multicolumn{5}{c}{$K=5$} \\
 &  & 1 & 2 & 1 & 2 & 3 & 1 & 2 & 3 & 4 & 1 & 2 & 3 & 4 & 5 \\ \hline
\multirow{2}{*}{Overall} & Per fold & 3.09 & 2.40 & 2.31 & 2.67 & 2.87 & 3.37 & 2.21 & 2.31 & 2.22 & 2.68 & 3.11 & 1.57 & 3.36 & 3.40 \\
 & Average & \multicolumn{2}{c|}{2.75 $\pm$ 0.35} & \multicolumn{3}{c|}{2.62 $\pm$ 0.23} & \multicolumn{4}{c|}{2.53 $\pm$ 0.49} & \multicolumn{5}{c}{2.82 $\pm$ 0.68} \\ \hline
\multirow{2}{*}{\makecell{Self.-press/\\{Relax}}} & Per fold & 2.54 & 2.07 & 2.09 & 2.04 & 2.31 & 3.03 & 2.19 & 2.08 & 1.94 & 1.95 & 2.76 & 1.51 & 2.88 & 2.96 \\
 & Average & \multicolumn{2}{c|}{2.31 $\pm$ 0.24} & \multicolumn{3}{c|}{2.14 $\pm$ 0.12} & \multicolumn{4}{c|}{2.31 $\pm$ 0.42} & \multicolumn{5}{c}{2.41 $\pm$ 0.58} \\ \hline
\multirow{2}{*}{Active-press.} & Per fold & 2.98 & 2.38 & 2.16 & 2.72 & 1.93 & 1.85 & 1.97 & 1.78 & 1.97 & 2.41 & 2.13 & 1.38 & 3.36 & 2.15 \\
 & Average & \multicolumn{2}{c|}{2.68 $\pm$ 0.30} & \multicolumn{3}{c|}{2.27 $\pm$ 0.33} & \multicolumn{4}{c|}{1.89 $\pm$ 0.08} & \multicolumn{5}{c}{2.29 $\pm$ 0.64} \\ \hline
\multirow{2}{*}{Venting} & Per fold & 3.09 & 0.73 & 0.69 & 0.69 & 3.24 & 0.72 & 0.70 & 0.70 & 0.68 & 0.72 & 0.74 & 0.71 & 0.79 & 0.74 \\
 & Average & \multicolumn{2}{c|}{1.91 $\pm$ 1.18} & \multicolumn{3}{c|}{1.54 $\pm$ 1.20} & \multicolumn{4}{c|}{0.70 $\pm$ 0.01} & \multicolumn{5}{c}{0.74 $\pm$ 0.03} \\ \hline
\multirow{2}{*}{Sloshing} & Per fold & 3.90 & 3.06 & 3.42 & 3.47 & 4.05 & 5.71 & 2.29 & 3.38 & 3.09 & 4.05 & 4.28 & 2.02 & 4.51 & 4.93 \\
 & Average & \multicolumn{2}{c|}{3.48 $\pm$ 0.42} & \multicolumn{3}{c|}{3.65 $\pm$ 0.29} & \multicolumn{4}{c|}{3.62 $\pm$ 1.27} & \multicolumn{5}{c}{3.96 $\pm$ 1.01} \\ \hline
\end{tabular}

}
\label{tab:fold_nrmse}
\end{table*}

This trend is further quantified in Table \ref{tab:fold_loss_info}, which summarizes the minimum training and out-of-fold losses obtained for each fold and each investigated value of $K$, together with the corresponding averaged values and standard deviations. The results show that, although the average minimum loss decreases with increasing $K$, the dispersion of the results between folds increases, suggesting that while the additional training data improves the average predictive capability, the obtained performance becomes more dependent on the specific data partition. In particular, the average minimum out-of-fold loss decreases from $10.3\times10^{-3}$ for $K=2$ to $7.3\times10^{-3}$ for $K=5$, corresponding to an improvement of approximately $29\%$. However, the standard deviation between folds increases from $0.1\times10^{-3}$ to $3.2\times10^{-3}$.

This behavior is partly attributed to the limited size and heterogeneity of the available database. While increasing $K$ increases the number of training data available, it simultaneously reduces the size of the out-of-fold validation set. 
As a result, if a validation fold contains time series corresponding to regions of the experimental parameter space that are less represented in the training data, the resulting prediction errors have a stronger influence on the average out-of-fold loss. This may lead to larger errors for individual folds and increased variability between folds. Therefore, the measured out-of-fold loss becomes more sensitive to the specific train-validation partition, particularly when the validation subset contains only a limited number of time series.

To further assess whether the observed differences in the loss functional translate into improved predictions, the normalized root mean square error (NRMSE) is evaluated. For each state component, the root mean square error is first computed as
\begin{equation}
\mathrm{RMSE}(\bm{s}) =
\sqrt{\frac{1}{n_t}\sum_{k=0}^{n_t-1}
\left( \check{\bm{s}}(t_k) - \bm{s}(t_k;\bm{w}) \right)^2}
\in \mathbb{R}^{n_s}
\,,
\end{equation}
where $\check{\bm{s}}$ and $\bm{s}$ denote the reference and predicted state vectors, respectively, and $n_t$ and $n_s$ represent the number of time samples and state variables.
The RMSE is subsequently normalized by the time-averaged magnitude of each reference state component:
\begin{equation}
\mathrm{NRMSE}(\bm{s}) =
\frac{\mathrm{RMSE}(\bm{s})}
{\displaystyle \frac{1}{n_t}\sum_{k=0}^{n_t-1} \check{\bm{s}}(t_k)}
\in \mathbb{R}^{n_s}
\,.
\end{equation}
This formulation provides a component-wise measure of the prediction error, allowing the contribution of individual state variables to be assessed independently. The averaging in the denominator is computed over the samples contained in the database under consideration, e.g., over the entire database when reporting overall NRMSE values, or restricted to the corresponding subset of samples when NRMSE is reported per scenario.

In addition to the component-wise NRMSE values, a global NRMSE is defined by averaging the normalized errors across all $n_s$ state components:
\begin{equation}
\mathrm{NRMSE}_{\mathrm{global}} =
\frac{1}{n_s} \sum_{j=0}^{n_s-1}
\mathrm{NRMSE}(s_j)
\,.
\end{equation}
The global NRMSE provides a single metric for comparing the overall predictive performance of the different $K$-fold configurations, while the component-wise values allow the prediction accuracy of individual states to be examined. 

The resulting global NRMSE values for each fold and each value of $K$ are reported in Table \ref{tab:fold_nrmse}. The results are reported in five categories. The first, labeled `overall', computes the NRMSE using all available points in the database. The remaining four correspond to the key operating scenarios of interest: (1) static, undisturbed evolution (self-pressurization and relaxation), (2) active-pressurization, (3) venting, and (4) sloshing.

The table indicates that, over all scenarios, the trends observed in the out-of-fold loss are not directly reflected in the NRMSE values. Although $K=5$ achieves the lowest average minimum out-of-fold loss, it does not result in the lowest prediction error, with $K=4$ yielding the smallest overall NRMSE ($2.53\%$) compared to $K=5$ ($2.82\%$). This apparent discrepancy is expected, as the loss functional and the NRMSE quantify different aspects of the model performance. Moreover, the differences between these values remain relatively small, indicating that all investigated values of $K$ lead to comparable predictive accuracy. The fold-to-fold variability observed in the results further supports the previous findings regarding the influence of the training-validation partition. In particular, $K=5$ exhibits the largest spread in the overall NRMSE ($2.82 \pm 0.68\%$).

Focusing now on the individual operating scenarios, the corresponding NRMSE values are analyzed.
The self-pressurization/relaxation cases exhibit limited sensitivity to the cross-validation strategy, with all values of $K$ yielding comparable NRMSE values. This indicates that these dynamics are sufficiently represented in the database, such that increasing the amount of training data provides only marginal improvements.
On the other hand, the active-pressurization cases show a clearer improvement with increasing $K$, with the lowest error obtained for $K=4$. This suggests that the additional training samples improve the representation of the transient pressurization dynamics. However, the increase in variability observed for $K=5$ suggests that, for smaller validation sets, the estimated performance becomes more dependent on the specific experiments included in each fold.

The venting cases present the strongest dependence on the fold composition, with a substantial reduction in both the average error and fold-to-fold variability as $K$ increases. This trend suggests that a sufficient representation of the venting dynamics within the training folds is important for achieving consistent predictions.
Lastly, the sloshing cases consistently exhibit the largest prediction errors across all cross-validation configurations, highlighting the increased difficulty associated with these dynamics. Moreover, increasing $K$ does not lead to a systematic reduction in error, suggesting that the prediction accuracy is limited not only by the amount of available training data but also by the intrinsic complexity and variability of the sloshing phenomena represented in the database.

\ref{app:component_out_of_fold_K5} complements this analysis by presenting the component-wise out-of-fold performance for $K=5$, further illustrating the variability in predictive performance across folds for each state component.

To complement these performance metrics, we now examine representative out-of-fold time series from the $K=5$ configuration. Two trained models are examined: the model obtained from fold 3, which achieved the lowest overall global NRMSE, and the model obtained from fold 5, which exhibited the highest overall global NRMSE (Table \ref{tab:fold_nrmse}). For each model, two representative validation time series are presented, showing the corresponding evolution of the state variables and learned closure parameters.

\begin{figure*}[!htb]
    \centering
    \begin{subfigure}[t]{0.48\textwidth}
        \centering
        \includegraphics[width=\linewidth]{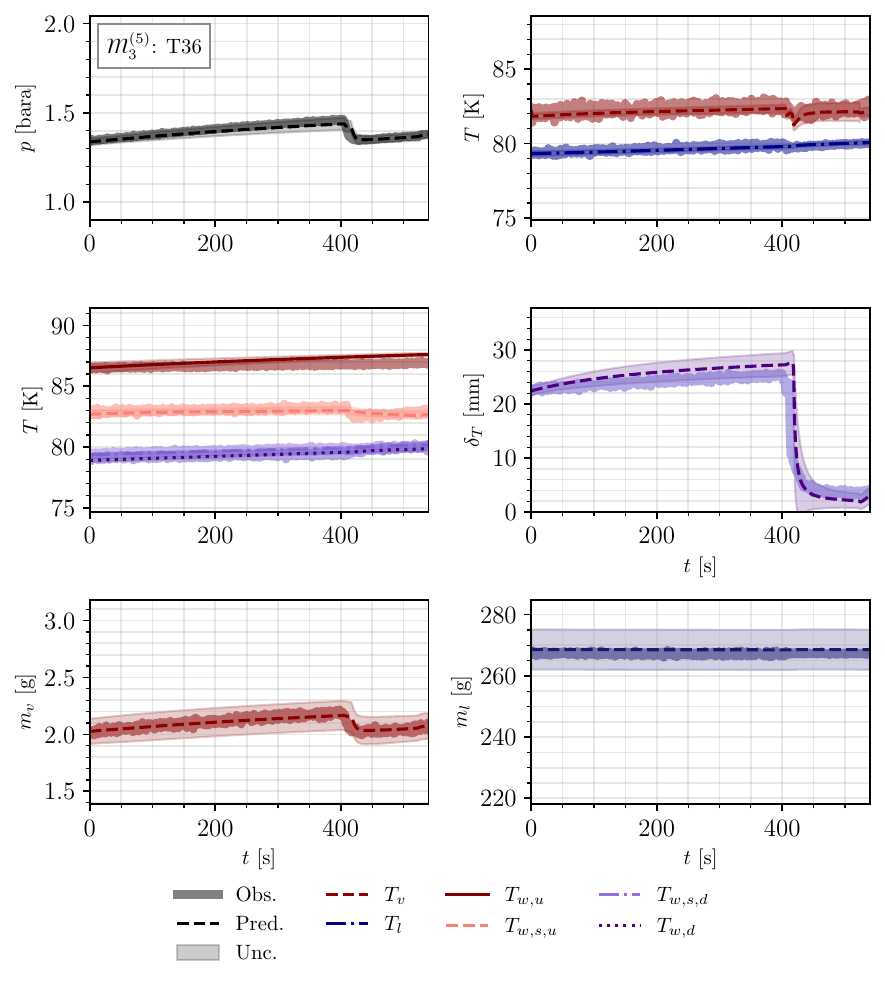}
        \caption{
        Time series of the experimental data (thick solid line) and predictions generated by model $\mathcal{M}_{3}^{(5)}$ (dashed line) for the state vector components in experimental case T36.
        }
        \label{fig:M35_T1}
    \end{subfigure}
    \hfill
    \begin{subfigure}[t]{0.48\textwidth}
        \centering
        \includegraphics[width=\linewidth]{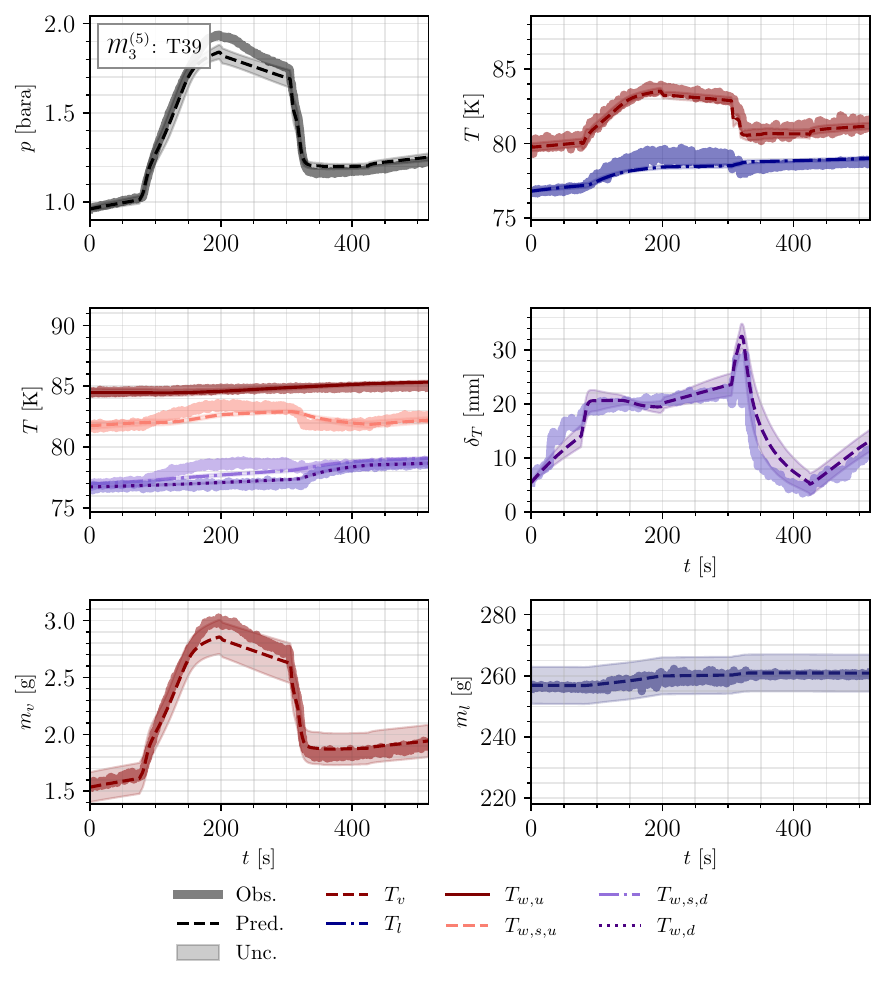}
        \caption{
        Time series of the experimental data (thick solid line) and predictions generated by model $\mathcal{M}_{3}^{(5)}$ (dashed line) for the state vector components in experimental case T39.
        }
        \label{fig:M35_T2}
    \end{subfigure}
    \vspace{1em}
    \begin{subfigure}[t]{0.48\textwidth}
        \centering
        \includegraphics[width=\linewidth]{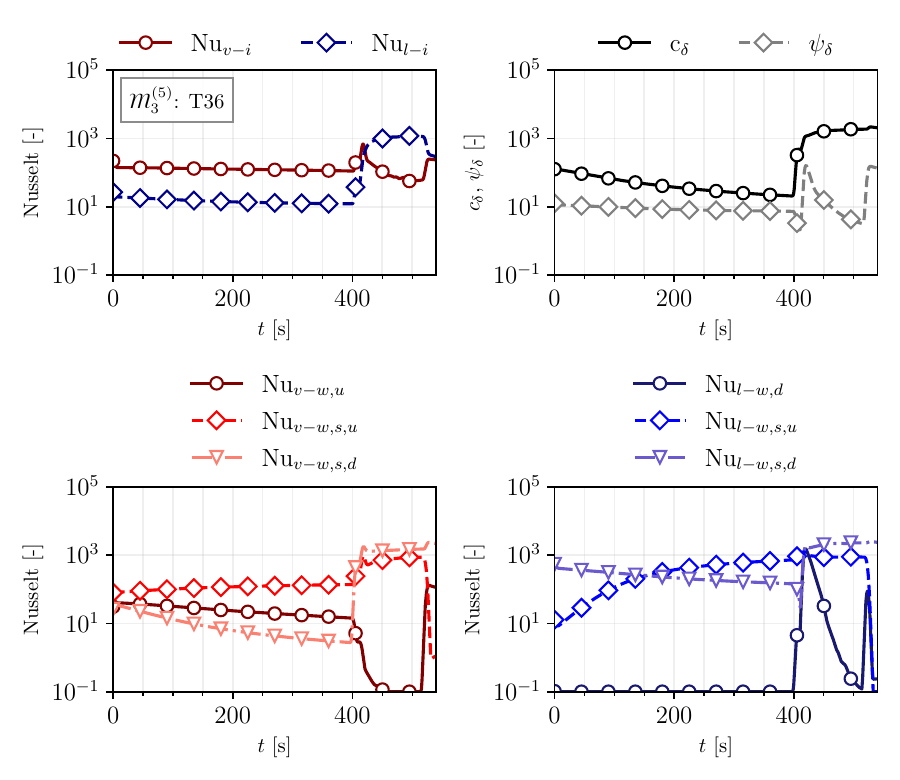}
        \caption{
        Time series of the closure parameters generated by model $\mathcal{M}_{3}^{(5)}$ in case T36.
        }
        \label{fig:M35_C1}
    \end{subfigure}
    \hfill
    \begin{subfigure}[t]{0.48\textwidth}
        \centering
        \includegraphics[width=\linewidth]{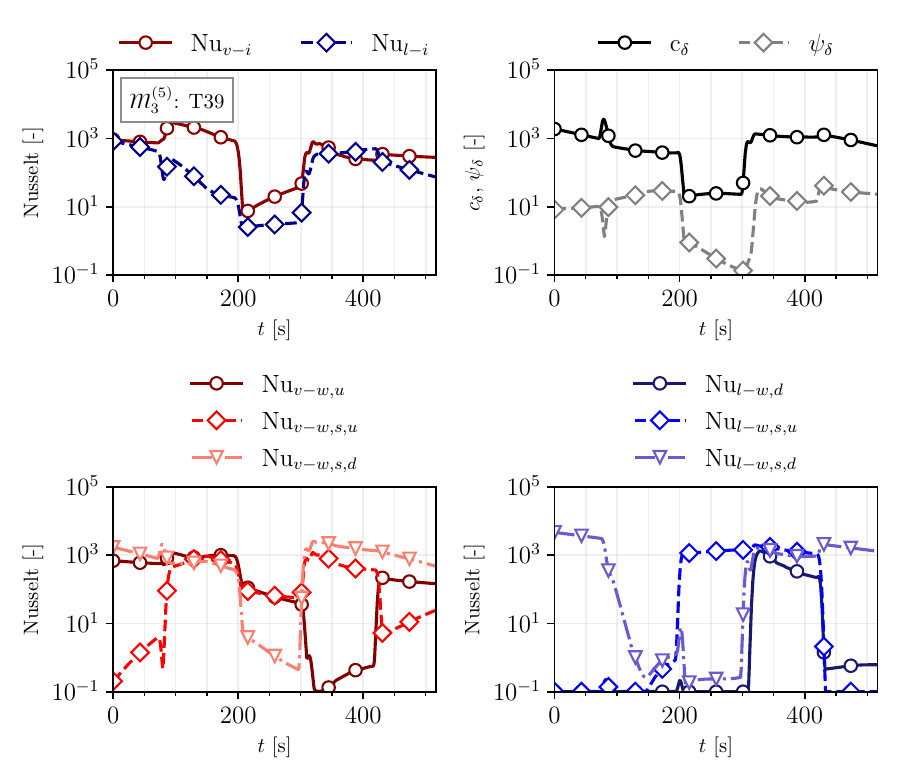}
        \caption{
        Time series of the closure parameters generated by model $\mathcal{M}_{3}^{(5)}$ in case T39.
        }
        \label{fig:M35_C2}
    \end{subfigure}
    
    \caption{Comparison of model $\mathcal{M}_{3}^{(5)}$ predictions against experimental data for two out-of-fold validation cases: T36 (left) and T39 (right). Top row shows the state vector time series; bottom row shows the corresponding closure parameters.}
    \label{fig:M35_quad}
\end{figure*}

In the following, we denote the model obtained from fold $j$ of the $K-$fold cross-validation procedure as $\mathcal{M}_{j}^{(K)}$. Accordingly, the two models considered in the following analysis are $\mathcal{M}_3^{(5)}$ and $\mathcal{M}_5^{(5)}$.
Figure \ref{fig:M35_quad} illustrates representative out-of-fold predictions of model $\mathcal{M}_3^{(5)}$ for validation cases T36 (\ref{fig:M35_T1} and \ref{fig:M35_C1}) and T39 (\ref{fig:M35_T2} and \ref{fig:M35_C2}).

The experimental state is shown as a thick solid line, while the out-of-fold predictions are shown as thin dashed, dotted, or dash-dotted lines. The prediction spread, depicted in the figures, is estimated by performing 100 realizations in which: (1) the initial conditions are perturbed with normally distributed noise with a standard deviation of $5\%$ of each state component, and (2) the neural networks are evaluated with a dropout rate of $10\%$. The shaded regions in the plots represent this spread, defined as twice the standard deviation of all realizations at each time instant (a $2\sigma$ spread), while the dashed line denotes the mean prediction across realizations.
The present approach does not capture all sources of neural-network uncertainty, such as weight initialization, optimizer stochasticity, or hyperparameter choice. A more comprehensive assessment would require an ensemble strategy across varying configurations. Nevertheless, the current approach captures variability in initial conditions and the prediction process, providing an estimate of the model sensitivity to these factors. Furthermore, each cross-validation fold was trained from a single random realization; the reported fold-to-fold spread (Section~\ref{sec:res_sensitivity}) therefore also contains unquantified optimization stochasticity that was not isolated from the effect of data partitioning.

We first examine case T36, shown in Figures \ref{fig:M35_T1} and \ref{fig:M35_C1}. The experiment initially undergoes self-pressurization for approximately the first 400~s due to the imposed heater temperatures, with average values of $\overline{T}_{h,s}\approx 84.9$~K and $\overline{T}_{h,u}\approx 88.2$~K. This results in a gradual pressure increase to a peak value of approximately $1.5$~bara. Following this static period, the tank is subjected to swirl sloshing with $A_e/R\approx0.026$ and $f_e/f_{1,1}\approx1.06$, inducing a rapid pressure drop that reduces the tank pressure to approximately 1.36~bara, close to its initial value at $t=0$.
The model accurately predicts the pressure evolution, including both the self-pressurization-induced pressure rise and the subsequent sloshing-induced pressure drop. It also accurately represents the measured fluid and solid temperatures. Compared with active-pressurization, self-pressurization is a considerably slower process, resulting in only modest variations in the measured temperatures. Nevertheless, the model successfully captures the subtle decrease in vapor temperature at the onset of sloshing, as well as the corresponding cooling of the upper side-wall temperature. The thermal boundary layer thickness is also well predicted. During the static phase, this variable exhibits a gradual increase, reflecting the progressive development of thermal stratification within the liquid. Upon the onset of sloshing, the thermal boundary layer is rapidly disrupted due to the enhanced mixing induced by swirl motion, causing $\delta_T$ to collapse towards zero. Finally, the model accurately captures the boil-off dynamics during the static phase, with the vapor mass increasing by approximately $10\%$ before decreasing during sloshing, indicating the occurrence of condensation as a result of vapor cooling.
Focusing now on the evolution of the closure parameters, the vapor–interface $\mathrm{Nu}_{v-i}$ and liquid–interface $\mathrm{Nu}_{l-i}$ Nusselt numbers remain relatively low during the static phase, dominated by free convection. With the onset of sloshing, forced convection enhances interfacial heat transfer, with swirl-induced mixing increasing $\mathrm{Nu}_{l-i}$ by two orders of magnitude and $\mathrm{Nu}_{v-i}$ by one. The thermal boundary layer parameters reflect this transition: during the static phase, $c_\delta$ remains positive but gradually decreases, indicating a slowdown of stratification growth as temperature gradients stabilize. Conversely, $\psi_\delta$ increases by approximately one order of magnitude at sloshing onset, indicating the rapid destruction of the stratification layer and collapse of $\delta_T$. The wall-fluid Nusselt numbers further highlight the change in heat transfer mechanisms: during self-pressurization, exchanges with the side walls and top cover dominate, driving wall-induced buoyancy fluxes, whereas during sloshing, $\mathrm{Nu}_{l-w,d}$ increases by four orders of magnitude due to the enhanced mixing. The moderate increase of the remaining wall Nusselt numbers further confirms the effectiveness of swirl-induced mixing.

Case T39, shown in Figures \ref{fig:M35_T2} and \ref{fig:M35_C2}, consists of a four-stage experiment. The test begins with an initial static phase lasting approximately $80$~s, followed by an active-pressurization stage with $\overline{\dot{m}}_p \approx 0.03$~g/s, until the vapor pressure reaches $1.93$~bara. At this point, the pressurant port is closed, and the system enters a relaxation phase lasting approximately two minutes. During this period, the pressure decreases due to thermal exchanges and condensation between the warmer vapor and the surrounding colder regions of the tank, including the side walls and liquid phase. The experiment then transitions to swirl sloshing under excitation conditions identical to those of Case T36. However, unlike T36, the system in T39 undergoes active-pressurization prior to sloshing, resulting in stronger thermal gradients and a system further from thermodynamic equilibrium. Consequently, the onset of sloshing produces a sharp pressure decrease, from approximately $1.7$~bara to $1.2$~bara.
The model shows overall good agreement with the experimental data. However, the stronger transient conditions in this case result in larger discrepancies compared with the previous scenario. The maximum pressure is underpredicted, leading to a similar underestimation of the relaxation pressure drop rate, while the sloshing-induced pressure decrease is accurately captured. Similar behavior is observed for the vapor mass. The vapor temperature is well predicted, whereas the liquid temperature increase during pressurization and relaxation is underestimated. The largest deviations and uncertainty bands are observed for $\delta_T$. Nevertheless, the uncertainty bounds generally encompass the reference data. The evolution of $\delta_T$ follows the expected trends, with an initial increase during pressurization, continued growth during relaxation, and a sharp peak at sloshing onset due to the redistribution of thermal gradients. As swirl-induced mixing develops and the liquid temperature becomes more homogeneous, $\delta_T$ rapidly collapses. The subsequent increase for $t>400$~s results from the renewed development of thermal gradients driven by heat leakage from the tank walls.
The closure parameters exhibit physically consistent trends throughout the experiment. The vapor–interface, liquid–interface, and wall–vapor Nusselt numbers increase during active-pressurization, decrease during the relaxation phase, and increase again during sloshing. This behavior is consistent with the underlying heat transfer mechanisms, as both active-pressurization and sloshing enhance convective fluxes due to the induced fluid motion. In contrast, the relaxation phase is primarily governed by natural convection. The magnitude of the inferred Nusselt numbers is also comparable to that obtained in the previous case, despite the different operating conditions and initial state. This consistency, particularly under identical sloshing conditions, indicates that the data-driven closure parameters remain physically meaningful across different scenarios.

\begin{figure*}[!htb]
    \centering
    \begin{subfigure}[t]{0.48\textwidth}
        \centering
        \includegraphics[width=\linewidth]{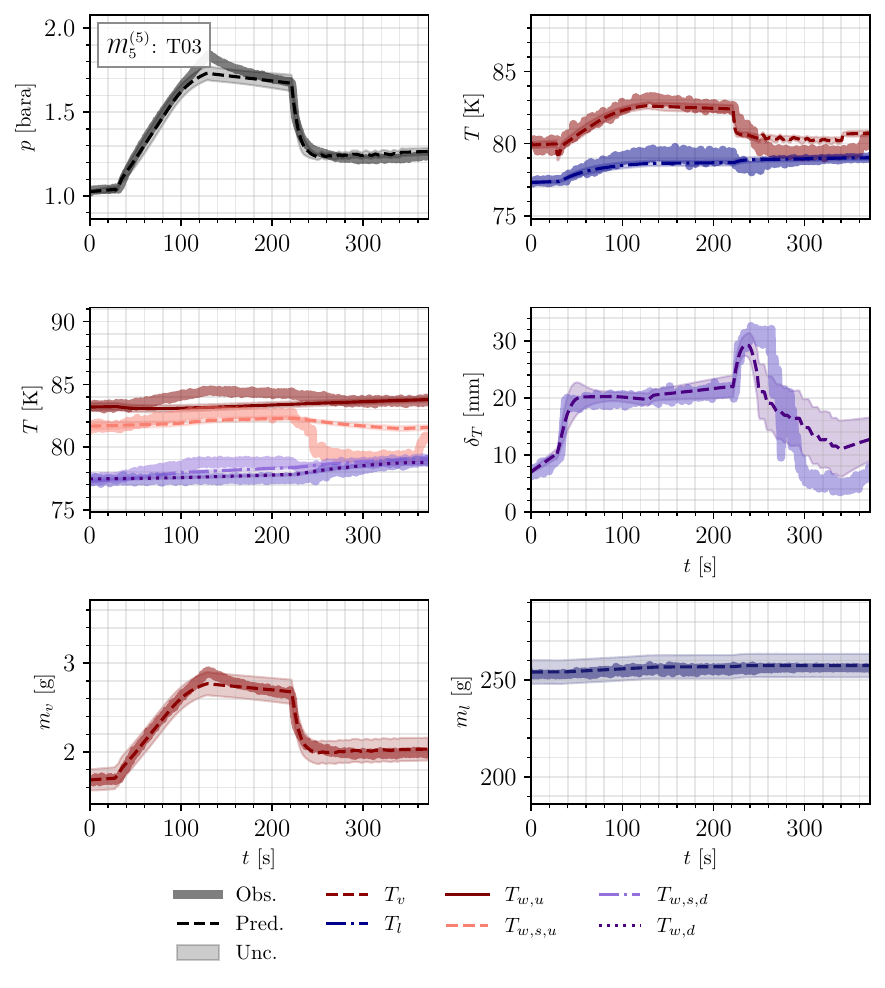}
        \caption{
        Time series of the experimental data (thick solid line) and predictions generated by model $\mathcal{M}_{5}^{(5)}$ (dashed line) for the state vector components in experimental case T03.
        }
        \label{fig:M55_T1}
    \end{subfigure}
    \hfill
    \begin{subfigure}[t]{0.48\textwidth}
        \centering
        \includegraphics[width=\linewidth]{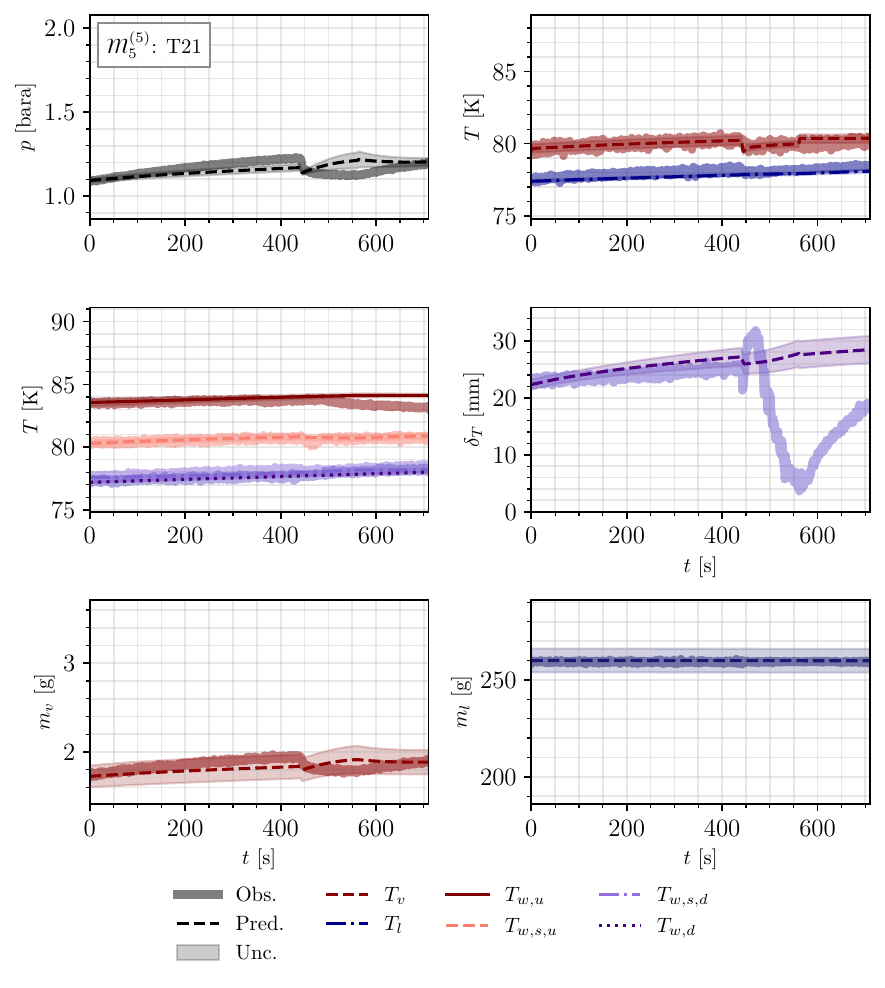}
        \caption{
        Time series of the experimental data (thick solid line) and predictions generated by model $\mathcal{M}_{5}^{(5)}$ (dashed line) for the state vector components in experimental case T21.
        }
        \label{fig:M55_T2}
    \end{subfigure}
    \vspace{1em}
    \begin{subfigure}[t]{0.48\textwidth}
        \centering
        \includegraphics[width=\linewidth]{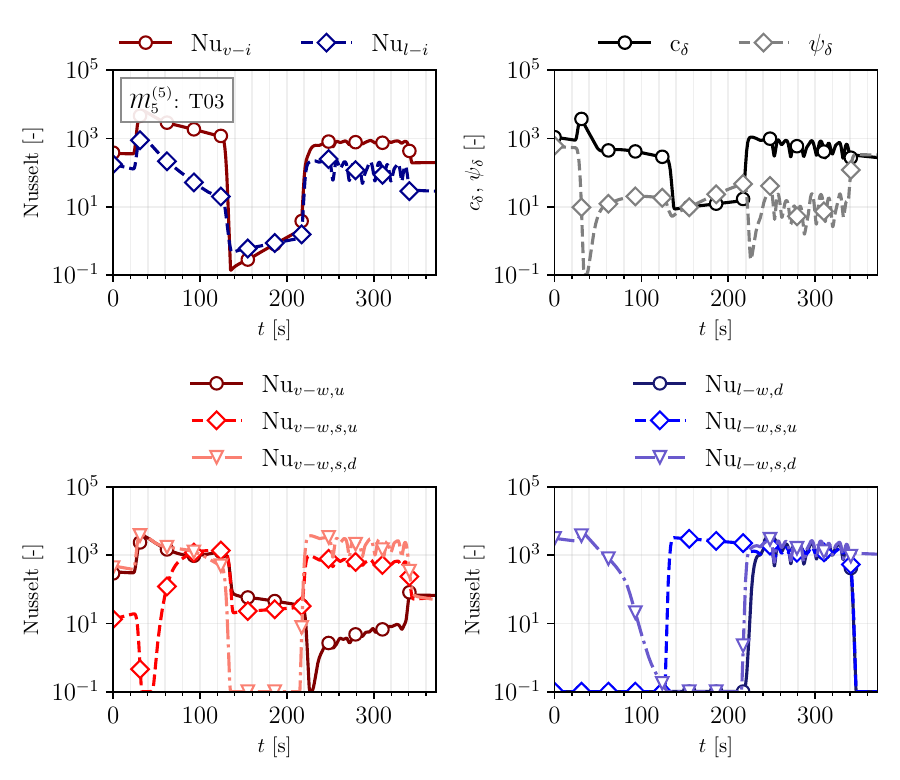}
        \caption{
        Time series of the closure parameters generated by model $\mathcal{M}_{5}^{(5)}$ in case T03.
        }
        \label{fig:M55_C1}
    \end{subfigure}
    \hfill
    \begin{subfigure}[t]{0.48\textwidth}
        \centering
        \includegraphics[width=\linewidth]{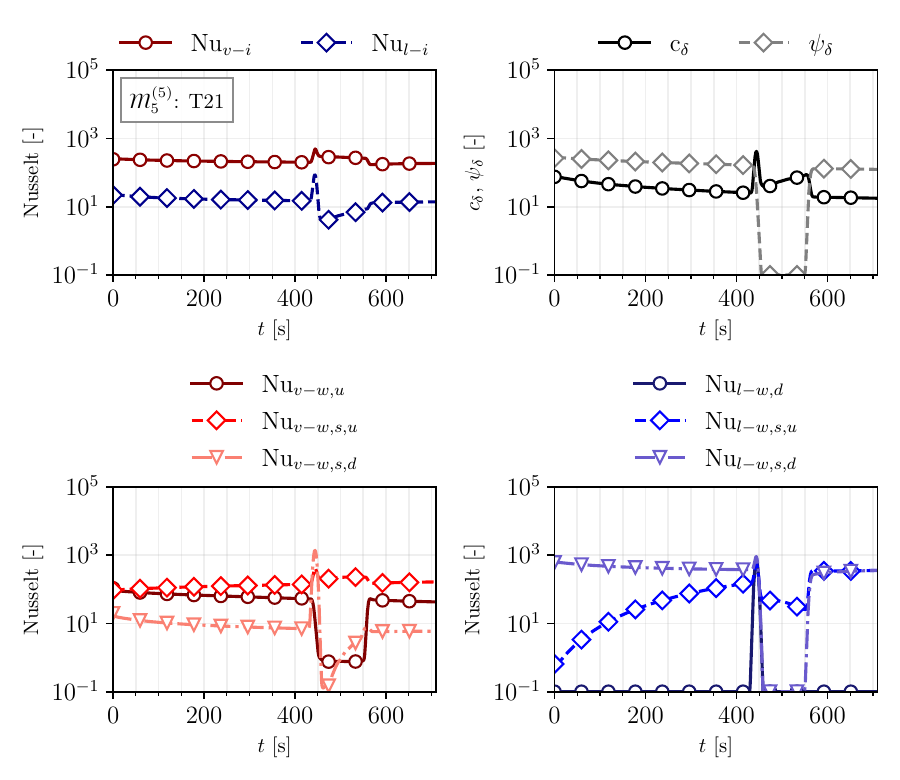}
        \caption{
        Time series of the closure parameters generated by model $\mathcal{M}_{5}^{(5)}$ in case T21.
        }
        \label{fig:M55_C2}
    \end{subfigure}
    
    \caption{Comparison of model $\mathcal{M}_{5}^{(5)}$ predictions against experimental data for two out-of-fold validation cases: T03 (left) and T21 (right). Top row shows the state vector time series; bottom row shows the corresponding closure parameters.}
    \label{fig:M55_quad}
\end{figure*}

Figure \ref{fig:M55_quad} illustrates out-of-fold predictions of model $\mathcal{M}_5^{(5)}$ for validation cases T03 (\ref{fig:M55_T1} and \ref{fig:M55_C1}) and T21 (\ref{fig:M55_T2} and \ref{fig:M55_C2}).

Case T03 presents an initial static phase followed by active-pressurization with an average pressurant flow rate of $\overline{\dot{m}}_p\approx0.03$~g/s. This stage produces steady pressure increase from approximately $1.03$~bara to $1.86$~bara, accompanied by a similar rise in both the vapor and bulk liquid temperatures, as well as a rapid growth of the thermal boundary layer. During the subsequent relaxation phase, the thermal gradients established during pressurization drive a pressure decay lasting approximately five minutes. Over this period, the pressure decreases to approximately $1.68$~bara at $t\approx220$~s, while the liquid thermal stratification continues to develop, as reflected by the increasing $\delta_T$. Finally, sloshing ($A_e/R\approx0.025$ and $f_e/f_{1,1}\approx0.94$) is initiated, producing an exponential-like pressure drop, the collapse of $\delta_T$, and the homogenization of both the fluid and wall temperature fields.
The model agrees well with the experimental measurements across the different operating scenarios. The pressure evolution is well reproduced, although the peak value, at the end of the pressurization, is slightly underestimated. The vapor temperature is accurately predicted during the pressurization and relaxation stages, whereas the cooling induced by sloshing is underestimated. In contrast, the liquid temperature is well captured throughout the experiment.
The wall temperatures are also generally well predicted, although the upper side-wall temperature exhibits a weaker response to sloshing in the model. Experimentally, $\check{T}_{w,s,u}$ rapidly decreases to nearly the same value as $\check{T}_{w,s,d}$, whereas the predicted temperature undergoes only a modest cooling. Finally, the thermal boundary layer thickness is well reproduced, following trends similar to those observed in Case T39 (Figure \ref{fig:M35_T2}), albeit with relatively large uncertainty bands during the sloshing phase.
The closure parameters exhibit trends consistent with the changing scenarios. Qualitatively, the vapor–interface, liquid–interface, and wall-fluid Nusselt numbers increase during active-pressurization, decrease during relaxation, and increase again once sloshing begins.

Case T21 corresponds to a self-pressurization experiment followed by planar sloshing. During the initial static phase, the imposed wall heating leads to a gradual pressure increase as heat is transferred into the fluid, with the vapor pressure rising from approximately $1.1$~bara to $1.25$~bara over the duration of the test. Following this period, sloshing is initiated with $A_e/R\approx 0.025$ and $f_e/f_{1,1}\approx 0.91$, producing a steady pressure decrease.
Overall,  the predictions and associated uncertainty bands encompass most of the experimental data. The largest discrepancies are observed for the thermal boundary layer thickness, $\delta_T$. At the onset of sloshing, i.e., $t\approx480$~s, the model predicts a sharp increase in $\delta_T$, suggesting that the planar excitation initially promotes the growth of the thermal boundary layer. However, the experimental data shows this initial increase is followed by a rapid collapse and subsequent recovery, indicating that the model does not fully capture the competing stratification and destratification dynamics induced by this planar sloshing excitation. A similar trend is observed in the pressure response, where the sloshing-induced pressure drop is underestimated, while the subsequent pressure recovery is overpredicted. This suggests that the model underestimates the intensity of the mixing generated by the sloshing motion, leading to the observed discrepancies in $p_v$, $m_v$, and $\delta_T$. Nevertheless, the fluid and wall temperatures are generally well reproduced, with the exception of the upper cover temperature, whose cooling following the onset of sloshing is also underestimated. These discrepancies may indicate that low-amplitude sloshing conditions are underrepresented in the training database.
The inferred closure parameters follow trends similar to those observed for Case T36 (Figure \ref{fig:M35_C1}). However, lower values are obtained for $\mathrm{Nu}_{l-i}$, $\mathrm{Nu}_{v-w,s,u}$, $\mathrm{Nu}_{l-w,s,u}$, and $\mathrm{Nu}_{l-w,s,d}$, reflecting the lower-amplitude sloshing dynamics observed in the state evolution.

Overall, both models exhibit good predictive performance across unseen validation cases, capturing the dominant thermodynamic and thermal-hydraulic transients despite the diversity of operating conditions. Although $\mathcal{M}_5^{(5)}$ shows larger discrepancies than $\mathcal{M}_3^{(5)}$, particularly for planar sloshing, the inferred closure parameters remain physically consistent and evolve according to the expected heat transfer mechanisms. This highlights the ability of the proposed framework to generalize to previously unseen experiments while maintaining physically meaningful predictions.


\subsection{Full-database model performance}
\label{sec:full_model_performance}

The previous analysis highlighted the impact of data partitioning on model performance and generalization capability. Building on these findings, this section evaluates the model trained on the complete experimental database, focusing on its reconstruction accuracy over the full experimental domain.

\begin{figure}[!htb]
    \centering
    \includegraphics[width=\linewidth]{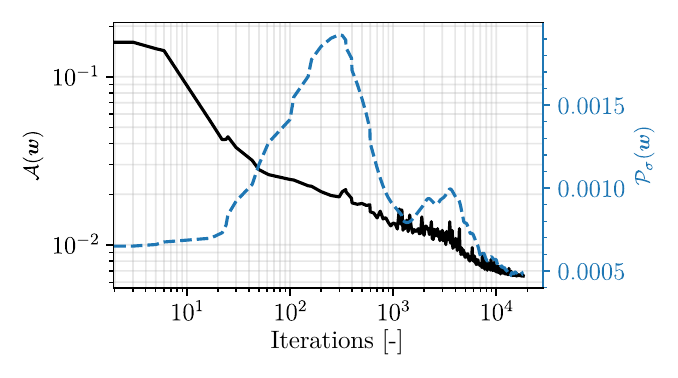}
    \caption{
    Loss functional $\mathcal{A}(\bm{w})$ for the model evaluated on the complete database, alongside the isolated entropy-based penalty term $\mathcal{P}_\sigma$.
    }
    \label{fig:loss}
\end{figure}

Figure \ref{fig:loss} shows the evolution of the loss functional for the training set over the course of the optimization. As training progresses, the augmented loss decreases by nearly two orders of magnitude, matching the order of magnitude of the values observed in Figure \ref{fig:loss_train_fold}, and reaching a final value of $( 6.58 \pm 0.01 )\times 10^{-3}$.
The figure also shows, on the same shared x-axis, the isolated entropy penalty term $\mathcal{P}_\sigma$ as a dashed blue line, with its corresponding values indicated on the right-hand vertical axis. The dashed curve for $\mathcal{P}_\sigma$ starts from a small but non-zero value. As training progresses, the penalty term rises to a maximum around iterations $10^{2}$–$10^{3}$, roughly coinciding with the plateau observed in $\mathcal{A}(\mathbf{w})$, before decreasing again as the optimization converges, settling to a value comparable to its initial level. This behavior is consistent with the notion that the penalty plays an active role in discouraging negative entropy production during training. The non-monotonic evolution may reflect the optimizer exploring configurations that risk violating the second law, while the subsequent decrease could indicate a correction toward more physically admissible solutions.

Since the entire database is incorporated in training the model, the stopping criterion is defined in the absence of an independent validation set. Rather, the model is considered converged when the loss variations are limited to the fifth decimal place over 1000 iterations, which occurs at approximately iteration 15000.

\begin{table*}[!htb]
\centering
\renewcommand{\arraystretch}{1.4}
\caption{
Normalized root mean square error (NRMSE) for the model trained on the complete experimental database. Columns show the global error (mean over all state vector components) followed by errors for each component: pressure $p_v$, vapor and liquid temperatures ($T_v$, $T_l$), wall temperatures ($T_{w,u}$, $T_{w,{s,u}}$, $T_{w,{s,d}}$, $T_{w,d}$), vapor and liquid masses ($m_v$, $m_l$), liquid volume ($V_l$), and thermal boundary layer thickness ($\delta_T$). Rows indicate the scenario: overall (all data), self-pressurization/relaxation, active-pressurization, venting, and sloshing.
}

\begin{tabular}{c|c|ccccccccccc}
\hline
& 
\makecell{Global \\ {[\%]}}
& \makecell{$p_v$ \\ {[\%]}}
& \makecell{$T_v$ \\ {[\%]}} 
& \makecell{$T_l$ \\ {[\%]}} 
& \makecell{$T_{w,u}$ \\ {[\%]}} 
& \makecell{$T_{w,{s,u}}$ \\ {[\%]}} 
& \makecell{$T_{w,{s,d}}$ \\ {[\%]}} 
& \makecell{$T_{w,d}$ \\ {[\%]}} 
& \makecell{$m_v$ \\ {[\%]}} 
& \makecell{$m_l$ \\ {[\%]}} 
& \makecell{$V_l$ \\ {[\%]}} 
& \makecell{$\delta_T$ \\ {[\%]}} \\ \hline
Overall & 1.60 & 1.79 & 0.31 & 0.36 & 0.51 & 0.56 & 0.38 & 0.27 & 2.03 & 0.21 & 0.84 & 10.30 \\
Self-press./Relax. & 1.30 & 1.38 & 0.28 & 0.34 & 0.47 & 0.41 & 0.37 & 0.28 & 1.62 & 0.21 & 0.75 & 8.20 \\
Active-Press. & 1.57 & 2.42 & 0.30 & 0.50 & 0.69 & 0.63 & 0.61 & 0.18 & 2.44 & 0.27 & 0.98 & 8.21 \\
Venting & 0.65 & 0.23 & 0.29 & 0.13 & 0.59 & 0.15 & 0.14 & 0.12 & 1.31 & 0.11 & 0.96 & 3.16 \\
Sloshing & 2.30 & 2.49 & 0.40 & 0.40 & 0.48 & 0.85 & 0.30 & 0.33 & 2.79 & 0.19 & 0.91 & 16.13 \\
\hline
\end{tabular}
\label{tab:model_NRMSE}
\end{table*}

\begin{figure*}[!htb]
    \centering
    \includegraphics[width=0.95\linewidth]{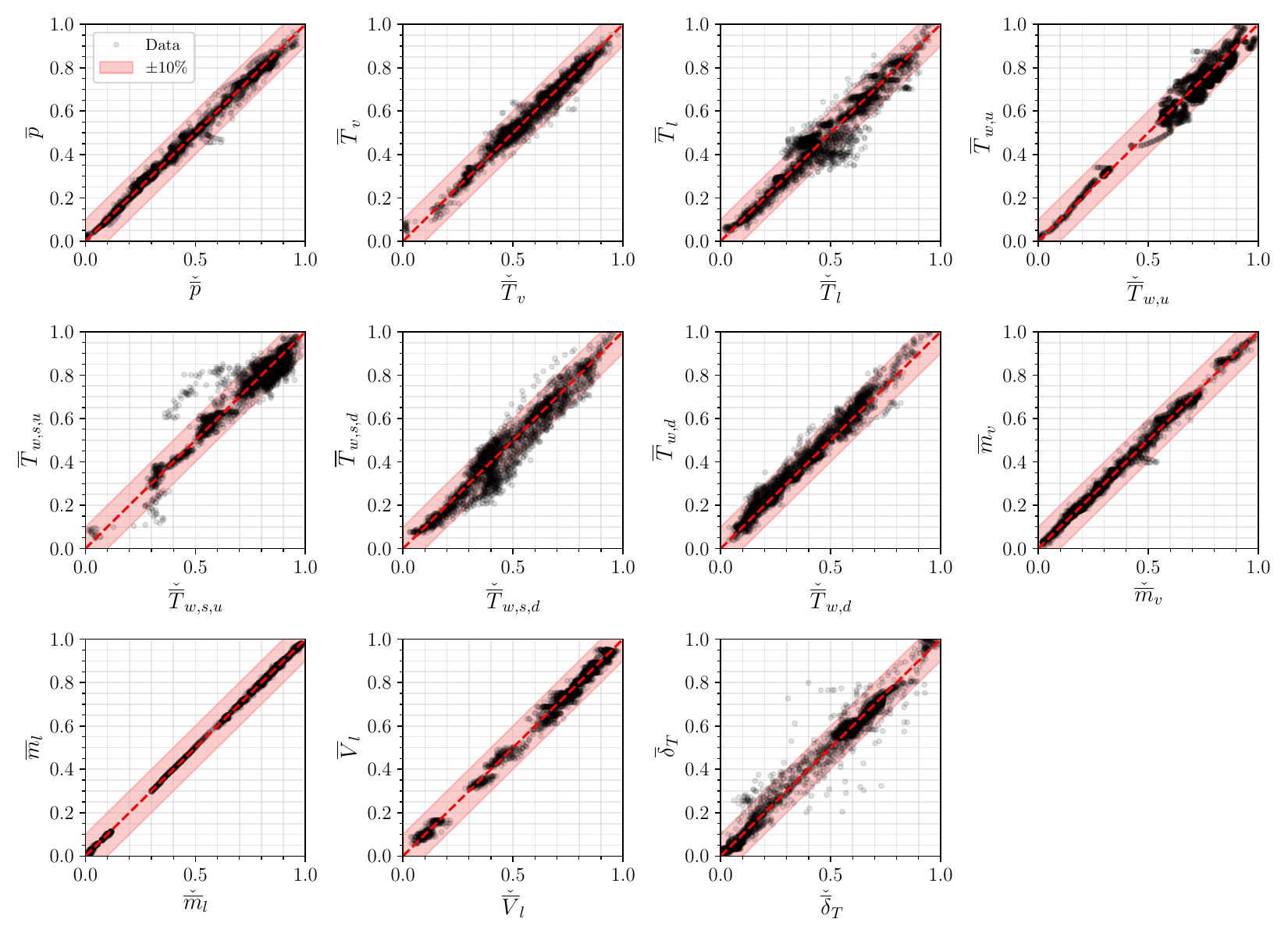}
    \caption{
    Comparison between the reconstructed state $\bm{s}=[p_v\,, T_v\,, T_l\,, T_{w_u}\,, T_{w_{s,u}}\,, T_{w_{s,d}}\,, T_{w_d}\,, m_v\,, m_l\,, V_l\,, \delta_T]$ and reference state $\check{\bm{s}}$ for the model trained on the complete database. The red shaded region denotes a deviation of $\pm 0.1$ in min-max-normalized coordinates (10\% of the normalization range). All state variables are normalized using min–max scaling to the range $[0,1]$.
    }
    \label{fig:state_comparison}
\end{figure*}

\begin{figure}[!htb]
    \centering
    \includegraphics[width=\linewidth]{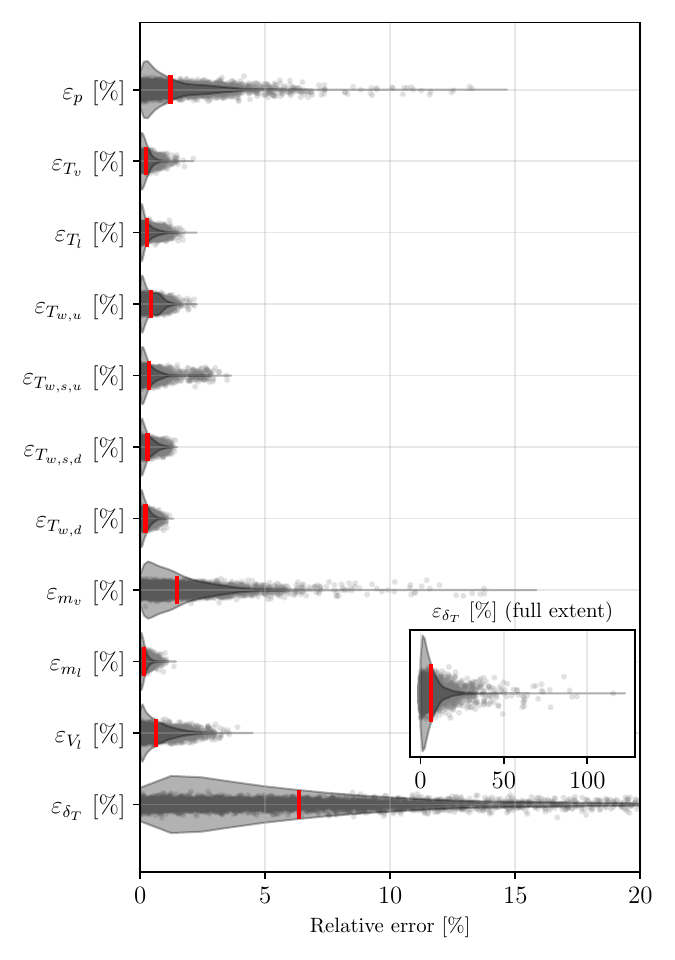}
    \caption{
    Distribution of the relative absolute error for each state vector component. Each subplot corresponds to one component, with the reconstruction error distribution shown as black violins. Scatter points are overlaid to illustrate the spread of individual data points.
    }
    \label{fig:violin}
\end{figure}

Table \ref{tab:model_NRMSE} reports the model performance in terms of the NRMSE. The global and component-wise errors are evaluated on the complete experimental dataset. Since these data points are entirely in-sample, the reported values should be interpreted as reconstruction errors rather than as a measure of predictive performance. 
The results are reported following an identical structure to that of Table \ref{tab:fold_nrmse}.

Focusing on the general model performance, the global reconstruction errors across all scenarios remain acceptable, staying under $2.3\%$. These values suggest that the reconstructed state achieves good agreement with the experimental data. These errors are lower than those reported in Table \ref{tab:fold_nrmse}. However, this is to be expected, since the latter correspond to out-of-fold prediction errors, whereas the values reported here are in-sample reconstruction errors.
Turning to the component-wise values, the minimum errors are achieved for the temperatures of the fluid and solid nodes, while the thermal boundary layer thickness $\delta_T$ consistently displays the largest errors.
The relatively large errors in $\delta_T$, reaching roughly $16\%$, may be attributed to the inherent difficulty of modeling the thermal boundary layer thickness. Unlike conserved quantities, $\delta_T$ is a geometric measure that depends on the thermal distribution within the liquid phase and on the fill level. Under static conditions, it can be captured by describing the growth of thermal stratification. However, when different operating scenarios are chained together, abrupt changes in thermal gradients (e.g., due to rapid pressurizations, or sloshing destratification) lead to sharp variations in $\delta_T$, which are challenging to represent in \eqref{eq:ddeltadt}. Nevertheless, monitoring this parameter provides valuable insight into the thermal state and inertia of the system. Moreover, it crucially encodes information about the spatial distribution of temperature with a single liquid node.

The key variable of interest is the tank pressure, which achieves a maximum scenario-specific pressure NRMSE of $2.49\%$ in the sloshing scenario. The order of magnitude of this error matches that of the vapor mass across the different scenarios. This trend reflects the coupling between pressure and vapor mass previously noted in \cite{ludwig_pressure_2013}. Concretely, when pressure dynamics are dominated by phase change, the fidelity of the pressure reconstruction becomes directly dependent on the accuracy of the vapor mass reconstruction.

Shifting focus to the performance across specific scenarios, the lowest errors occur in the static cases driven solely by heat ingress, i.e., self-pressurization and relaxation, as well as in the venting scenario. In contrast, active-pressurization and sloshing lead to higher errors, indicating that the fast time scales characteristic of these conditions are more challenging for the model to capture.

Figure \ref{fig:state_comparison} illustrates the performance of the trained model by comparing the true measured state $\check{\bm{s}}(t)$ with the reconstructed one $\bm{s}(t,\bm{w})$. Each state component was normalized using min–max scaling (see \eqref{eq:s_MinMax}). The red shaded region indicates a deviation of $\pm 0.1$ in min-max-normalized coordinates (10\% of the normalization range). Most points fall within this range, demonstrating good agreement between the reconstruction and the measurements. Nevertheless, some outliers are observed, particularly for the liquid temperature, the lower-lateral wall node, and the thermal boundary-layer thickness. 

This analysis is complemented by Figure \ref{fig:violin}, which depicts the distribution of the component-wise relative absolute error $\varepsilon_{\bm{s}}$ using violin plots with overlaid scatter data points.
For the pressure and vapor mass reconstructions, most data points remain within a relative error of $5\%$, showcasing the model's ability to accurately capture the dominant system dynamics. Nevertheless, some outliers reach errors of approximately $15\%$. These larger deviations are likely associated with transient conditions with stronger phase-change effects, where inaccuracies in the predicted heat and mass transfer rates can amplify deviations in the state.
The largest errors are observed for the thermal boundary layer thickness, for which the majority of reconstructions remain below $50\%$, but for some points can reach errors close to $140\%$. These large relative errors are amplified whenever the reconstructed layer thickness approaches zero, since the relative-error denominator becomes small; part of the apparent error therefore reflects this metric sensitivity rather than modeling difficulty alone.

Overall, the results obtained by training on the full experimental database confirm the trends observed throughout this study. The in-sample reconstruction errors remain low across all scenarios, and the training loss converges to values comparable to those achieved on the smaller cross-validation folds, suggesting that the optimization is not adversely affected by the increased size or diversity of the dataset.



\section{Conclusions}\label{sec:conclusions}

This work presented a data-driven closure modeling framework for a ground-based upright cylindrical cryogenic tank operating with liquid nitrogen. The experimental database comprised measurements obtained across a wide range of operating scenarios, including self-pressurization, active-pressurization, relaxation, venting, and sloshing.

The framework combines a zero-dimensional thermo-hydraulic model of the cryogenic tank with a set of closure parameters inferred directly from experimental observations. These parameters include eight Nusselt numbers describing the heat transfer interactions between vapor-liquid and fluid-solid nodes, two additional parameters characterizing the evolution of the liquid-side thermal boundary layer, and a pressurant vapor fraction accounting for condensation effects within the injection line. These terms are modeled using artificial neural networks, whose inputs consist of dimensionless quantities derived from the tank's thermo-hydraulic state and external inputs applied to the system.

The main objective of this work was to assess the feasibility of deploying such a physics-integrated data-driven framework to the problem of cryogenic liquid storage, while identifying its limitations and quantifying its predictive capability. To this end, a $K$-fold cross-validation analysis was first performed. This allowed for assessing the framework's ability to learn the closure terms from a limited experimental database and to characterize the sensitivity of the learned model to the amount and distribution of training data. Building on these results, a final model was then trained on the complete experimental database, acting as a digital twin of the tank. Its reconstruction capability was evaluated across the full range of operating scenarios.

The cross-validation analysis showed that the training performance of the framework is essentially insensitive to the amount of available data, with all values of $K \in \{2,3,4,5\}$ converging to training losses that differ only slightly relative to the observed fold-to-fold variability. This indicates that the closure parametrization is sufficiently general to accommodate additional training data without any loss of fitting capability. Across $K=2$-$5$, predictive accuracy remained comparable: the lowest mean overall NRMSE occurred at $K=4$, while $K=5$ exhibited the largest fold-to-fold spread, indicating sensitivity to the composition of the smaller held-out sets. This effect was attributed to the limited size and heterogeneity of the experimental database. Scenario-specific NRMSE results further showed that self-pressurization, relaxation, and venting dynamics are consistently well captured regardless of the data partition, whereas active-pressurization and, in particular, sloshing scenarios exhibit greater sensitivity to the specific train-validation split. Representative out-of-fold predictions further confirmed that the framework generalizes well to unseen conditions, while the inferred closure parameters evolved consistently with the underlying physical mechanisms.

Building on these findings, the model trained on the complete experimental database achieved low in-sample reconstruction errors across all state variables and operating scenarios, with a training loss consistent with the values obtained during cross-validation. This confirms that the proposed closure parametrization can incorporate the full diversity of the database without compromising its fitting capability. Component-wise analysis showed that the fluid and solid temperatures, as well as the pressure and vapor mass, are reconstructed with the lowest errors, while the thermal boundary layer thickness consistently exhibited the largest deviations, reflecting the inherent difficulty of representing this geometric, spatially-averaged quantity, particularly under rapidly changing operating conditions. 

Several research directions can be explored from the present contribution. First, extending the framework to incorporate experimental data from other tank configurations, including different fluids, geometries, and scales, would provide a more rigorous assessment of the generalizability and predictive capability of the learned closure laws. Second, while the present study focused on establishing the feasibility and reliability of the framework and on quantifying its out-of-fold predictive performance, a systematic benchmarking and hyperparameter tuning could be performed to identify more efficient network architectures. Finally, the current formulation relies on a hard switch between operating scenarios. Incorporating fuzzy-logic-based or soft-switching strategies could allow the framework to more naturally handle transitional or overlapping regimes.
Nevertheless, the presented results demonstrate that the proposed framework already provides a reliable and physically consistent tool for modeling cryogenic tank dynamics, offering a promising foundation for future developments.

\section*{Acknowledgments}
This work has received financial support from various sources. The experimental setups at VKI were developed in the framework of a project with the European Space Agency (ESA) under contract number 4000129315/19/NL/MG. The views expressed herein can in no way be taken to reflect the official opinion of the European Space Agency. P. Marques was supported by a FRIA grant (Ref. FC47297) from the ‘Fonds de la Recherche Scientifique (F.R.S.-FNRS)’. M. A. Mendez is supported by the European Research Council (ERC, grant agreement No 101165479 RE-TWIST StG). Views and opinions expressed are however those of the authors only and do not necessarily reflect those of the European Union or the European Research Council. Neither the European Union nor the granting authority can be held responsible for them.

\section*{Declaration of generative AI and AI-assisted technologies in the manuscript preparation process}

During the preparation of this work, the authors used Claude (Anthropic) and ChatGPT (OpenAI) in order to improve the language and readability of the manuscript. After using this tool/service, the authors reviewed and edited the content as needed and take full responsibility for the content of the published article.

\bibliographystyle{elsarticle-num-names}
\bibliography{bibliography}



\appendix

\section{Component-wise out-of-fold performance for $K=5$}
\label{app:component_out_of_fold_K5}

\begin{figure*}[!htb]
    \centering
    \includegraphics[width=\linewidth]{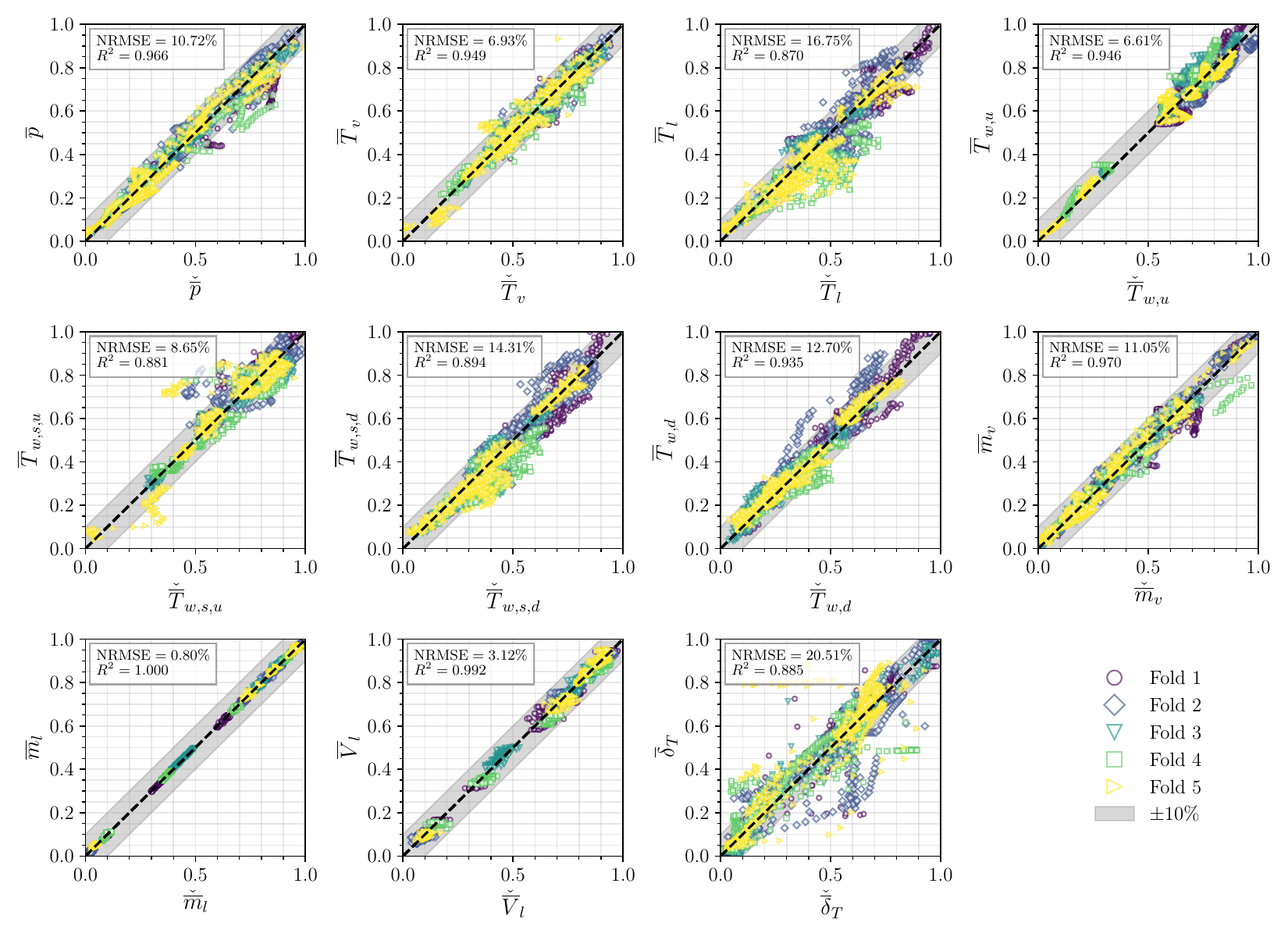}
    \caption{
    Out-of-fold predicted state $\bm{s}=[p_v\,, T_v\,, T_l\,, T_{w_u}\,, T_{w_{s,u}}\,, T_{w_{s,d}}\,, T_{w_d}\,, m_v\,, m_l\,, V_l\,, \delta_T]$ and reference state $\check{\bm{s}}$ for $K=5$ cross-validation folds. The predictions from each fold are shown with a distinct marker and color. The black dashed line denotes the $x=y$ reference, and the shaded band indicates a deviation of $\pm 0.1$ in min-max-normalized coordinates (10\% of the normalization range). For each state component, the normalized root-mean-square error (NRMSE) and coefficient of determination ($R^2$), averaged across all folds, are reported in the corresponding panel. All variables are shown normalized to the range $[0,1]$.
    }
    \label{fig:state_out_of_fold}
\end{figure*}

\begin{figure*}[!htb]
    \centering
    \includegraphics[width=1\linewidth]{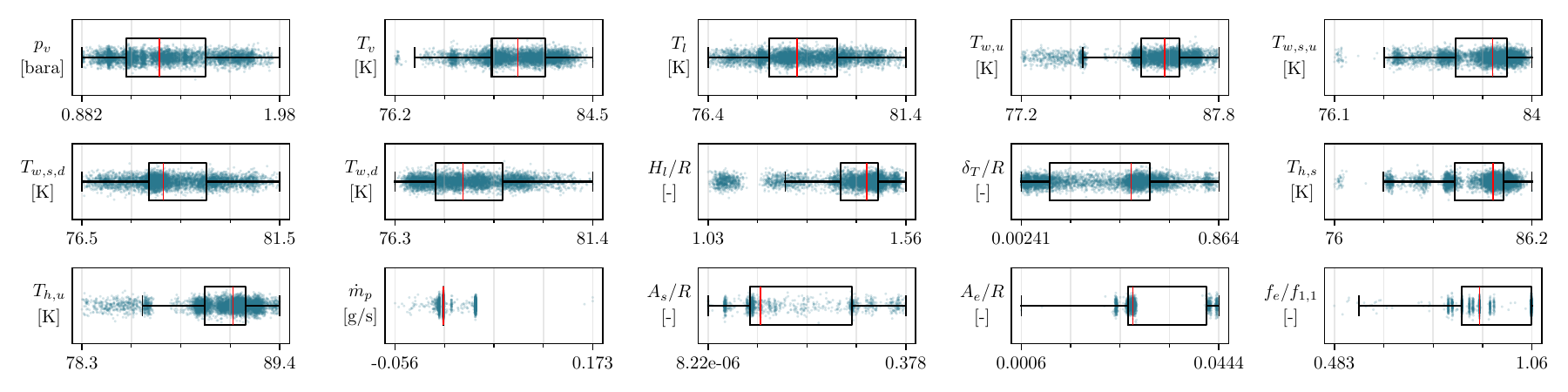}
    \caption{
    Boxplots with scatter distributions showing the spread of system states and exogenous inputs across the experimental database. State variables include pressure ($p_v$), node temperatures, scaled liquid fill-ratio ($H_l/R$), and scaled thermal boundary layer thickness ($\delta_T/R$). Exogenous inputs include heater temperatures ($T_{h,u}$ and $T_{h,s}$), mass flow rate ($\dot{m}_p$), scaled sloshing excitation amplitude ($A_e/R$) and frequency ($f_e/f_{1,1}$), and scaled measured wave amplitude ($A_s/R$). Red vertical lines denote medians.
    }
    \label{fig:variable_ranges}
\end{figure*}

The component-wise prediction accuracy is further examined in Figure \ref{fig:state_out_of_fold}, which presents the out-of-fold predictions for each state component for the $K=5$ cross-validation configuration. The predicted values from each fold are compared against the corresponding reference states, with the deviation from the reference data through the NRMSE and coefficient of determination ($R^2$) for each component. 
Each state component is normalized independently using min-max scaling, such that its values are mapped between 0 and 1.

Overall, most prediction points remain within $\pm 10\%$ of the reference data, confirming the framework's good generalization capability. However, the spread between folds is not uniform across the state variables, with some quantities exhibiting significantly larger differences depending on the validation subset. This behavior is consistent with the trends observed in the global NRMSE analysis, where increasing $K$ resulted in increased sensitivity to the specific data partition.

The smallest fold-to-fold variations are observed for the pressure, vapor temperature, and mass states, which remain closely clustered around the reference values for all folds. This indicates that these quantities are robustly captured independently of the specific train-validation split. Conversely, larger differences between folds are observed for the liquid and wall temperature states, and the thermal boundary-layer thickness. These variables are more sensitive to the representation of the underlying thermal conditions within each fold, resulting in larger variations in the prediction error.
The thermal boundary-layer thickness presents the largest dispersion among folds, with the associated prediction error exceeding that of the remaining state variables. 

\section{Experimental database for ground-based cryogenic model}
\label{app:database}

The experiments span a wide range of conditions, summarized in Fig.~\ref{fig:variable_ranges}, covering the limit operating conditions of the experimental setup described in Section~\ref{sec:experimental_setup}.
The database includes 44 static (self-pressurization or relaxation) scenarios, in which the system is left undisturbed, and the main drivers for the system evolution are the heating elements and the residual heat ingress from the cryostat insulation. The upper heater temperature was varied in the range $T_{h,u}\in[78.3, 89.4]$~K, while for the lateral heater $T_{h,s}\in[76.0, 86.2]$~K, thereby producing distinct system responses to these triggers.
The database includes 31 pressurization events, with mass flow rates from 0.01 (T32) to 0.17~g/s (T34), reaching a maximum system pressure of 1.98~bara in T47. Upstream conditions were kept nearly constant. The main variation across cases is the mass flow rate, which produces differences in the condensed vapor mass and thermal stratification rates, as investigated in a previous contribution~\cite{marques_experimental_2026}.

Venting is the most limited scenario in the database, with only five recorded time series, i.e., T12-T16. Each spans around 15~minutes, combining cases of steady venting (T14-T16) with sharp transitions from pressurized to atmospheric conditions (T12 and T13). The latter produce fast transient responses in the tank, and are the only cases where the liquid becomes temporarily superheated as pressure drops to ambient.
The database contains 40 sloshing experiments, in which the pressurized ullage exchanges heat and mass with the subcooled liquid. Excitation amplitude and frequency were varied to produce planar (e.g., T00-T02, T07, T08), chaotic (e.g., T03-T06, T09-T11), and swirl sloshing (e.g., T20, T22-T36). The considered sloshing experiments systematically vary the fill-level ratio $H_l/R$, the thermal stratification $\delta_T/R$, and the initial vapor pressure prior to sloshing $p_v$. The tank's pressurized state prior to sloshing was achieved through self-pressurization (e.g., T20, T21, T28, T31) and active-pressurization. Two isothermal, planar sloshing cases (T37, T38) were also included to isolate dynamic effects in the absence of heat and mass transfer.

\begin{table*}[!htb]
\centering
\caption{
Overview of the experimental database. For each case, the initial state at the start of the time series is reported, including the vapor pressure $p_v$, vapor temperature $T_v$, liquid temperature $T_l$, upper-cover temperature $T_{w,u}$, upper and lower lateral-wall temperatures $T_{w,s,u}$ and $T_{w,s,d}$, bottom-cover temperature $T_{w,d}$, liquid fill-ratio $H_l/R$, and thermal boundary layer thickness $\delta_T$. 
For each case, the four rightmost columns list the exogenous inputs active during each scenario, together with the duration $\Delta t$ over which they are applied. Self-pressurization is characterized by the heater temperatures $\overline{T}_{h,s}$ and $\overline{T}_{h,u}$, active-pressurization is characterized by the mean pressurant mass flow rate $\overline{m}_p$, upstream pressure $\overline{p}_u$ and temperature $\overline{T}_u$, sloshing is characterized by the excitation and free-surface amplitudes normalized by the tank radius, $A_e/R$ and $A_s/R$, and the scale excitation frequency $f_e/f_{1,1}$.
}
\renewcommand{\arraystretch}{1.4}

\resizebox{\textwidth}{!}{
\begin{tabular}{c|ccccccccc|cccc}
\hline
Case & 
\makecell{$p_v$ \\ {[bara]}} & 
\makecell{$T_v$ \\ {[K]}} & 
\makecell{$T_l$ \\ {[K]}} & 
\makecell{$T_{w,u}$ \\ {[K]}} & 
\makecell{$T_{w,s,u}$ \\ {[K]}} & 
\makecell{$T_{w,s,d}$ \\ {[K]}} & 
\makecell{$T_{w,d}$ \\ {[K]}} & 
\makecell{$H_l/R$ \\ {[-]}} & 
\makecell{$\delta_T/R$ \\ {[-]}} & 
Self-press & Active-press & Venting & Sloshing \\ \hline
T00
& 1.25 & 80.4 & 78.9 & 81.9 & 81.7 & 79.0 & 78.8 & 1.39 & 0.28 &
\makecell{$\Delta t = 750.3$~s \\ {$\overline{T}_{h,s}=83.1$~K} \\ {{$\overline{T}_{h,u}=84.3$~K}}} & 
\makecell{$\Delta t = 416.1$~s \\ {$\overline{m}_{p}=0.04$~g/s} \\ {{$\overline{p}_{u}=3.87$~bara}} \\ {{$\overline{T}_{u}=293.13$~K}} } & 
$-$ &
\makecell{$\Delta t = 119.5$~s \\ {$A_e/R=0.022$} \\ {{$A_s/R=0.050$}} \\ {{$f_e/f_{1,1}=0.82$}} }
\\ \hline
T01
& 1.17 & 80.8 & 78.6 & 83.4 & 81.9 & 78.9 & 78.6 & 1.40 & 0.12 &
\makecell{$\Delta t = 365.9$~s \\ {$\overline{T}_{h,s}=83.7$~K} \\ {{$\overline{T}_{h,u}=86.1$~K}}} & 
\makecell{$\Delta t = 85.1$~s \\ {$\overline{m}_{p}=0.03$~g/s} \\ {{$\overline{p}_{u}=3.79$~bara}} \\ {{$\overline{T}_{u}=293.13$~K}} } & 
$-$ &
\makecell{$\Delta t = 119.9$~s \\ {$A_e/R=0.022$} \\ {{$A_s/R=0.050$}} \\ {{$f_e/f_{1,1}=0.82$}} }
\\ \hline
T02
& 1.22 & 81.1 & 78.8 & 83.0 & 82.0 & 79.1 & 79.1 & 1.37 & 0.17 &
\makecell{$\Delta t = 372.4$~s \\ {$\overline{T}_{h,s}=83.6$~K} \\ {{$\overline{T}_{h,u}=85.6$~K}}} & 
\makecell{$\Delta t = 128.0$~s \\ {$\overline{m}_{p}=0.04$~g/s} \\ {{$\overline{p}_{u}=3.75$~bara}} \\ {{$\overline{T}_{u}=293.14$~K}} } & 
$-$ &
\makecell{$\Delta t = 119.6$~s \\ {$A_e/R=0.022$} \\ {{$A_s/R=0.059$}} \\ {{$f_e/f_{1,1}=0.82$}} }
\\ \hline
T03
& 1.03 & 80.1 & 77.3 & 83.2 & 81.7 & 77.3 & 77.5 & 1.39 & 0.17 &
\makecell{$\Delta t = 371.7$~s \\ {$\overline{T}_{h,s}=83.8$~K} \\ {{$\overline{T}_{h,u}=86.1$~K}}} & 
\makecell{$\Delta t = 105.4$~s \\ {$\overline{m}_{p}=0.03$~g/s} \\ {{$\overline{p}_{u}=3.82$~bara}} \\ {{$\overline{T}_{u}=293.15$~K}} } & 
$-$ &
\makecell{$\Delta t = 119.7$~s \\ {$A_e/R=0.025$} \\ {{$A_s/R=0.254$}} \\ {{$f_e/f_{1,1}=0.94$}} }
\\ \hline
T04
& 1.14 & 80.4 & 78.3 & 84.0 & 82.2 & 78.3 & 78.2 & 1.41 & 0.16 &
\makecell{$\Delta t = 340.3$~s \\ {$\overline{T}_{h,s}=84.0$~K} \\ {{$\overline{T}_{h,u}=87.1$~K}}} & 
\makecell{$\Delta t = 91.4$~s \\ {$\overline{m}_{p}=0.03$~g/s} \\ {{$\overline{p}_{u}=3.85$~bara}} \\ {{$\overline{T}_{u}=293.14$~K}} } & 
$-$ &
\makecell{$\Delta t = 119.8$~s \\ {$A_e/R=0.025$} \\ {{$A_s/R=0.274$}} \\ {{$f_e/f_{1,1}=0.94$}} }
\\ \hline
T05
& 1.13 & 80.6 & 78.2 & 84.5 & 82.1 & 78.3 & 78.3 & 1.35 & 0.13 &
\makecell{$\Delta t = 360.8$~s \\ {$\overline{T}_{h,s}=84.2$~K} \\ {{$\overline{T}_{h,u}=87.3$~K}}} & 
\makecell{$\Delta t = 86.4$~s \\ {$\overline{m}_{p}=0.03$~g/s} \\ {{$\overline{p}_{u}=3.79$~bara}} \\ {{$\overline{T}_{u}=293.21$~K}} } & 
$-$ &
\makecell{$\Delta t = 119.8$~s \\ {$A_e/R=0.043$} \\ {{$A_s/R=0.254$}} \\ {{$f_e/f_{1,1}=0.88$}} }
\\ \hline
T06
& 1.15 & 80.8 & 78.3 & 84.2 & 82.6 & 78.6 & 78.4 & 1.34 & 0.13 &
\makecell{$\Delta t = 359.1$~s \\ {$\overline{T}_{h,s}=84.2$~K} \\ {{$\overline{T}_{h,u}=87.1$~K}}} & 
\makecell{$\Delta t = 85.4$~s \\ {$\overline{m}_{p}=0.03$~g/s} \\ {{$\overline{p}_{u}=3.81$~bara}} \\ {{$\overline{T}_{u}=293.16$~K}} } & 
$-$ &
\makecell{$\Delta t = 119.8$~s \\ {$A_e/R=0.043$} \\ {{$A_s/R=0.295$}} \\ {{$f_e/f_{1,1}=0.88$}} }
\\ \hline
T07
& 1.28 & 81.6 & 79.5 & 85.0 & 82.9 & 80.1 & 79.7 & 1.06 & 0.04 &
\makecell{$\Delta t = 337.0$~s \\ {$\overline{T}_{h,s}=84.0$~K} \\ {{$\overline{T}_{h,u}=87.0$~K}}} & 
\makecell{$\Delta t = 66.9$~s \\ {$\overline{m}_{p}=0.03$~g/s} \\ {{$\overline{p}_{u}=3.72$~bara}} \\ {{$\overline{T}_{u}=293.18$~K}} } & 
$-$ &
\makecell{$\Delta t = 119.7$~s \\ {$A_e/R=0.022$} \\ {{$A_s/R=0.050$}} \\ {{$f_e/f_{1,1}=0.83$}} }
\\ \hline
T08
& 1.23 & 81.1 & 79.3 & 85.3 & 83.2 & 79.9 & 79.4 & 1.05 & 0.03 &
\makecell{$\Delta t = 318.3$~s \\ {$\overline{T}_{h,s}=84.2$~K} \\ {{$\overline{T}_{h,u}=87.3$~K}}} & 
\makecell{$\Delta t = 71.1$~s \\ {$\overline{m}_{p}=0.03$~g/s} \\ {{$\overline{p}_{u}=3.70$~bara}} \\ {{$\overline{T}_{u}=293.11$~K}} } & 
$-$ &
\makecell{$\Delta t = 119.7$~s \\ {$A_e/R=0.022$} \\ {{$A_s/R=0.046$}} \\ {{$f_e/f_{1,1}=0.83$}} }
\\ \hline
T09
& 1.17 & 80.9 & 78.6 & 85.5 & 82.6 & 79.1 & 78.8 & 1.08 & 0.10 &
\makecell{$\Delta t = 372.3$~s \\ {$\overline{T}_{h,s}=84.2$~K} \\ {{$\overline{T}_{h,u}=87.1$~K}}} & 
\makecell{$\Delta t = 84.1$~s \\ {$\overline{m}_{p}=0.03$~g/s} \\ {{$\overline{p}_{u}=3.77$~bara}} \\ {{$\overline{T}_{u}=293.18$~K}} } & 
$-$ &
\makecell{$\Delta t = 119.5$~s \\ {$A_e/R=0.025$} \\ {{$A_s/R=0.261$}} \\ {{$f_e/f_{1,1}=0.95$}} }
\\ \hline
T10
& 0.96 & 79.8 & 76.9 & 84.3 & 82.1 & 77.3 & 76.8 & 1.07 & 0.22 &
\makecell{$\Delta t = 402.7$~s \\ {$\overline{T}_{h,s}=84.0$~K} \\ {{$\overline{T}_{h,u}=86.7$~K}}} & 
\makecell{$\Delta t = 118.2$~s \\ {$\overline{m}_{p}=0.03$~g/s} \\ {{$\overline{p}_{u}=3.81$~bara}} \\ {{$\overline{T}_{u}=293.09$~K}} } & 
$-$ &
\makecell{$\Delta t = 119.9$~s \\ {$A_e/R=0.043$} \\ {{$A_s/R=0.209$}} \\ {{$f_e/f_{1,1}=0.89$}} }
\\ \hline
T11
& 1.26 & 81.2 & 79.2 & 84.4 & 82.6 & 79.6 & 79.3 & 1.06 & 0.20 &
\makecell{$\Delta t = 348.1$~s \\ {$\overline{T}_{h,s}=83.9$~K} \\ {{$\overline{T}_{h,u}=86.6$~K}}} & 
\makecell{$\Delta t = 73.2$~s \\ {$\overline{m}_{p}=0.03$~g/s} \\ {{$\overline{p}_{u}=3.74$~bara}} \\ {{$\overline{T}_{u}=293.17$~K}} } & 
$-$ &
\makecell{$\Delta t = 119.9$~s \\ {$A_e/R=0.043$} \\ {{$A_s/R=0.252$}} \\ {{$f_e/f_{1,1}=0.89$}} }
\\ \hline
\end{tabular}
}
\label{app:database_training}
\end{table*}

\begin{table*}[!htb]
\ContinuedFloat
\centering
\caption{(Continued)
}
\renewcommand{\arraystretch}{1.4}

\resizebox{\textwidth}{!}{
\begin{tabular}{c|ccccccccc|cccc}
\hline
Case & 
\makecell{$p_v$ \\ {[bara]}} & 
\makecell{$T_v$ \\ {[K]}} & 
\makecell{$T_l$ \\ {[K]}} & 
\makecell{$T_{w,u}$ \\ {[K]}} & 
\makecell{$T_{w,s,u}$ \\ {[K]}} & 
\makecell{$T_{w,s,d}$ \\ {[K]}} & 
\makecell{$T_{w,d}$ \\ {[K]}} & 
\makecell{$H_l/R$ \\ {[-]}} & 
\makecell{$\delta_T/R$ \\ {[-]}} & 
Self-press & Active-press & Venting & Sloshing \\ \hline
T12
& 1.57 & 82.4 & 81.0 & 85.3 & 83.1 & 80.9 & 81.0 & 1.36 & 0.42 &
\makecell{$\Delta t = 46.3$~s \\ {$\overline{T}_{h,s}=84.0$~K} \\ {{$\overline{T}_{h,u}=87.6$~K}}} & 
$-$ &
\makecell{$\Delta t = 925.2$~s \\ {$\overline{m}_{p}=-0.02$~g/s} } & 
$-$
\\ \hline
T13
& 1.04 & 80.5 & 77.9 & 84.8 & 82.4 & 78.2 & 78.1 & 1.30 & 0.01 &
$-$ &
$-$ &
\makecell{$\Delta t = 925.2$~s \\ {$\overline{m}_{p}=-0.01$~g/s} } & 
$-$
\\ \hline
T14
& 1.04 & 79.7 & 77.7 & 84.6 & 82.3 & 78.2 & 78.1 & 1.27 & 0.05 &
$-$ &
$-$ &
\makecell{$\Delta t = 925.1$~s \\ {$\overline{m}_{p}=-0.00$~g/s} } & 
$-$
\\ \hline
T15
& 1.03 & 80.7 & 77.7 & 85.1 & 82.4 & 78.4 & 78.0 & 1.22 & 0.03 &
$-$ &
$-$ &
\makecell{$\Delta t = 925.1$~s \\ {$\overline{m}_{p}=-0.00$~g/s} } & 
$-$
\\ \hline
T16
& 1.03 & 80.6 & 77.7 & 85.8 & 82.3 & 78.3 & 78.0 & 1.21 & 0.00 &
$-$ &
$-$ &
\makecell{$\Delta t = 925.1$~s \\ {$\overline{m}_{p}=-0.00$~g/s} } & 
$-$
\\ \hline
T17
& 0.89 & 76.3 & 76.5 & 77.4 & 76.4 & 76.8 & 76.8 & 1.46 & 0.03 &
\makecell{$\Delta t = 749.9$~s \\ {$\overline{T}_{h,s}=78.5$~K} \\ {{$\overline{T}_{h,u}=79.1$~K}}} & 
$-$ &
$-$ &
$-$
\\ \hline
T18
& 0.97 & 77.8 & 76.9 & 78.3 & 79.0 & 76.9 & 76.9 & 1.46 & 0.19 &
\makecell{$\Delta t = 749.9$~s \\ {$\overline{T}_{h,s}=80.4$~K} \\ {{$\overline{T}_{h,u}=80.3$~K}}} & 
$-$ &
$-$ &
$-$
\\ \hline
T19
& 1.07 & 78.5 & 77.5 & 79.3 & 79.4 & 77.6 & 77.4 & 1.46 & 0.46 &
\makecell{$\Delta t = 749.9$~s \\ {$\overline{T}_{h,s}=80.7$~K} \\ {{$\overline{T}_{h,u}=81.1$~K}}} & 
$-$ &
$-$ &
$-$
\\ \hline
T20
& 1.03 & 79.3 & 77.0 & 83.1 & 80.2 & 76.9 & 76.8 & 1.48 & 0.55 &
\makecell{$\Delta t = 1779.9$~s \\ {$\overline{T}_{h,s}=81.8$~K} \\ {{$\overline{T}_{h,u}=85.1$~K}}} & 
$-$ &
$-$ &
\makecell{$\Delta t = 118.8$~s \\ {$A_e/R=0.026$} \\ {{$A_s/R=0.392$}} \\ {{$f_e/f_{1,1}=1.06$}} }
\\ \hline
T21
& 1.09 & 79.5 & 77.4 & 83.5 & 80.3 & 77.6 & 77.2 & 1.42 & 0.54 &
\makecell{$\Delta t = 709.9$~s \\ {$\overline{T}_{h,s}=81.8$~K} \\ {{$\overline{T}_{h,u}=85.0$~K}}} & 
$-$ &
$-$ &
\makecell{$\Delta t = 119.6$~s \\ {$A_e/R=0.025$} \\ {{$A_s/R=0.160$}} \\ {{$f_e/f_{1,1}=0.91$}} }
\\ \hline
T22
& 1.06 & 79.4 & 77.6 & 83.2 & 80.3 & 77.6 & 77.6 & 1.41 & 0.14 &
\makecell{$\Delta t = 659.9$~s \\ {$\overline{T}_{h,s}=81.8$~K} \\ {{$\overline{T}_{h,u}=84.8$~K}}} & 
\makecell{$\Delta t = 19.3$~s \\ {$\overline{m}_{p}=0.02$~g/s} \\ {{$\overline{p}_{u}=1.51$~bara}} \\ {{$\overline{T}_{u}=295.17$~K}} } & 
$-$ &
\makecell{$\Delta t = 112.8$~s \\ {$A_e/R=0.026$} \\ {{$A_s/R=0.289$}} \\ {{$f_e/f_{1,1}=1.06$}} }
\\ \hline
T23
& 1.06 & 79.3 & 77.4 & 83.3 & 80.2 & 77.5 & 77.3 & 1.47 & 0.34 &
\makecell{$\Delta t = 666.9$~s \\ {$\overline{T}_{h,s}=81.8$~K} \\ {{$\overline{T}_{h,u}=84.9$~K}}} & 
\makecell{$\Delta t = 21.1$~s \\ {$\overline{m}_{p}=0.02$~g/s} \\ {{$\overline{p}_{u}=1.50$~bara}} \\ {{$\overline{T}_{u}=295.47$~K}} } & 
$-$ &
\makecell{$\Delta t = 119.0$~s \\ {$A_e/R=0.026$} \\ {{$A_s/R=0.332$}} \\ {{$f_e/f_{1,1}=1.06$}} }
\\ \hline
T24
& 1.01 & 79.2 & 77.3 & 83.6 & 80.1 & 77.4 & 77.3 & 1.48 & 0.10 &
\makecell{$\Delta t = 700.9$~s \\ {$\overline{T}_{h,s}=82.0$~K} \\ {{$\overline{T}_{h,u}=85.2$~K}}} & 
\makecell{$\Delta t = 47.9$~s \\ {$\overline{m}_{p}=0.03$~g/s} \\ {{$\overline{p}_{u}=1.91$~bara}} \\ {{$\overline{T}_{u}=295.76$~K}} } & 
$-$ &
\makecell{$\Delta t = 119.8$~s \\ {$A_e/R=0.026$} \\ {{$A_s/R=0.289$}} \\ {{$f_e/f_{1,1}=1.06$}} }
\\ \hline
T25
& 0.98 & 79.2 & 77.0 & 83.2 & 79.8 & 77.0 & 77.0 & 1.49 & 0.09 &
\makecell{$\Delta t = 529.9$~s \\ {$\overline{T}_{h,s}=82.0$~K} \\ {{$\overline{T}_{h,u}=84.8$~K}}} & 
\makecell{$\Delta t = 94.6$~s \\ {$\overline{m}_{p}=0.03$~g/s} \\ {{$\overline{p}_{u}=2.06$~bara}} \\ {{$\overline{T}_{u}=296.07$~K}} } & 
$-$ &
\makecell{$\Delta t = 119.8$~s \\ {$A_e/R=0.026$} \\ {{$A_s/R=0.392$}} \\ {{$f_e/f_{1,1}=1.06$}} }
\\ \hline
T26
& 1.02 & 79.3 & 77.4 & 83.5 & 80.0 & 77.4 & 77.3 & 1.49 & 0.10 &
\makecell{$\Delta t = 536.9$~s \\ {$\overline{T}_{h,s}=82.1$~K} \\ {{$\overline{T}_{h,u}=85.0$~K}}} & 
\makecell{$\Delta t = 65.4$~s \\ {$\overline{m}_{p}=0.03$~g/s} \\ {{$\overline{p}_{u}=2.06$~bara}} \\ {{$\overline{T}_{u}=296.04$~K}} } & 
$-$ &
\makecell{$\Delta t = 119.6$~s \\ {$A_e/R=0.026$} \\ {{$A_s/R=0.332$}} \\ {{$f_e/f_{1,1}=1.06$}} }
\\ \hline
T27
& 1.02 & 79.7 & 77.3 & 83.4 & 80.2 & 77.3 & 77.2 & 1.49 & 0.11 &
\makecell{$\Delta t = 904.9$~s \\ {$\overline{T}_{h,s}=82.2$~K} \\ {{$\overline{T}_{h,u}=85.3$~K}}} & 
\makecell{$\Delta t = 155.5$~s \\ {$\overline{m}_{p}=0.04$~g/s} \\ {{$\overline{p}_{u}=2.70$~bara}} \\ {{$\overline{T}_{u}=296.33$~K}} } & 
$-$ &
\makecell{$\Delta t = 119.2$~s \\ {$A_e/R=0.026$} \\ {{$A_s/R=0.392$}} \\ {{$f_e/f_{1,1}=1.06$}} }
\\ \hline
T28
& 1.27 & 80.7 & 79.2 & 84.6 & 80.9 & 79.2 & 79.3 & 1.49 & 0.19 &
\makecell{$\Delta t = 814.9$~s \\ {$\overline{T}_{h,s}=84.2$~K} \\ {{$\overline{T}_{h,u}=86.3$~K}}} & 
$-$ &
$-$ &
\makecell{$\Delta t = 118.9$~s \\ {$A_e/R=0.026$} \\ {{$A_s/R=0.289$}} \\ {{$f_e/f_{1,1}=1.06$}} }
\\ \hline
\end{tabular}
}

\end{table*}

\begin{table*}[!htb]
\ContinuedFloat
\centering
\caption{(Continued)
}
\renewcommand{\arraystretch}{1.4}

\resizebox{\textwidth}{!}{
\begin{tabular}{c|ccccccccc|cccc}
\hline
Case & 
\makecell{$p_v$ \\ {[bara]}} & 
\makecell{$T_v$ \\ {[K]}} & 
\makecell{$T_l$ \\ {[K]}} & 
\makecell{$T_{w,u}$ \\ {[K]}} & 
\makecell{$T_{w,s,u}$ \\ {[K]}} & 
\makecell{$T_{w,s,d}$ \\ {[K]}} & 
\makecell{$T_{w,d}$ \\ {[K]}} & 
\makecell{$H_l/R$ \\ {[-]}} & 
\makecell{$\delta_T/R$ \\ {[-]}} & 
Self-press & Active-press & Venting & Sloshing \\ \hline
T29
& 1.02 & 80.1 & 77.1 & 85.0 & 81.7 & 77.6 & 77.2 & 1.47 & 0.11 &
\makecell{$\Delta t = 1557.9$~s \\ {$\overline{T}_{h,s}=84.5$~K} \\ {{$\overline{T}_{h,u}=86.2$~K}}} & 
\makecell{$\Delta t = 100.0$~s \\ {$\overline{m}_{p}=0.03$~g/s} \\ {{$\overline{p}_{u}=3.31$~bara}} \\ {{$\overline{T}_{u}=296.44$~K}} } & 
$-$ &
\makecell{$\Delta t = 118.2$~s \\ {$A_e/R=0.026$} \\ {{$A_s/R=0.329$}} \\ {{$f_e/f_{1,1}=1.06$}} }
\\ \hline
T30
& 0.97 & 80.3 & 77.0 & 84.9 & 81.5 & 77.1 & 77.1 & 1.45 & 0.08 &
\makecell{$\Delta t = 999.9$~s \\ {$\overline{T}_{h,s}=84.5$~K} \\ {{$\overline{T}_{h,u}=86.6$~K}}} & 
\makecell{$\Delta t = 101.7$~s \\ {$\overline{m}_{p}=0.03$~g/s} \\ {{$\overline{p}_{u}=3.20$~bara}} \\ {{$\overline{T}_{u}=296.59$~K}} } & 
$-$ &
\makecell{$\Delta t = 119.3$~s \\ {$A_e/R=0.026$} \\ {{$A_s/R=0.289$}} \\ {{$f_e/f_{1,1}=1.06$}} }
\\ \hline
T31
& 1.38 & 81.7 & 79.6 & 85.2 & 82.9 & 79.7 & 79.3 & 1.48 & 0.56 &
\makecell{$\Delta t = 664.9$~s \\ {$\overline{T}_{h,s}=84.5$~K} \\ {{$\overline{T}_{h,u}=86.8$~K}}} & 
$-$ &
$-$ &
\makecell{$\Delta t = 119.8$~s \\ {$A_e/R=0.026$} \\ {{$A_s/R=0.289$}} \\ {{$f_e/f_{1,1}=1.06$}} }
\\ \hline
T32
& 1.01 & 80.0 & 77.2 & 85.8 & 81.7 & 77.3 & 77.1 & 1.44 & 0.18 &
\makecell{$\Delta t = 1632.9$~s \\ {$\overline{T}_{h,s}=84.8$~K} \\ {{$\overline{T}_{h,u}=87.9$~K}}} & 
\makecell{$\Delta t = 498.4$~s \\ {$\overline{m}_{p}=0.01$~g/s} \\ {{$\overline{p}_{u}=3.12$~bara}} \\ {{$\overline{T}_{u}=297.16$~K}} } & 
$-$ &
\makecell{$\Delta t = 118.5$~s \\ {$A_e/R=0.026$} \\ {{$A_s/R=0.332$}} \\ {{$f_e/f_{1,1}=1.06$}} }
\\ \hline
T33
& 1.54 & 82.7 & 80.5 & 86.7 & 83.5 & 80.5 & 80.1 & 1.47 & 0.62 &
\makecell{$\Delta t = 1138.9$~s \\ {$\overline{T}_{h,s}=85.0$~K} \\ {{$\overline{T}_{h,u}=88.5$~K}}} & 
$-$ &
$-$ &
\makecell{$\Delta t = 118.7$~s \\ {$A_e/R=0.026$} \\ {{$A_s/R=0.289$}} \\ {{$f_e/f_{1,1}=1.06$}} }
\\ \hline
T34
& 0.97 & 80.5 & 76.9 & 86.2 & 81.8 & 77.1 & 77.0 & 1.50 & 0.12 &
\makecell{$\Delta t = 582.9$~s \\ {$\overline{T}_{h,s}=84.9$~K} \\ {{$\overline{T}_{h,u}=88.1$~K}}} & 
\makecell{$\Delta t = 22.4$~s \\ {$\overline{m}_{p}=0.11$~g/s} \\ {{$\overline{p}_{u}=3.01$~bara}} \\ {{$\overline{T}_{u}=297.35$~K}} } & 
$-$ &
\makecell{$\Delta t = 119.6$~s \\ {$A_e/R=0.026$} \\ {{$A_s/R=0.392$}} \\ {{$f_e/f_{1,1}=1.06$}} }
\\ \hline
T35
& 1.45 & 82.3 & 79.7 & 85.5 & 83.1 & 79.6 & 79.1 & 1.54 & 0.62 &
\makecell{$\Delta t = 170.9$~s \\ {$\overline{T}_{h,s}=84.8$~K} \\ {{$\overline{T}_{h,u}=87.1$~K}}} & 
$-$ &
$-$ &
\makecell{$\Delta t = 120.0$~s \\ {$A_e/R=0.026$} \\ {{$A_s/R=0.289$}} \\ {{$f_e/f_{1,1}=1.06$}} }
\\ \hline
T36
& 1.34 & 82.2 & 79.3 & 86.5 & 82.7 & 79.4 & 78.9 & 1.51 & 0.54 &
\makecell{$\Delta t = 539.9$~s \\ {$\overline{T}_{h,s}=84.9$~K} \\ {{$\overline{T}_{h,u}=88.2$~K}}} & 
$-$ &
$-$ &
\makecell{$\Delta t = 119.3$~s \\ {$A_e/R=0.026$} \\ {{$A_s/R=0.392$}} \\ {{$f_e/f_{1,1}=1.06$}} }
\\ \hline
T37
& 1.14 & 78.5 & 78.4 & 80.4 & 78.3 & 78.2 & 78.3 & 1.47 & 0.06 &
\makecell{$\Delta t = 173.6$~s \\ {$\overline{T}_{h,s}=78.8$~K} \\ {{$\overline{T}_{h,u}=81.9$~K}}} & 
$-$ &
$-$ &
\makecell{$\Delta t = 120.0$~s \\ {$A_e/R=0.025$} \\ {{$A_s/R=0.138$}} \\ {{$f_e/f_{1,1}=0.91$}} }
\\ \hline
T38
& 1.16 & 78.7 & 78.4 & 80.6 & 78.7 & 78.4 & 78.5 & 1.46 & 0.06 &
\makecell{$\Delta t = 195.6$~s \\ {$\overline{T}_{h,s}=78.9$~K} \\ {{$\overline{T}_{h,u}=81.7$~K}}} & 
$-$ &
$-$ &
\makecell{$\Delta t = 119.9$~s \\ {$A_e/R=0.044$} \\ {{$A_s/R=0.112$}} \\ {{$f_e/f_{1,1}=0.86$}} }
\\ \hline
T39
& 0.96 & 79.9 & 76.8 & 84.5 & 81.8 & 77.0 & 76.7 & 1.40 & 0.13 &
\makecell{$\Delta t = 516.4$~s \\ {$\overline{T}_{h,s}=84.6$~K} \\ {{$\overline{T}_{h,u}=86.3$~K}}} & 
\makecell{$\Delta t = 125.3$~s \\ {$\overline{m}_{p}=0.03$~g/s} \\ {{$\overline{p}_{u}=2.97$~bara}} \\ {{$\overline{T}_{u}=297.05$~K}} } & 
$-$ &
\makecell{$\Delta t = 119.7$~s \\ {$A_e/R=0.026$} \\ {{$A_s/R=0.392$}} \\ {{$f_e/f_{1,1}=1.06$}} }
\\ \hline
T40
& 0.97 & 80.2 & 76.7 & 85.8 & 81.9 & 76.8 & 76.6 & 1.41 & 0.12 &
\makecell{$\Delta t = 449.0$~s \\ {$\overline{T}_{h,s}=85.0$~K} \\ {{$\overline{T}_{h,u}=87.6$~K}}} & 
\makecell{$\Delta t = 103.7$~s \\ {$\overline{m}_{p}=0.03$~g/s} \\ {{$\overline{p}_{u}=2.94$~bara}} \\ {{$\overline{T}_{u}=297.29$~K}} } & 
$-$ &
\makecell{$\Delta t = 119.8$~s \\ {$A_e/R=0.025$} \\ {{$A_s/R=0.138$}} \\ {{$f_e/f_{1,1}=0.91$}} }
\\ \hline
T41
& 0.97 & 80.3 & 76.8 & 85.2 & 81.8 & 76.8 & 76.8 & 1.42 & 0.15 &
\makecell{$\Delta t = 460.2$~s \\ {$\overline{T}_{h,s}=84.7$~K} \\ {{$\overline{T}_{h,u}=87.1$~K}}} & 
\makecell{$\Delta t = 101.7$~s \\ {$\overline{m}_{p}=0.03$~g/s} \\ {{$\overline{p}_{u}=2.90$~bara}} \\ {{$\overline{T}_{u}=297.11$~K}} } & 
$-$ &
\makecell{$\Delta t = 119.8$~s \\ {$A_e/R=0.025$} \\ {{$A_s/R=0.137$}} \\ {{$f_e/f_{1,1}=0.91$}} }
\\ \hline
T42
& 1.04 & 81.0 & 77.4 & 86.2 & 82.0 & 77.6 & 77.3 & 1.45 & 0.12 &
\makecell{$\Delta t = 453.4$~s \\ {$\overline{T}_{h,s}=85.1$~K} \\ {{$\overline{T}_{h,u}=88.1$~K}}} & 
\makecell{$\Delta t = 89.6$~s \\ {$\overline{m}_{p}=0.03$~g/s} \\ {{$\overline{p}_{u}=2.87$~bara}} \\ {{$\overline{T}_{u}=297.02$~K}} } & 
$-$ &
\makecell{$\Delta t = 119.9$~s \\ {$A_e/R=0.044$} \\ {{$A_s/R=0.102$}} \\ {{$f_e/f_{1,1}=0.86$}} }
\\ \hline
\end{tabular}
}

\end{table*}

\begin{table*}[!htb]
\ContinuedFloat
\centering
\caption{(Continued)
}
\renewcommand{\arraystretch}{1.4}

\resizebox{\textwidth}{!}{
\begin{tabular}{c|ccccccccc|cccc}
\hline
Case & 
\makecell{$p_v$ \\ {[bara]}} & 
\makecell{$T_v$ \\ {[K]}} & 
\makecell{$T_l$ \\ {[K]}} & 
\makecell{$T_{w,u}$ \\ {[K]}} & 
\makecell{$T_{w,s,u}$ \\ {[K]}} & 
\makecell{$T_{w,s,d}$ \\ {[K]}} & 
\makecell{$T_{w,d}$ \\ {[K]}} & 
\makecell{$H_l/R$ \\ {[-]}} & 
\makecell{$\delta_T/R$ \\ {[-]}} & 
Self-press & Active-press & Venting & Sloshing \\ \hline
T43
& 0.96 & 80.4 & 76.7 & 85.2 & 81.4 & 76.8 & 76.7 & 1.46 & 0.12 &
\makecell{$\Delta t = 466.5$~s \\ {$\overline{T}_{h,s}=84.8$~K} \\ {{$\overline{T}_{h,u}=87.0$~K}}} & 
\makecell{$\Delta t = 101.5$~s \\ {$\overline{m}_{p}=0.03$~g/s} \\ {{$\overline{p}_{u}=2.84$~bara}} \\ {{$\overline{T}_{u}=297.06$~K}} } & 
$-$ &
\makecell{$\Delta t = 119.4$~s \\ {$A_e/R=0.044$} \\ {{$A_s/R=0.102$}} \\ {{$f_e/f_{1,1}=0.86$}} }
\\ \hline
T44
& 0.92 & 79.2 & 76.5 & 86.2 & 81.3 & 76.7 & 76.4 & 1.47 & 0.10 &
\makecell{$\Delta t = 557.8$~s \\ {$\overline{T}_{h,s}=85.2$~K} \\ {{$\overline{T}_{h,u}=87.5$~K}}} & 
\makecell{$\Delta t = 98.6$~s \\ {$\overline{m}_{p}=0.03$~g/s} \\ {{$\overline{p}_{u}=2.80$~bara}} \\ {{$\overline{T}_{u}=296.96$~K}} } & 
$-$ &
\makecell{$\Delta t = 119.6$~s \\ {$A_e/R=0.026$} \\ {{$A_s/R=0.332$}} \\ {{$f_e/f_{1,1}=1.06$}} }
\\ \hline
T45
& 0.99 & 80.5 & 76.6 & 85.6 & 82.1 & 76.8 & 76.6 & 1.51 & 0.29 &
\makecell{$\Delta t = 461.1$~s \\ {$\overline{T}_{h,s}=85.0$~K} \\ {{$\overline{T}_{h,u}=87.5$~K}}} & 
\makecell{$\Delta t = 98.8$~s \\ {$\overline{m}_{p}=0.03$~g/s} \\ {{$\overline{p}_{u}=2.77$~bara}} \\ {{$\overline{T}_{u}=296.92$~K}} } & 
$-$ &
\makecell{$\Delta t = 119.6$~s \\ {$A_e/R=0.026$} \\ {{$A_s/R=0.289$}} \\ {{$f_e/f_{1,1}=1.06$}} }
\\ \hline
T46
& 0.96 & 80.7 & 76.8 & 86.3 & 82.0 & 76.9 & 76.6 & 1.41 & 0.12 &
\makecell{$\Delta t = 499.0$~s \\ {$\overline{T}_{h,s}=85.2$~K} \\ {{$\overline{T}_{h,u}=88.2$~K}}} & 
\makecell{$\Delta t = 91.6$~s \\ {$\overline{m}_{p}=0.03$~g/s} \\ {{$\overline{p}_{u}=2.74$~bara}} \\ {{$\overline{T}_{u}=296.84$~K}} } & 
$-$ &
\makecell{$\Delta t = 119.9$~s \\ {$A_e/R=0.025$} \\ {{$A_s/R=0.166$}} \\ {{$f_e/f_{1,1}=0.91$}} }
\\ \hline
T47
& 0.97 & 80.5 & 76.8 & 87.0 & 82.2 & 77.1 & 76.6 & 1.43 & 0.18 &
\makecell{$\Delta t = 478.6$~s \\ {$\overline{T}_{h,s}=85.4$~K} \\ {{$\overline{T}_{h,u}=89.2$~K}}} & 
\makecell{$\Delta t = 97.8$~s \\ {$\overline{m}_{p}=0.03$~g/s} \\ {{$\overline{p}_{u}=2.71$~bara}} \\ {{$\overline{T}_{u}=296.84$~K}} } & 
$-$ &
\makecell{$\Delta t = 119.4$~s \\ {$A_e/R=0.044$} \\ {{$A_s/R=0.116$}} \\ {{$f_e/f_{1,1}=0.86$}} }
\\ \hline
\end{tabular}
}

\end{table*}

\end{document}